\documentclass[journal=jacsat,manuscript=article,layout=twocolumn]{achemso}

\usepackage[version=3]{mhchem} 
\usepackage{amssymb,amsmath,mleftright,mathtools}
\usepackage{bm}
\usepackage{float}
\usepackage{graphicx}
\usepackage{subcaption}
\usepackage{physics}
\usepackage{algorithm}
\usepackage{algpseudocode}
\usepackage{color}
\usepackage{comment}

\SectionNumbersOn

\usepackage{acro}
\acsetup{barriers/use=true, barriers/reset=true}

\DeclareAcronym{CPU}{
short = {CPU},
long = {central processing unit}
}

\DeclareAcronym{DFT}{
short = {DFT},
long = {density functional theory}
}

\DeclareAcronym{GPU}{
short = {GPU},
long = {graphics processing unit}
}

\DeclareAcronym{KS}{
short = {KS},
long = {Kohn-Sham}
}

\DeclareAcronym{LDA}{
short = {LDA},
long = {Local density approximation}
}

\DeclareAcronym{SDFT}{
short = {SDFT},
long = {spin density functional theory}
}

\DeclareAcronym{TDDFT}{
short = {TDDFT},
long = {time-dependent density functional theory}
}

\DeclareAcronym{R-TDDFT}{
short = {R-TDDFT},
long = {relativistic time-dependent density functional theory}
}

\DeclareAcronym{PAW}{
short = {PAW},
long = {projected augmented wave}
}

\DeclareAcronym{XC}{
short = {XC},
long = {exchange correlation}
}

\DeclareAcronym{MPI}{
short = {MPI},
long = {message passing interface}
}

\newcommand{\matr}[1]{\mathit{#1}}

\newcommand{\llnl}{Quantum Simulations Group, Lawrence Livermore National Laboratory, Livermore, California 94550, USA}

\author{Jacopo Simoni}
\affiliation{Department of Materials Science and Engineering, University of Wisconsin-Madison, Madison, WI, 53706, USA}
\email{jsimoni@wisc.edu}
\author{Xavier Andrade}
\affiliation{\llnl}
\author{Wuzhang Fang}
\affiliation{Department of Materials Science and Engineering, University of Wisconsin-Madison, Madison, WI, 53706, USA}
\author{Andrew C. Grieder}
\affiliation{Department of Materials Science and Engineering, University of Wisconsin-Madison, Madison, WI, 53706, USA}
\author{Alfredo A. Correa}
\affiliation{\llnl}
\author{Tadashi Ogitsu}
\affiliation{\llnl}
\author{Yuan Ping}
\affiliation{Department of Materials Science and Engineering, University of Wisconsin-Madison, Madison, WI, 53706, USA}
\alsoaffiliation{Department of Physics, University of Wisconsin-Madison, Madison, WI, 53706, USA}
\alsoaffiliation{Department of Chemistry, University of Wisconsin-Madison, Madison, WI, 53706, USA}

\title{Spin non-Collinear Real-Time Time-Dependent Density-Functional Theory and Implementation in the Modern GPU-Accelerated \textsc{INQ} code}

\abbreviations{IR,NMR,UV}

\begin{document}





\begin{abstract}
  
  \Ac{TDDFT} describes the time evolution of quantum mechanical many-electron systems under the influence of external time-dependent electric and magnetic fields.
  \textsc{INQ} is especially designed to efficiently solve the real-time \ac{TDDFT} equations on \acp{GPU}, which aims to overcome the computational limitation of time and size scales of non-equilibrium quantum dynamics.
  In this work, we present an implementation of non-collinear \ac{TDDFT} for the \textsc{INQ} code to simulate spin dynamics in real time.
  We discuss the implementation of non-collinear magnetic effects, \ac{XC} magnetic fields, spin-orbit coupling, and the interaction between the electronic system and external magnetic fields, with plane-wave basis and poseudopotential approximations. We consider several prototypical examples of spin dynamics in magnetic clusters and solids after light excitation.
\end{abstract}

\acbarrier

\section{Introduction}

Magnetism is a fundamental interaction in nature that is responsible for the attractive and repulsive forces observed in magnetic materials.
Localized magnetic moments can interact and cause macroscopic magnetic ordering.
Ferromagnetism and anti-ferromagnetism are the simplest forms of such macroscopic ordering;
their origin is quantum mechanical and determined by the exchange interaction and the Pauli exclusion principle.
Magnetic anisotropy arises instead from the coupling between spin and spatial degrees of freedom in a material.
This is determined at the atomic level by the spin-orbit interaction.
A correct ab-initio description of the magnetic properties of materials must be able to capture these different interactions\cite{Sanvito2019}.

The success of \ac{DFT} in describing the magnetic properties of materials can be ascribed to its ability to account for exchange interactions, via the \ac{XC} potential, and magnetic anisotropy through the spin-orbit interaction\cite{Bihlmayer2018}.
Although \ac{DFT} is a formally exact approach to the calculation of the ground state of electronic systems, in practice, the exact \ac{XC} functional is unknown and approximations must be considered.
Local-density approximations have been shown to be quite successful in describing the equilibrium magnetic properties\cite{BROOKS20012059} and the slow adiabatic dynamics in solids\cite{PhysRevLett.75.729,PhysRevB.54.1019}.

The aim of \ac{TDDFT}, i.e. \ac{DFT}'s time-dependent counterpart, is to extend this success to linear-response properties and to the nonperturbative real-time domain\cite{Onida2002,Marques2004,Casida2012,JornetSomoza2015,Correa2018,Byun2020,Kononov2022,Xu2024}.
Under non-equilibrium conditions, the exchange interactions become dynamical, and the simple static picture of Heisenberg exchange breaks down.
To overcome this problem, electronic degrees of freedom should also evolve dynamically.

Out-of-equilibrium dynamics in magnetic systems is important for many different applications including
spintronics\cite{Igor2004} and orbitronics\cite{Gu2024,Jo2024}, magnonics \cite{magnonics} and quantum magnonics \cite{YUAN20221,ZHANG2023100044},
quantum computation\cite{10.1063/PT.3.4270},
skyrmion dynamics for magnetic storage technologies\cite{Nagaosa2013,GOBEL20211,Fert2017},
ultrafast magnetism\cite{UFM2018},
the chiral-induced spin selectivity (CISS) effect\cite{Bloom2024,doi:10.1126/science.adi9601},
and collinear spin and orbital Edelstein effect in solids\cite{Annika2024}.

Ultrafast magnetism refers to the sub-picosecond dynamics of magnetic systems under the application of intense laser pulses.
For example, the complete demagnetization of a \ce{Ni} film has been experimentally demonstrated\cite{PhysRevLett.76.4250}.
Further experiments have confirmed such findings and have shown the possibility of laser-induced spin reorientation and modification of the magnetic structure\cite{doi:10.1126/sciadv.1603117,kimel2005,PhysRevLett.93.197403}.
Femtomagnetism is a research field that focuses on the control of the magnetic order in materials using femtosecond laser pulses and has two main goals:
(i) a better understanding of the physics of light-induced spin dynamics on very short time scales;
(ii) control and manipulation of the spin degrees of freedom on sub-picosecond time scales.
The uncertainty on the underlying physical mechanism responsible for this phenomenon makes \ac{TDDFT} a particularly attractive approach to studying such phenomena.

Several implementations of real-time \ac{TDDFT} are already available\cite{RMG,NWChem,Baczewski2014,Draeger2017,Noda2019,Khne2020,Mortensen2024} and some of these can perform spin-dynamics. \textsc{Octopus}\cite{tancogne_dejean_octopus_2020,C5CP00351B} works on a real-space grid and can be used to perform real-time simulation of spin polarized systems.
\textsc{Elk} \cite{Elk_code} is a full-potential linearized augmented plane-waves (FP-LAPW) code working for crystalline solids that can also simulate real-time evolution of magnetic systems.
The \textsc{INQ} code\cite{INQ2021} is a plane-wave basis code, designed to provide a coherent and computationally efficient framework for the implementation of \ac{TDDFT} that can make use of large GPU-based supercomputers.
\textsc{INQ} is a very compact streamlined implementation of \ac{DFT} and \ac{TDDFT}, only 15,000 lines of code, that is easy to maintain and further develop.
Here we discuss the implementation of non-collinear spin \ac{DFT} and \ac{TDDFT} in the code, which provides an accurate and efficient tool to study various spin-related phenomena in and out of equilibrium.

In section~\ref{sec:theory} we discuss the theoretical background required for the non-collinear spin \ac{DFT} implementation.
In section~\ref{sec:implement} we present the details of the implementation.
In section~\ref{sec:res} we present results and a comparison with other codes by looking in particular at the dynamics under external electric and magnetic pulses.

\section{Theoretical Background} \label{sec:theory}

Before introducing the non-collinear version of spin \ac{DFT} in subsection \ref{sec:noncolSDFT} we briefly discuss the collinear version in \ref{sec:spinpol}.
\ac{TDDFT} and its non-collinear version are discussed in sections \ref{sec:tddft} and \ref{sec:nc-tddft}.
In the remaining part of the theory section, we analyze the approximations used to implement the non-collinear equations in \textsc{INQ}.

\subsection{Density Functional Theory for spin polarized systems} \label{sec:spinpol}

\ac{DFT} has been a widely successful theory for the calculation of electronic structures in both molecules and materials\cite{Kohn1996}.
The basic formalism of \ac{DFT} is given by the \ac{KS} set of equations\cite{PhysRev.140.A1133} (atomic units are used throughout)
\begin{equation}\label{eq:KSDFT}
    \left[-\frac{1}{2}\nabla^2 + v_\text{KS}[n]({\bf r})\right]\psi_{i}^\text{KS}({\bf r}) = \varepsilon_{i}^\text{KS}\psi_{i}^\text{KS}({\bf r}) ,
\end{equation}
where \(\varepsilon_{i}^\text{KS}\) and \(\psi_{i}^\text{KS}({\bf r})\) correspond to the \ac{KS} energy eigenvalues and eigenstates. The index \(i=(n,{\bf k})\) corresponds to a combination of the band index \(n\) and {\bf k}-point index.
\(v_\text{KS}({\bf r})\) is the \ac{KS} potential, a functional of the electron density given by the sum of three terms:
the external potential due to the surrounding ions,
the Hartree,
and the \ac{XC} potentials:
\begin{align} \label{eq:KSpot}
    &v_\text{KS}[n]({\bf r})
    = v_\text{ext}({\bf r}) + v_\text{H}[n]({\bf r}) + v_\text{xc}[n]({\bf r})\nonumber
    \\
    &=-\sum_{\rm a}\frac{Z_{\rm a}}{|{\bf r} - {\bf R}_{\rm a}|} + \int d{\bf r'}\frac{n({\bf r'})}{|{\bf r} - {\bf r'}|} + v_\text{xc}[n]({\bf r})\ .
\end{align}

Eq.~(\ref{eq:KSDFT}) and (\ref{eq:KSpot}) cannot be directly applied to magnetic systems.
\Ac{SDFT}\cite{vonBarth_1972} requires the extension of Eq.~(\ref{eq:KSDFT}) for spin polarized systems through the introduction of a two-component field variable \(n_{\alpha=0,1}({\bf r})\) that corresponds to the spin-up and spin-down densities.
This is the so-called \emph{spin-unrestricted formulation} of \ac{DFT}\cite{https://doi.org/10.1002/qua.24309}.
The direction of the spin polarization is arbitrary, but it is usually taken along the \(z\)-axis.
In the presence of an external magnetic field \(b_\text{ext}({\bf r})\) \ac{SDFT} yields, for each component $\alpha$, the equation
\begin{multline}\label{eq:SDFT}
    \bigg[-\frac{1}{2}\nabla^2 + v_{\rm KS}^\alpha[n_0,n_1]({\bf r})\\
    +(-1)^\alpha\mu_{\rm B} b_{\rm ext}({\bf r})\bigg]\psi_{i\alpha}^\text{KS}({\bf r}) = \varepsilon_{i\alpha}^\text{KS}\psi_{i\alpha}^\text{KS}({\bf r})\ .
\end{multline}
%
Here, the two density components are obtained from \(n_\alpha({\bf r}) = \sum_i f_{i\alpha}|\psi_{i\alpha}^\text{KS}({\bf r})|^2\), with \(f_{i\alpha}\) indicating the state occupation.
\(\varepsilon_{i\alpha}^\text{KS}\) are the \ac{KS} eigenvalues for the spin-up and down states.
\(\mu_\text{B}\) is the Bohr magneton (\(1/2\) in atomic units).

Eq.~(\ref{eq:SDFT}) is usually expressed in terms of the total charge density $n({\bf r})=n_0({\bf r})+n_1({\bf r})$, and the spin density along the quantization axis of the system, $m({\bf r})=n_0({\bf r}) - n_1({\bf r})$.
This results in
\begin{align}
    \bar{v}_\text{KS}[n,m]({\bf r}) &= \frac{\sum_{\alpha}v^\alpha_{\rm KS}({\bf r})}{2} \nonumber\\
    &=v_{\rm ext}({\bf r}) + v_{\rm H}[n]({\bf r}) + \bar{v}_{\rm xc}[n,m]({\bf r})\\
    \bar{v}_{\rm xc}[n,m]({\bf r}) &= \frac{v_{\rm xc}^0({\bf r}) + v_{\rm xc}^1({\bf r})}{2}\\
    b_{\rm xc}[n,m]({\bf r}) &= \frac{\sum_{\alpha}(-1)^\alpha v^\alpha_{\rm KS}({\bf r})}{2} \nonumber\\
    &=\frac{v^0_{\rm xc}[n,m]({\bf r}) - v^1_{\rm xc}[n,m]({\bf r})}{2}\\
    b_{\rm s}[n,m]({\bf r}) &= \mu_\text{B} b_\text{ext}({\bf r}) + b_\text{xc}[n,m]({\bf r})\ .
\end{align}
By using this new set of variables, the \ac{KS} Hamiltonian can then be expressed in the spin space as a \(2\times 2\) matrix.
\begin{multline}\label{eq:HSDFT}
    \matr{\mathcal{H}}_{\rm KS}({\bf r}) = \Big[-\frac{1}{2}\nabla^2 + v_{\rm ext}({\bf r}) + v_{\rm H}[n]({\bf r})\\ 
    +\bar{v}_{\rm xc}[n,m]({\bf r})\Big]\matr{I}_2 + b_{\rm s}[n,m]({\bf r})\matr{\sigma}_\text{z},
\end{multline}
\(\matr{\sigma}_z\) is the z-component of the Pauli matrices and \(\matr{I}_2\) is the \(2\times2\) identity matrix.
Eq.~(\ref{eq:HSDFT}) has a strong limitation: the spins are always aligned along the quantization axis.
This means that we need to generalize the equations of collinear \ac{SDFT} if we want to perform spin dynamics. Magnetic states can be obtained by energy minimization; however, they cannot rotate dynamically and change over time.
We discuss how this can be done in the next section.

\subsection{Non-collinear formulation of spin \ac{DFT}} \label{sec:noncolSDFT}

In the Hamiltonian of Eq.~(\ref{eq:HSDFT}) the spin is always projected along the \(z\) axis.
This means that the spin cannot rotate and that the Hamiltonian of the system commutes with \(\matr{\sigma}_z\), \(\left[\matr{\mathcal{H}}_\text{KS}({\bf r}), \matr{\sigma}_z\right] = 0\).

The \ac{KS} eigenstates, \(\psi_{i\alpha}^\text{KS}({\bf r})\), are then also eigenstates of \(\matr{\sigma}_z\), and are labeled according to the index \(i\) and the spin index \(\alpha = 0,1\).
The generalization of Eq.~(\ref{eq:HSDFT}) to spin non-collinear cases requires the spin quantization axis not to be fixed along the z axis and to be free to rotate.
In such a case, the spin density becomes a vector field
\begin{equation}
    {\bf m}({\bf r}) = \sum_i f_i\psi_i^{\rm KS}({\bf r})^\dagger\boldsymbol{\matr{\sigma}}\psi_i^\text{KS}({\bf r})\ ,
\end{equation}
where the \ac{KS} wave function is represented by a two-dimensional spinorial field.
The spinor wave function is not an eigenstate of the \(\matr{\sigma}_\text{z}\) Pauli matrix.
This means that the spinor is not polarized along the z-axis, contrary to the eigenstates of ordinary collinear spin \ac{DFT} that are polarized either up or down (\(0\), \(1\)).
The spinor state can be in a superposition of up and down spin states.

The spin non-collinear Hamiltonian can then be written as
\begin{multline}\label{eq:HNC-SDFT}
    \matr{\mathcal{H}}_\text{KS}({\bf r}) = \Big[-\frac{1}{2}\nabla^2 + v_\text{ext}({\bf r}) + v_\text{H}[n]({\bf r})\\
    +\bar{v}_{\rm xc}^\text{NC}[n, {\bf m}]({\bf r})\Big]\matr{I}_2 + \boldsymbol{b}_{\rm s}^\text{NC}[n,{\bf m}]({\bf r})\cdot\boldsymbol{\matr{\sigma}}\ ,
\end{multline}
where
\begin{equation}
    {\bf b}_\text{s}^\text{NC}[n,{\bf m}]({\bf r}) = \mu_\text{B}{\bf b}_\text{ext}({\bf r}) + {\bf b}_\text{xc}^\text{NC}[n,{\bf m}]({\bf r})\ .
\end{equation}
\(\bar{v}_\text{xc}^\text{NC}[n,{\bf m}]\) and \({\bf b}_\text{xc}^\text{NC}[n,{\bf m}]\) are functionals of the charge density and of the spin density vector fields and generalize \(\bar{v}_\text{xc}[n,m]({\bf r})\) and \(b_\text{xc}[n,m]({\bf r})\) in the non-collinear case.

\subsection{Time Dependent Density Functional Theory} \label{sec:tddft}

As an extension of \ac{DFT} to real time, \ac{TDDFT} is based on the Runge-Gross theorem \cite{PhysRevLett.52.997}.
This theorem states that if two external potentials, \(v_\text{ext}({\bf r},t)\) and \(v_\text{ext}'({\bf r},t)\), both time-dependent and Taylor-expandable (around an initial time \(t_0\)) differ by more than a simple time-dependent function \(C(t)\), then the evolving time-dependent densities \(n({\bf r},t)\) and \(n'({\bf r},t)\) starting from the same ground state must also be different.
This is equivalent to saying that there is always a one-to-one mapping between the external time-dependent potential and the electron density.
The van Leeuwen theorem guarantees instead that the time-dependent density \(n({\bf r},t)\) evolving from an initial state \(\Psi_0\) under an external potential \(v_\text{ext}({\bf r},t)\) can also be reproduced by a fictitious non-interacting system \cite{PhysRevLett.82.3863}.

The external potential can be expressed in the following general form
\begin{equation}
    v_\text{ext}({\bf r},t) = v_\text{ext}^0({\bf r}) + \theta(t-t_0)v_\text{ext}^1({\bf r},t)\ .
\end{equation}
Where \(v_\text{ext}^0({\bf r})\) is the ground state external potential, corresponding to the bare electron-ion Coulomb interaction.
\(v_\text{ext}^1({\bf r},t)\) is the external scalar time-dependent perturbation.
The time-dependent \ac{KS} equations are\cite{10.1093/acprof:oso/9780199563029.001.0001}
\begin{multline}\label{eq:TDDFT}
    \mathrm{i}\frac{\partial}{\partial t}\psi_i^\text{KS}({\bf r},t)=\Bigg[-\frac{1}{2}\nabla^2 + v_\text{ext}({\bf r},t) + v_{\rm H}[n]({\bf r},t)\\
    +v_\text{xc}^\text{NA}[n,\Psi_0,\Phi_0]({\bf r},t)\Bigg]\psi_i^\text{KS}({\bf r},t)\ .
\end{multline}
\(v_\text{xc}^\text{NA}[n,\Psi_0,\Phi_0]({\bf r},t)\) is the non-adiabatic \ac{XC} potential of the time-dependent density.
This is also a functional of the initial many-body state \(\Psi_0\) and of the initial non-interacting \ac{KS} state \(\Phi_0\).
In the case we start the evolution from the system's ground-state, \(\Psi_0\) and \(\Phi_0\) are both uniquely defined functionals of the initial density, and \(v_\text{xc}^\text{NA}\) is simply a functional of the complete time-dependent density.
The static \ac{XC} potential used to compute the ground-state electronic density must match the non-adiabatic \ac{XC} potential at time \(t=0\)
\begin{equation}
    v_\text{xc}^\text{NA}[n]({\bf r}, t = 0) = v_\text{xc}[n]({\bf r})\ .
\end{equation}
However, at arbitrary times, the two functionals do not necessarily match.
The extension of this formalism to spin polarized systems produces the time-dependent extension of collinear spin \ac{DFT}
\begin{multline}\label{eq:TDSDFT0}
    \mathrm{i}\frac{\partial\psi_{i\sigma}^\text{KS}({\bf r},t)}{\partial t} =
    \Bigg[\Bigg(-\frac{1}{2}\nabla^2 + v_\text{ext}({\bf r},t)+v_\text{H}[n]({\bf r},t)\\
    +\bar{v}_{\rm xc}^{\rm NA}[n,m]({\bf r},t)\Bigg)\matr{I}_2+ b_\text{s}^\text{NA}[n,m]({\bf r},t)\matr{\sigma}_{\rm z}\Bigg]\psi_{i\sigma}^\text{KS}({\bf r},t).
\end{multline}
Here, \(\matr{\sigma}\) is the spin quantum number associated with the wave function. 

\subsection{The non-collinear spin \ac{TDDFT} equations} \label{sec:nc-tddft}
The generalization from collinear to non-collinear dynamical systems is obtained by replacing the \ac{XC} functionals in Eq.~(\ref{eq:TDSDFT0}) with their non-collinear extension
\begin{multline}\label{eq:TDSDFT}
    \mathrm{i}\frac{\partial\psi_i^\text{KS}({\bf r},t)}{\partial t} =\Bigg[\Bigg(-\frac{1}{2}\nabla^2 + v_\text{ext}({\bf r},t) \\
    +v_\text{H}[n]({\bf r},t)+\bar{v}_\text{xc}^\text{NA+NC}[n,{\bf m}]({\bf r},t)\Bigg)\matr{I}_2\\
    +{\bf b}_\text{s}^\text{NA+NC}[n,{\bf m}]({\bf r},t)\cdot\matr{\boldsymbol{\sigma}}\Bigg]\psi_{i}^\text{KS}({\bf r},t)\ .
\end{multline}
The effective magnetic field for the KS system is given by
\begin{multline}
    {\bf b}_\text{s} ^\text{NA+NC}[n,{\bf m}]({\bf r},t) = \\
    {\bf b}_\text{xc}^\text{NA+NC}[n,{\bf m}]({\bf r},t)
        +\mu_\text{B}{\bf b}_\text{ext}({\bf r},t),\
\end{multline}
where $\textbf{b}_\text{ext}({\bf r},t)$\ is a time-dependent external magnetic field.
The spin density satisfies the continuity equation
\begin{multline} \label{eq:spincont}
    \frac{d}{dt}{\bf m}({\bf r},t) = -\nabla\cdot\bm{\mathcal{J}}_\text{s}({\bf r},t)\\
    +2{\bf b}_\text{s}^\text{NA+NC}[n,{\bf m}]({\bf r},t)\times {\bf m}({\bf r},t)\ ,
\end{multline}
where \(\bm{\mathcal{J}}_\text{s}({\bf r},t)\) is the spin current of the \ac{KS} system.
The second term is a precessional contribution due to the coupling with the effective magnetic field.
If we integrate over the entire volume of the system, \(\Omega\), the first term (\(\nabla\cdot\bm{\mathcal{J}}_\text{s}({\bf r},t)\)) produces a surface term that is equal to zero, given that there is no loss of charge, 
\begin{equation} \label{eq:GMT}
    \frac{d{\bf M}_\Omega}{dt} = 2\int_\Omega d{\bf r}\,{\bf b}_{\rm s}^\text{NA+NC}[n,{\bf m}]({\bf r},t)\times{\bf m}({\bf r},t)\ .
\end{equation}
Eq.~(\ref{eq:GMT}) gives the change in the total magnetization of the system over time.
The contribution arising from the \ac{XC} magnetic field, \(\int_\Omega d{\bf r}\,{\bf b}_\text{xc}^\text{NA+NC}[n,{\bf m}]({\bf r},t)\times{\bf m}({\bf r},t) = 0\), due to the \emph{zero torque} theorem\cite{PhysRevLett.87.206403,PhysRevB.107.115134}.
Note that the \textit{local} torque (\(\mathbf{b}_\text{xc}^\text{NA+NC}[n,{\bf m}]({\bf r},t)\times{\bf m}({\bf r},t)\)) can be non-vanishing.
In the presence of a homogeneous external magnetic field, the equation simply reduces to
\begin{equation} \label{eq:magrot}
    \frac{\mathrm{d}{\bf M}_\Omega}{\mathrm{d}t} = 2\mu_\text{B}{\bf b}_\text{ext}(t)\times{\bf M}_\Omega(t)\ .
\end{equation}
Eq.~(\ref{eq:magrot}) describes a simple precession of the total spin around the axis of the externally applied magnetic field.

\subsection{Approximations of the \ac{XC} potential}

Eq.~(\ref{eq:TDSDFT}) provides a formally-exact evolution of the spin \({\bf m}({\bf r},t)\) and electronic density \(n({\bf r},t)\) under external time-dependent fields.
The key quantities are the two \ac{XC} functionals \(\bar{v}_\text{xc}^\text{NA+NC}[n,{\bf m}]\) and \({\bf b}_\text{xc}^\text{NA+NC}[n,{\bf m}]\) whose exact form is unknown.
Usually, two major approximations are used:
(i) the adiabatic approximation;
(ii) the locally-collinear approximation.
We briefly discuss these two approximations in the next subsections.
Improvements beyond the adiabatic approximation have been proposed \cite{PhysRevLett.55.2850,PhysRevB.35.3003,PhysRevB.75.245127,PhysRevB.50.8170,PhysRevB.57.14569,PhysRevB.34.4989,PhysRevB.105.035123,PhysRevLett.120.166402,PhysRevLett.79.1905,PhysRevLett.79.4878,PhysRevB.65.235121} that we will consider in future implementations.

\subsubsection*{The adiabatic approximation}

The adiabatic approximation in \ac{TDDFT} assumes that the functional dependence of the \ac{XC} potential at time \(t\) can be approximated only by knowledge of the density at time \(t\), which means that there is no memory and no dependence on the full electron density history.
The non-adiabatic functional is then replaced by the functional of the ground-state \ac{DFT} with the instantaneous time-dependent density and magnetization.
In the non-collinear case (where `(A)' stands for adiabatic approximation)
\begin{equation}
    \bar{v}_\text{xc}^\text{NA+NC}[n,{\bf m}]({\bf r},t)\stackrel{(\text{A})}{\simeq}\bar{v}_\text{xc}^\text{GS+NC}[n(t),{\bf m}(t)]({\bf r})\ ,
\end{equation}
and
\begin{equation}
    {\bf b}_\text{xc}^\text{NA+NC}[n,{\bf m}]({\bf r},t) \stackrel{(\text{A})}{\simeq}{\bf b}^\text{GS+NC}_\text{xc}[n(t),{\bf m}(t)]({\bf r})\ ,
\end{equation}
the instantaneous density and magnetization are used inside the ground-state adiabatic functional.
The adiabatic approximation is fully justified only when the external perturbation is slow enough to keep the physical system in its instantaneous eigenstate during evolution and neglect the memory dependence in the \ac{XC} functional.
In linear response, this is equivalent to neglecting the frequency dependence in the \ac{XC} kernel \cite{Lacombe2023}.
This approximation is widely employed in \ac{TDDFT} in the form of the so-called adiabatic local (spin) density approximation (AL(S)DA)\cite{PhysRevLett.97.203001}.
\begin{multline} \label{eq:ALSDA1}
    \bar{v}_\text{xc}^\text{ALDA+NC}[n,{\bf m}]({\bf r},t)\\
    = \frac{\mathrm{\delta}}{\mathrm{\delta}\bar{n}}{E_\text{xc}^\text{HEG}[\bar{n}=n({\bf r},t),\bar{\bf m}={\bf m}({\bf r},t)]}\ ,
\end{multline}
where \(E_\text{xc}^\text{HEG}[n,{\bf m}]\) is the \ac{XC} energy functional of the homogeneous electron gas.
The value of the density \(n\) at each point of the spatial grid is given by the density field \(n({\bf r},t)\).
The use of GGA functionals \cite{PhysRevLett.77.3865} relaxes the spatial-locality condition by adding an additional dependence on the local gradient of the densities \(\nabla{n}\) and \(\nabla{\bf m}\), but these will still be computed at each point in space.
The \ac{XC} magnetic field within the AL(S)DA is
\begin{multline} \label{eq:ALSDA2}
    {\bf b}_\text{xc}^\text{ALDA+NC}[n,{\bf m}]({\bf r},t)\\
    = \frac{\mathrm{\delta}}{\mathrm{\delta}\bar{\bf m}}{E_\text{xc}^\text{HEG}[\bar{n}=n({\bf r},t),\bar{\bf m}=\bar{\bf m}({\bf r},t)]}\ .
\end{multline}
The limitations of the AL(S)DA approach have been discussed in the literature \cite{A910321J,Raghunathan2011} and will not be analyzed further here. In the next sections we will label the potentials with \text{A} to indicate that the adiabatic approximation is taken.

\subsubsection*{The locally-collinear approximation}

The functionals in Eq.~(\ref{eq:ALSDA1}) and (\ref{eq:ALSDA2}) depend on both the electron density and the magnetization-density vector of the KS system~\cite{PhysRevB.107.115134}.
The locally-collinear approach approximates the full non-collinear functionals using standard collinear functionals that depend only on the local spin-up \((n_0)\) and spin-down \((n_1)\) densities\cite{Kubler1988,Sandratskii01011998}.
In this approximation, the \ac{XC} energy functional of the non-collinear system is given by \(E_\text{xc}[n, {\bf m}]\stackrel{(\text{LC})}{\simeq}E_\text{xc}^\text{SDFT}[n, |{\bf m}|]\),
where \(E_\text{xc}^\text{SDFT}\) is the \ac{XC} energy functional of the collinear \ac{SDFT}.
The \ac{XC} fields in the locally-collinear approximation of ALDA are
\begin{equation}
    \bar{v}_\text{xc}^\text{ALDA+LC}[n, {\bf m}]({\bf r},t) = \bar{v}_\text{xc}^\text{ALSDA}[n, |{\bf m}|]({\bf r},t)\ ,\label{eq:LC+ALDA2}
\end{equation}
and
\begin{multline}
    {\bf b}_\text{xc}^\text{ALDA+LC}[n, {\bf m}]({\bf r},t)\\
    = b_\text{xc}^\text{ALSDA}[n,|{\bf m}|]({\bf r},t)\cdot\frac{{\bf m}({\bf r},t)}{|{\bf m}({\bf r},t)|}\label{eq:LC+ALDA}\ .
\end{multline}
Note that Eq.~(\ref{eq:LC+ALDA}) is ill-defined at \({\bf m}({\bf r}, t)=0\).
In this case, the \ac{XC} field is also set to zero.

Beyond the zero-magnetization issue, the locally-collinear functional has additional limitations\cite{PhysRevResearch.5.013036}:
(i) non-local functionals have an incorrect collinear limit;
(ii) the locally-collinear functionals are invariant under global rotation and satisfy the zero-torque theorem, but are insensitive to the local rotation of the magnetization vector; (iii) the locally-collinear functionals have no well-defined functional derivatives.
Improvements on this approximation have been proposed\cite{PhysRevLett.98.196405}: existing local and non-local functionals have been modified to account for spin non-collinearity \cite{PhysRevB.67.140406,Scalmani2012}, new gradient-corrected functionals have been proposed based on the spin spiral state of the electron gas \cite{PhysRevLett.111.156401,PhysRevB.88.245102}, and meta-GGA \ac{XC} functionals have been introduced \cite{PhysRevB.96.035141,PhysRevB.107.165111}.
We will not consider these functionals here;
we will use locally-collinear functionals in the simulations and focus on more advanced functionals in future studies.

We observe that Eq.~(\ref{eq:LC+ALDA2}) is directly applicable in the case of the ALDA functional, but the extension to more complicated functionals like GGA or meta-GGA makes a direct application of the locally collinear approximation potentially problematic. For instance it is known the a direct application of local collinearity in GGA functionals via
\begin{multline}
    {\bf b}_\text{xc}^\text{AGGA+LC}[n, \nabla{n}, {\bf m}, \nabla{\bf m}]({\bf r},t)\\
    = b_\text{xc}^\text{AGGA}[n,\nabla{n},|{\bf m}|,\nabla\abs{\bf m}]({\bf r},t)\cdot\frac{{\bf m}({\bf r},t)}{|{\bf m}({\bf r},t)|}\label{eq:LC+AGGA}\ .
\end{multline}
can produce numerical instabilities\cite{10.1063/1.5121713,C8CS00175H}. In the implementation section we will assume a general adiabatic + locally collinear approximation for the exchange-correlation potentials, while in the results section we will only consider LDA potentials and leave more advanced functionals like meta-GGA for future implementations.

Under the locally-collinear approximation and due to the zero local torque of the \ac{XC} field, the spin dynamics equation, Eq.~(\ref{eq:spincont}) becomes
\begin{multline} \label{eq:spincont2}
    \frac{\mathrm{d}{\bf m}({\bf r},t)}{\mathrm{d}t}=-\nabla\cdot\boldsymbol{\mathcal{J}}_\text{s}({\bf r},t)\\
    + 2\mu_{\rm B}{\bf b}_\text{ext}({\bf r},t)\times{\bf m}({\bf r},t),
\end{multline}
which corresponds to the effective equation for the dynamics of the local magnetization vector.

\subsection{Spin-orbit interaction and relativistic corrections}

The Hamiltonian of the non-collinear \ac{SDFT} written in Eq.~(\ref{eq:HNC-SDFT}) is non-relativistic.
Relativistic corrections are necessary for the description of the magnetic anisotropy and the interplay between spin and orbital degrees of freedom.
A proper discussion of the origin of the spin-orbit interaction, from fundamental quantum mechanical principles, requires starting from the Dirac equation \cite{Dyall2007IntroductionTR}. There are two main classes of approaches to account for relativistic effects in quantum chemistry. The first group is based on the four-components (4C) fully relativistic Dirac-Coulomb Hamiltonian \cite{VISSCHER1994120,https://doi.org/10.1002/cphc.201100682,sakurai1987advanced}. The second group of approaches is instead based on two-components (2C) semi-relativistic Hamiltonians where the positronic degrees of freedom are frozen. Among these, the Zeroth-Order Regular Approximation (ZORA) \cite{10.1063/1.467943} and the Douglas-Kroll-Hess Hamiltonians (DKH)\cite{DOUGLAS197489,https://doi.org/10.1002/wcms.67} are highly successful methods. The exact two-component (X2C) Hamiltonian \cite{Peng2012,10.1063/1.4803693} has been developed more recently and provides an exact description of the positive energy spectrum of the associated 4C Hamiltonian. The \ac{R-TDDFT} also belongs to this second group \cite{C8CS00175H}. Recently four-current \ac{R-TDDFT} has been applied to simulate chiral-induced spin selectivity in molecular systems\cite{zheng2025chiralitydrivenmagnetizationemergesrelativistic}.

\textsc{INQ} is a plane-wave code with pseudopotentials,  which require a different treatment of the spin orbit interaction. To explain the approximations we take, we start from the  Hamiltonian of relativistic DFT\cite{Dyall2007IntroductionTR} (with \(m = 1\) in atomic units) for a single electron in the presence of \ac{KS} potential \(v_\text{KS}[n]({\bf r})\). The KS potential of relativistic DFT is a functional of the four-component current $j$. If we neglect the dependence on the three-dimensional current density ${\bf j}({\bf r},t)$, and set the external vector potential to zero, the relativistic \ac{KS} equation is
\begin{equation}
    \mathrm{i}\partial_t\Psi^\text{KS}\!\!=\!\!\big[-ic\alpha^k\partial_k + \big(\beta - {\bf 1}_4\big) c^2 + v_\text{KS}[n]({\bf r},t) \big]\Psi^\text{KS}
\end{equation}
where \(\Psi^\text{KS}\) is a four-component spinor, and \(\alpha^k\) and \(\beta\) are the Dirac matrices.
\(\Psi^\text{KS}\) can be split into a small component, \(\Psi^\text{KS}_\text{S}\), and a large one, \(\Psi^\text{KS}_\text{L}\).
For applications in low-energy solid-state and molecular physics, the small component that is associated with the negative energy solutions of the Dirac equation\cite{Strange_1998} is often neglected.
The equation for the large component \cite{RevModPhys.65.733} after we expand perturbatively to order \( O(1/c^2)\)\cite{Dyall2007IntroductionTR} is
\begin{multline}
    \Big\{\frac{{\bf p}^2}{2}+v_\text{KS}[n]({\bf r}) -\frac{{\bf p}^4}{8c^2} + \frac{\nabla^2v_\text{KS}[n]({\bf r})}{8c^2} \\
    +\frac{1}{4c^2}\big(\nabla v_{\rm KS}[n]({\bf r})\times{\bf p}\big)\cdot\boldsymbol{\sigma}\Big\}\Tilde{\Psi}^\text{KS}_\text{L} = E_\text{(SR)}^\text{KS}\Tilde{\Psi}^\text{KS}_\text{L}.
\end{multline}
The first two terms in the semi-relativistic Hamiltonian are standard non-relativistic ones.
The third term is a relativistic kinetic correction, and the fourth term is the Darwin interaction, that becomes more important close to the nuclei.
The last term is the spin-orbit interaction, which depends on the effective electrostatic potential experienced by the electrons.
In the \ac{KS} system, this corresponds to the effective \ac{KS} scalar potential acting on each electron.

The spin-orbit interaction \(V_\text{SO}({\bf r})\) can be rearranged as
\begin{align}\label{eq:Vso}
    V_\text{SO}({\bf r}) &= \frac{1}{4c^2}\big(\nabla v_\text{KS}[n]({\bf r})\times{\bf p}\big)\cdot\boldsymbol{\sigma}\nonumber\\
    &= \frac{1}{4c^2}\big(\nabla v_\text{ext}({\bf r})\times{\bf p}\big)\cdot\boldsymbol{\sigma}\nonumber\\
    &+\frac{1}{4c^2}\big(\nabla v_\text{Hxc}[n]({\bf r})\times{\bf p}\big)\cdot\boldsymbol{\sigma}\ .
\end{align}
The first contribution is due to the bare ionic external potential, and the second to the electrostatic cloud formed by the other electrons in the system.
If we consider that the total electron density is the sum of atomic, \(n_\text{A}\), and valence interstitial contributions \(\Tilde{n}=n - n_\text{A}\), we can separate the effect of the atomic isolated contribution from the interstitial valence electrons in Eq.~(\ref{eq:Vso}):
\begin{multline}\label{eq:Vso-32}
    V_\text{SO}({\bf r}) = \frac{1}{4c^2}\big(\nabla\Tilde{v}[n_{\rm A}]({\bf r})\times{\bf p}\big)\cdot\boldsymbol{\sigma}\\
    +\frac{1}{4c^2}\big(\nabla \delta v_\text{Hxc}[n_\text{A},\Tilde{n}]({\bf r})\times{\bf p}\big)\cdot\boldsymbol{\sigma}\ .
\end{multline}
Here we have introduced the following effective potentials
\begin{equation}
    \Tilde{v}[n_\text{A}]({\bf r}) = v_\text{ext}({\bf r})+v_\text{Hxc}[n_\text{A},0]({\bf r})\ ,
\end{equation}
and
\begin{multline}
    \delta v_\text{Hxc}[n_\text{A},\tilde{n}]({\bf r})\\
     = v_\text{Hxc}[n_\text{A}, \tilde{n}]({\bf r}) - v_\text{Hxc}[n_\text{A},0]({\bf r})\ .
\end{multline}
The first term in \(V_\text{SO}\) of Eq.~(\ref{eq:Vso-32}) is the most important one and corresponds to the spin-orbit coupling due to the combined effect of the bare ionic potential and the atomic electron charge.
Under the assumption of non-overlapping atomic charges, \(n_\text{A}=\sum_{\rm a=1}^{N_{\rm a}}n_\text{A}^{\rm a}\), it can be written as a linear combination of effective potentials centered around each nucleus,
\begin{equation}
    \Tilde{v}[n_\text{A}]({\bf r}) = \sum_{\rm a=1}^{N_a}\Tilde{v}_{\rm a}[n_\text{A}^{\rm a}]({\bf r} - {\bf R}_{\rm a})\ .
\end{equation}
Meanwhile, the second term of Eq.~(\ref{eq:Vso-32}) depends on \(\delta v_\text{Hxc}({\bf r})\), which is the electrostatic and exchange-correlation potential due to the remaining interstitial valence charge, 
can be considered as a smaller correction to the atom-centered contribution. This term is dynamical given that it must be updated with the valence density of the system during the real-time evolution.

If we group together all the atomic spherically centered contributions, the total spin-orbit potential can then be expressed as
\begin{multline}
    V_\text{SO}({\bf r}) = \frac{1}{4c^2}\sum_{\rm a=1}^{N_{\rm a}}\frac{\partial_r\Tilde{v}_{\rm a}[n_{\text{A}}^{\rm a}](r)}{|{\bf r}-{\bf R}_{\rm a}|}\cdot\big({\bf r}-{\bf R}_{\rm a}\big)\times{\bf p}\cdot\boldsymbol{\matr{\sigma}}\\
    +\frac{1}{4c^2}\big(\nabla_{\bf r}\delta\Tilde{v}_{\rm Hxc}[n_\text{A}, \tilde{n}]({\bf r})\times{\bf p}\big)\cdot\boldsymbol{\matr{\sigma}}\ ,
\end{multline}
where the first term includes a spherically centered contribution around every atom, and the second includes an additional spherically non-centered term that depends on the electronic density in the interstitial regions.

If we neglect the second term, we are left with the atom-centered contribution only.
If we define the orbital momentum around atom \(a\) as \({\bf L}_{\rm a} = ({\bf r}-{\bf R}_{\rm a})\times{\bf p}\),  we obtain
\begin{equation} \label{Eq:VSO_ACA}
    V_\text{SO}({\bf r}) \stackrel{AC}{\simeq} \frac{1}{4c^2}\sum_{\rm a=1}^{N_{\rm a}}\frac{\partial_r\Tilde{v}_{\rm a}[n^{\rm a}_\text{A}](r)}{|{\bf r}-{\bf R}_{\rm a}|}{\bf L}_{\rm a}\cdot\boldsymbol{\matr{\sigma}}\ ,
\end{equation}
which we refer to as the atom-centered  (AC) approximation for the spin-orbit interaction. The detailed implementation is explained in Section 3.5 Eq.~\ref{Eq:Vps-soc}.
We note that the updated charge density during self-consistent field calculations updates the spin-orbit contribution to the total energy or spin-orbit splitting in band structures. 
This is a widely used and successful approximation in predicting spin-orbit splitting in materials with heavy elements \cite{PhysRevB.64.073106,Zhang2014}. The detailed numerical implementation of spin-orbit coupling is presented in section \ref{sec:SOC_impl}.
In the case of a single atom \(v_\text{ext}({\bf r})=-\frac{Z}{r}\) and if we neglect the effect of other electrons, we recover the ordinary form of the interaction
\begin{equation}
    V_\text{SO}({\bf r}) = \frac{Z}{4c^2 r^3}\big({\bf r}\times{\bf p}\big)\cdot\boldsymbol{\sigma} = \lambda {\bf L}\cdot\boldsymbol{\sigma}\ ,
\end{equation}
with $\lambda=\frac{Z}{4c^2 r^3}$.

By including the spin-orbit interaction in the spin continuity equation Eq.(\ref{eq:spincont2}) we obtain the following expression
\begin{multline} \label{eq:spincont3}
    \frac{\mathrm{d}{\bf m}({\bf r},t)}{\mathrm{d}t}=-\nabla\cdot\boldsymbol{\mathcal{J}}_{\rm s}({\bf r},t) \\
    +2\mu_\text{B}{\bf b}_{\rm ext}({\bf r},t)\times{\bf m}({\bf r},t) + \boldsymbol{\Gamma}_\text{SOC}({\bf r},t)\ ,
\end{multline}
where \(\boldsymbol{\Gamma}_\text{SOC}({\bf r},t)\) defines an additional spin-orbit torque.

\section{Numerical Implementation} \label{sec:implement}

In this section, we discuss the details of the implementation of the spin non-collinear formalism in the \textsc{INQ} code.

\subsection{Spinorial representation}

The single-particle wavefunctions of the non-collinear spin \ac{KS} Hamiltonian (\ref{eq:HNC-SDFT}) are in spinorial form.
Spinors are two-dimensional mathematical objects that transform according to the \texttt{SU(2)} representation of the three-dimensional rotation group.
This is
\begin{equation}
    \psi_{i}^\text{KS}({\bf r})\to e^{\mathrm{i}\frac{\phi}{2}(\hat{\bf n}\cdot\hat{\boldsymbol{\sigma}})}\psi_{i}^\text{KS}({\bf r})
\end{equation}
In practice, this means we need to include an additional index in the representations of the spinorial orbitals, that also have indices for the states and basis coefficients.

A key code-design consideration is how to include this new index both in terms of representation and memory layout to have a code that is at the same time simple, easy to read, and efficient.
All with the additional challenge of using the same code for spin unpolarized, collinear, and non-collinear simulations.

We have found that to obtain the most efficient memory layout, it is better to have the spinor index in the middle of the array, in between the basis coefficient (slowest index) and orbital indices (fastest index).
In this way, there are three possible views of the wave-function array.
The first one is to have the spinor index explicitly, for operations that involve spinor components like application to the potential or the calculation of the spin density.
The second view is a 2-dimensional array obtained by \emph{flattening} the spinor and state indices to effectively duplicate the number of states.
This view is useful for operations that act individually on each state and do not involve spinor information, like the application of the kinetic energy operator.
Finally, we can obtain a different 2-dimensional view of the array by compacting the basis coefficient and spinor indices.
In this case we can think that the spinor dimension duplicates the number of the basis coefficient.
This means that operations like orthogonalization can be easily done using the same code as the collinear case, that relies of accelerated linear algebra libraries like \texttt{BLAS}~\cite{Blackford2002} and \texttt{Lapack}~\cite{Anderson1999}.
All the different views of the wave-function array are managed by the \textsc{Multi}~\cite{Multi} C++ software library, that avoids any copies or access overhead.

When running in parallel over multiple nodes~\cite{Andrade2012}, it is important to ensure that for each state the two spinor components are kept in the same process.
Otherwise, the communication cost would be too high, as many operations mix the two components. 

In the spinorial representation, it is convenient to introduce the \ac{KS} spin-density matrix
\begin{multline}
    \rho_\text{s}^\text{KS}({\bf r})=\left [ {\begin{array}{cc}
    \rho_{\rm s}^{\uparrow\uparrow}({\bf r}) & \rho_\text{s}^{\uparrow\downarrow}({\bf r})\\
    \rho_\text{s}^{\downarrow\uparrow}({\bf r}) & \rho_\text{s}^{\downarrow\downarrow}({\bf r})\\
  \end{array} } \right ]\\
    =\sum_{i}f_{i}\left[ {\begin{array}{cc}
        \psi_{i\uparrow}^\text{KS}({\bf r}){\psi_{i\uparrow}^\text{KS}({\bf r})}^* &
        \psi_{i\uparrow}^\text{KS}({\bf r}){\psi_{i\downarrow}^{KS}({\bf r})}^* \\
        \psi_{i\downarrow}^\text{KS}({\bf r}){\psi_{i\uparrow}^\text{KS}({\bf r})}^* & \psi_{i\downarrow}^\text{KS}({\bf r}){\psi_{i\downarrow}^\text{KS}({\bf r})}^* \\
  \end{array} } \right]\,.
\end{multline}
From \(\rho_\text{s}^\text{KS}({\bf r})\) the magnetization density can be calculated as
\begin{equation}
    {\bf m}({\bf r}) = {\rm Tr}\big[\hat{\boldsymbol{\sigma}}\rho_\text{s}^\text{KS}({\bf r})\big]\ ,
\end{equation}
which reduces to the 3 different components
\begin{align}
    m_{\rm x}({\bf r}) &= \rho_{\rm s}^{\uparrow\downarrow}({\bf r}) + \rho_{\rm s}^{\downarrow\uparrow}({\bf r})\nonumber\\ 
    &= 2{\rm Re}[\rho_{\rm s}^{\uparrow\downarrow}({\bf r})]\label{eq:rho1}\\
    m_{\rm y}({\bf r}) &= \mathrm{i}\big(\rho_{\rm s}^{\uparrow\downarrow}({\bf r}) - \rho_{\rm s}^{\downarrow\uparrow}({\bf r})\big)\nonumber\\
    &= -2{\rm Im}[\rho_{\rm s}^{\uparrow\downarrow}({\bf r})]\label{eq:rho2}\\
    m_{\rm z}({\bf r}) &= \rho_{\rm s}^{\uparrow\uparrow}({\bf r}) - \rho_{\rm s}^{\downarrow\downarrow}({\bf r})\label{eq:rho3}
\end{align}
In the code, we store the spin-density matrix as a four-dimensional array of real values:
\begin{equation}\label{eq:density4}
    \rho_\text{v}({\bf r}) = \big\{ \rho_\text{s}^{\uparrow\uparrow}({\bf r}), \rho_\text{s}^{\downarrow\downarrow}({\bf r}), {\rm Re}[\rho_\text{s}^{\uparrow\downarrow}({\bf r})], {\rm Im}[\rho_\text{s}^{\uparrow\downarrow}({\bf r})]\big\}\ .
\end{equation}
This is more efficient than storing the full complex matrix that has redundant values due to its hermiticity.
The conversion between \(\rho_\text{v}({\bf r})\) and the magnetization density \({\bf m}({\bf r})\) is straightforward through Eqs.~(\ref{eq:rho1}), (\ref{eq:rho2}) and (\ref{eq:rho3}).

\subsection{Local-magnetization initialization procedure}

\begin{figure}
    \centering
    \caption{\textsc{DMInitialization} procedure for the spin-density matrix: The algorithm is organized in two main functions: (1) in \textsc{NearestAtom} given the grid point \texttt{i}, the list of atom coordinates \texttt{AtomsList} and the real-space grid \texttt{Spacegrid} we find the atom \texttt{a} closer to the grid point \texttt{i}; this is equivalent to a \textsc{Voronoi} partition of the real-space grid based on the distance from the atoms. (2) In the function \textsc{ComputeSpinDensity} we use the \texttt{MagneticMoment} vector in input associated with atom \texttt{a} and the local grid point density at distance \texttt{d} from the atom center to compute the spin density vector \(\rho_\text{v}\) at position \texttt{i} according to Eq.~(\ref{eq:density4}). The procedure is then repeated over all the points in the grid.}\label{alg:mag_init}
    \fbox{\begin{minipage}{\columnwidth}
    \begin{algorithmic}[1]
        \Procedure{DMInitialization}{ \texttt{Spacegrid}, \texttt{AtomsList}, \texttt{MagneticMoments}, \texttt{Density}}
        \Function{NearestAtom}{\(i\),\texttt{Spacegrid},\texttt{AtomsList}}
            \State \(r \gets \texttt{Spacegrid}[i]\)
            \State \(a \gets \Call{Voronoi}{r, \texttt{AtomsList}}\)
            \State \Return \(a\)
        \EndFunction
        \Function{ComputeSpinDensity}{\texttt{M},\texttt{n}}
            \State \(\rho_v\gets \{0, 0, 0, 0\}\)
            \If{\(n > 0\)}
                \State \(m\gets M/n\)
                \State \(\rho_\text{v}\gets \{(1 + m[2])/2,(1 - m[2])/2,m[0]/2,-m[1]/2\}\) 
            \EndIf
            \State \Return \(n\rho_\text{v}\)
        \EndFunction
        \State \(i \gets 0\); \(N \gets \texttt{len}(spacegrid)\)
        \For{\(i \in 0\ldots N-1\)}
            \State \(a\gets \)\Call{NearestAtom}{\(i\), \texttt{SpaceGrid}, \texttt{AtomsList}}
            \State \(d\gets \)\Call{Distance}{\texttt{Spacegrid}[i], \texttt{Spacegrid}[\texttt{AtomsList}[\texttt{a}]]}
            \State \(\rho_\text{v}[i]\gets\)\Call{ComputeSpinDensity}{ \texttt{MagneticMoments[a]}, \texttt{Density}[a](d)}
        \EndFor
        \State \Return \(\rho_\text{v}\)
        \EndProcedure
    \end{algorithmic}
    \end{minipage}}
\end{figure}

The initialization of a non-collinear calculation, in particular the density, is critical as the ground-state solver can get trapped in meta-stable spin configurations and not reach the lowest energy solution.
Conversely, in some cases we do want to converge to a meta-stable configuration, for example, an anti-ferromagnetic state, and setting the appropriate starting point for the simulation allows us to do just that.

The electron density can be initialized from a linear combination of atomic densities directly on the real-space grid.
In the case of spin polarized and non-collinear systems, we must initialize the full spin-density matrix \(\rho_\text{s}({\bf r})\).
The initial guess for the spin-density matrix of the system can be written as follows.
\begin{equation}
    \rho_\text{s}({\bf r}) = \sum_{\rm a=1}^{N_{\rm a}}\rho_\text{s}^{\rm a}({\bf r}-{\bf R}_{\rm a}),\
\end{equation}
where \(\rho_\text{s}^{\rm a}({\bf r}-{\bf R}_{\rm a})\) is the atomic density matrix centered around the atom {\rm a} and it is computed from the charge distribution, \(n_{\rm a}(|{\bf r}|)\), obtained from the pseudopotential, and from the initial atomic magnetic moment, \({\bf m}_{\rm a}\), which must be provided in input.
At each point on the spatial grid, the initial guess for the spin-density matrix is calculated according to the pseudo-algorithm in Fig.~(\ref{alg:mag_init}) and stored in the vector field variable \(\rho_\text{v}({\bf r})\), a field set defined on the spatial grid.
In the initialization algorithm, the function \textsc{NearestAtom} takes in input the grid point index \texttt{i}, the real space grid \texttt{Spacegrid} and the list of atoms \texttt{AtomsList} and finds the closest atom to the grid point \texttt{i}.
This is essentially a \textsc{Voronoi} algorithm\cite{doi:10.1142/8685} used to partition the real space grid w.r.t. atomic centers.
We then use the \textsc{ComputeSpinDensity} function to obtain the spin density matrix \(\rho_\text{v}(i)\) at the grid point \texttt{i} from the \texttt{MagneticMoments} vector and the atomic \texttt{Density} using Eq.~(\ref{eq:density4}).
The procedure is then repeated over the entire real-space grid.
\subsection{The \ac{XC} potential in spinorial representation}
Within the adiabatic and locally-collinear approximation in Eq.~(\ref{eq:LC+ALDA}) and Eq.~(\ref{eq:LC+ALDA2}), the \ac{XC} potential can be expressed as
\begin{multline}
    \mathcal{V}_{\rm xc}[n,{\bf m}]({\bf r},t) = \bar{v}_\text{xc}^\text{A+LC}[n,{\bf m}]({\bf r},t){\bf I}_2\\
    +{\bf b}_\text{xc}^\text{A+LC}[n,{\bf m}]({\bf r},t)\cdot\boldsymbol{\sigma}.
\end{multline}
The potential ${\bf b}_\text{xc}^\text{A+LC}$ is computed from Eq.~(\ref{eq:LC+ALDA}) evaluating the functional using the local electron density \((n)\) and magnetization magnitude \((\abs{\bf m})\), and projecting the result along the magnetization direction. In the case where the value of $\abs{\bf m}$ is below a given threshold value, the field ${\bf b}_\text{xc}^\text{A+LC}$ is set to zero to avoid divergence.
By omitting the labels, this is equivalent in spinorial representation to
\begin{equation}
    \hat{\mathcal{V}}_{\rm xc}=
    \left [ {\begin{array}{cc}
    \bar{v}_\text{xc} + b_\text{xc}^{\rm z} & b_\text{xc}^-\\
    b_\text{xc}^+ & \bar{v}_\text{xc} - b_\text{xc}^{\rm z}\\
  \end{array} } \right ],
\end{equation}
where we have used \(b_\text{xc}^\pm = b_\text{xc}^\text{x}\pm i b_\text{xc}^{\rm y}\).
This leads to the following expression for the \ac{XC} energy
\begin{multline}
    E_\text{xc} = \int_{\Omega}d{\bf r}\big[ \rho_\text{s}^{\uparrow\uparrow}({\bf r})\big(\bar{v}_\text{xc}({\bf r}) + b_\text{xc}^{\rm z}({\bf r})\big)\\
    +\rho_\text{s}^{\downarrow\downarrow}({\bf r})\big(\bar{v}_\text{xc}({\bf r}) - b_\text{xc}^{\rm z}({\bf r})\big)\\
    +2{\rm Re}[\rho_\text{s}^{\uparrow\downarrow}({\bf r})]b_\text{xc}^{\rm x}({\bf r})
    -2{\rm Im}[\rho_\text{s}^{\uparrow\downarrow}({\bf r})]b_\text{xc}^{\rm y}({\bf r})\big]\ .   
\end{multline}

Analogous to what we do for the density in Eq.~(\ref{eq:density4}), we can use a four-dimensional vectorial representation of the \ac{XC} field defined as
\begin{align}\label{eq:potential4}
    &\mathcal{V}_\text{v}^\text{xc}({\bf r}) = \big[\bar{v}_\text{xc}^+({\bf r}), \bar{v}_\text{xc}^-({\bf r}), 2b_\text{xc}^\text{x}({\bf r}),-2b_\text{xc}^{\rm y}({\bf r})\big]\nonumber\\
    &\bar{v}_\text{xc}^+({\bf r}) = \bar{v}_\text{xc}({\bf r})+b_\text{xc}^{\rm z}({\bf r})\nonumber\\
    &\bar{v}_\text{xc}^-({\bf r}) = \bar{v}_\text{xc}({\bf r})-b_\text{xc}^{\rm z}({\bf r})\ .
\end{align}
This gives us a simple expression for the energy:
\begin{equation}\label{eq:xc_energy4}
    E_\text{xc} = \int_\Omega d{\bf r}<\rho_\text{v}({\bf r}), \mathcal{V}_\text{v}^\text{xc}({\bf r})>\ ,
\end{equation}
where \(<,>\) is the canonical scalar product between the two 4-component representations.

\subsection{Zeeman coupling and external magnetic field}\label{sec:zeemancoupl}

Analogously to the \ac{XC} magnetic field, the Zeeman coupling to an external homogeneous magnetic field, \({\bf b}_{\rm ext}\), is also defined in the spinorial representation using
\begin{multline}\label{eq:zeeman_potential}
    \hat{\mathcal{V}}_{\rm z}=\mu_\text{B}\boldsymbol{\sigma}\cdot{\bf b}_\text{ext}\\
    =\mu_\text{B}\left[ {\begin{array}{cc}
    b_\text{ext}^z & b_{\rm ext}^x -{\rm i}b_{\rm ext}^y\\
    b_\text{ext}^x +{\rm i}b_\text{ext}^y & - b_\text{ext}^z\\
  \end{array} } \right]\ .
\end{multline}
In the collinear case, the two off-diagonal terms disappear, and the expression reduces to a fully diagonal form.
The Zeeman energy is computed as we do for the \ac{XC} energy, Eqs.~(\ref{eq:potential4}) and (\ref{eq:xc_energy4}), using \(\int_\Omega d{\bf r}<\rho_\text{v}({\bf r}), \mathcal{V}_\text{v}^{\rm z}({\bf r})>\).
We only implement the spin Zeeman term in this work. 
We will consider the orbital Zeeman contribution or orbital response to an external magnetic field~\cite{Lebedeva2019-er}, as well as second-order magnetic response~\cite{2022Kaneko} in future studies.

\subsection{Spin-orbit coupling implementation}\label{sec:SOC_impl}

Here we discuss the implementation of the spin-orbit interaction based on the pseudopotential method.
Other implementations are available in the literature based on the first-order perturbation theory variation of scalar relativistic wave functions \cite{PhysRevB.43.4286,PhysRevB.47.4238}, but will not be considered here.
A fully-relativistic treatment of the problem would require a four-component formalism based on Dirac spinors \cite{https://doi.org/10.1002/cphc.201100682}, or a two-component exact formalism like X2C\cite{10.1063/1.4803693} as discussed in the theory section.
We have also seen that all relativistic effects up to order \(\alpha^2\) (where \(\alpha\) is the fine structure constant) can be captured from a Scr{\"o}dinger-like equation with an additional spin-orbit term \cite{PhysRevB.21.2630,PhysRevB.64.073106}.
In this form, the total ionic pseudopotential can be written as~\cite{Kleinman1982}
\begin{equation} \label{eq:Vps}
    \hat{v}_\text{ps} = \sum_{\rm a=1}^{N_{\rm a}}\Bigg\{\hat{V}_{\rm L}^{\rm a} + \sum_{\ell,j,m_j}\frac{\ket*{\delta V_{\ell j}^{\rm a}\phi_{\ell j;m_j}^{\rm a}}\bra*{\phi_{\ell j,m_j}^{\rm a}\delta V_{\ell j}^{\rm a}}}{\mel*{\phi_{\ell j,m_j}^{\rm a}}{\delta V_{\ell j}^{\rm a}}{\phi_{\ell j,m_j}^{\rm a}}}\Bigg\},
\end{equation}
where the sum goes over each of the \(N_a\) atoms and over the quantum numbers \(\ell, j, m_j\).
These are the quantum numbers associated with the solution of the non-interacting Dirac equation in a centro-symmetric potential.
The contribution \(\hat{V}_\text{L}^{\rm a}\) is local in space and long-ranged.
The second term in the sum-over-atoms is the short-ranged non-local atomic contribution.
The potential \(\delta V_{lj}^{\rm a}({\bf r}) = V_{lj}^{\rm a}({\bf r}-{\bf R}_{\rm a})-V_\text{L}^{\rm a}({\bf r}-{\bf R}_{\rm a})\) is the difference between the pseudopotential channel \((l, j)\) and the local long-range pseudopotential \(V_\text{L}^{\rm a}\).
This term includes the contribution to the spin-orbit interaction;
it can be shown, in fact, that the effective pseudo-spin orbit interaction can be written as \cite{VANSETTEN201839}
\begin{equation} \label{Eq:Vps-soc}
    \hat{v}_\text{SO}^\text{ps} = \sum_{\rm a=1}^{N_{\rm a}}\sum_{\ell}\frac{2}{2\ell+1}\big(V^{\rm a}_{\ell+1/2} - V^{\rm a}_{\ell-1/2}\big)\hat{\bf L}_{\rm a}\cdot\hat{\bf S}\ ,
\end{equation}
which has the same form as the atom-centered approximation in Eq.~(\ref{Eq:VSO_ACA}) and sets the level of our current spin orbit coupling implementation.
The potential \(\hat{v}_\text{SO}^\text{ps}\) is not consistently updated with the density during the \ac{TDDFT} iterations. This is often considered a good approximation given that the core electrons are considered frozen during the dynamics of the system.

The set of wave functions \(\{\phi_{lj;m_j}^{\rm a}\}\) in Eq.~(\ref{eq:Vps}) corresponds to the solution of the KS problem for the isolated atom.
These wave functions are expressed as the product of a radial component and of the spinorial spherical harmonics
\begin{align}
    \ket*{\phi_{\ell j;m_j}^{\rm a}} &= \ket*{R_{\ell j}^{\rm a}}\otimes\ket*{\Phi_{\ell j}^{m_j}}\nonumber\\
    \braket*{{\bf r}}{\phi_{\ell j;m_j}^{\rm a}} &= R_{\ell j}^{\rm a}(|{\bf r}-{\bf R}_{\rm a}|)\Phi_{\ell j}^{m_j}(\hat{\bf\Omega}).
\end{align}
The first factor of the atomic pseudo wave function is the radial component, \(R_{lj}^{\rm a}\), and the second factor is the spinorial spherical harmonics
\begin{multline}
    \Phi_{\ell,j=\ell\pm\frac{1}{2}}^{m_j = m\pm\frac{1}{2}}(\hat{\bf\Omega})= \sqrt{\frac{\ell\pm m+1}{2\ell+1}}\Upsilon^{m_j-\frac{1}{2}}_\ell(\hat{\bf\Omega})\chi_\uparrow\\
    \pm\sqrt{\frac{\ell\mp m}{2\ell+1}}\Upsilon_\ell^{m_j+\frac{1}{2}}(\hat{\bf\Omega})\chi_\downarrow\ .
\end{multline}
Our implementation of the spinorial spherical harmonics is available as a standalone library called \textsc{sharmonic}~\cite{Andrade2025}.
This library also implements real and complex harmonics, supports GPUs, and it is quite efficient and accurate.

In the next section, we discuss simulations of different systems using this spin non-collinear implementation for both ground state \ac{DFT} and real-time \ac{TDDFT}.

\section{Results and Discussion} \label{sec:res}

This section is organized as follows.
We first discuss the spin density vector field (spin texture) and the spin non-collinearity in magnetic clusters and solids.
Then we look at the spin-orbit and Zeeman splitting in the presence of external magnetic fields in atoms and the magnetic response of paramagnetic solids.
Finally, we focus on the real-time dynamics of magnetic systems under external magnetic and electric pulses.

\subsection{Spin non-collinearity in magnetic clusters}\label{sec:noncol_clust}

\begin{figure*}%
    \centering
    \subfloat[\centering \ce{Fe2} spin density without spin-orbit interaction]{{\includegraphics[width=0.46\textwidth]{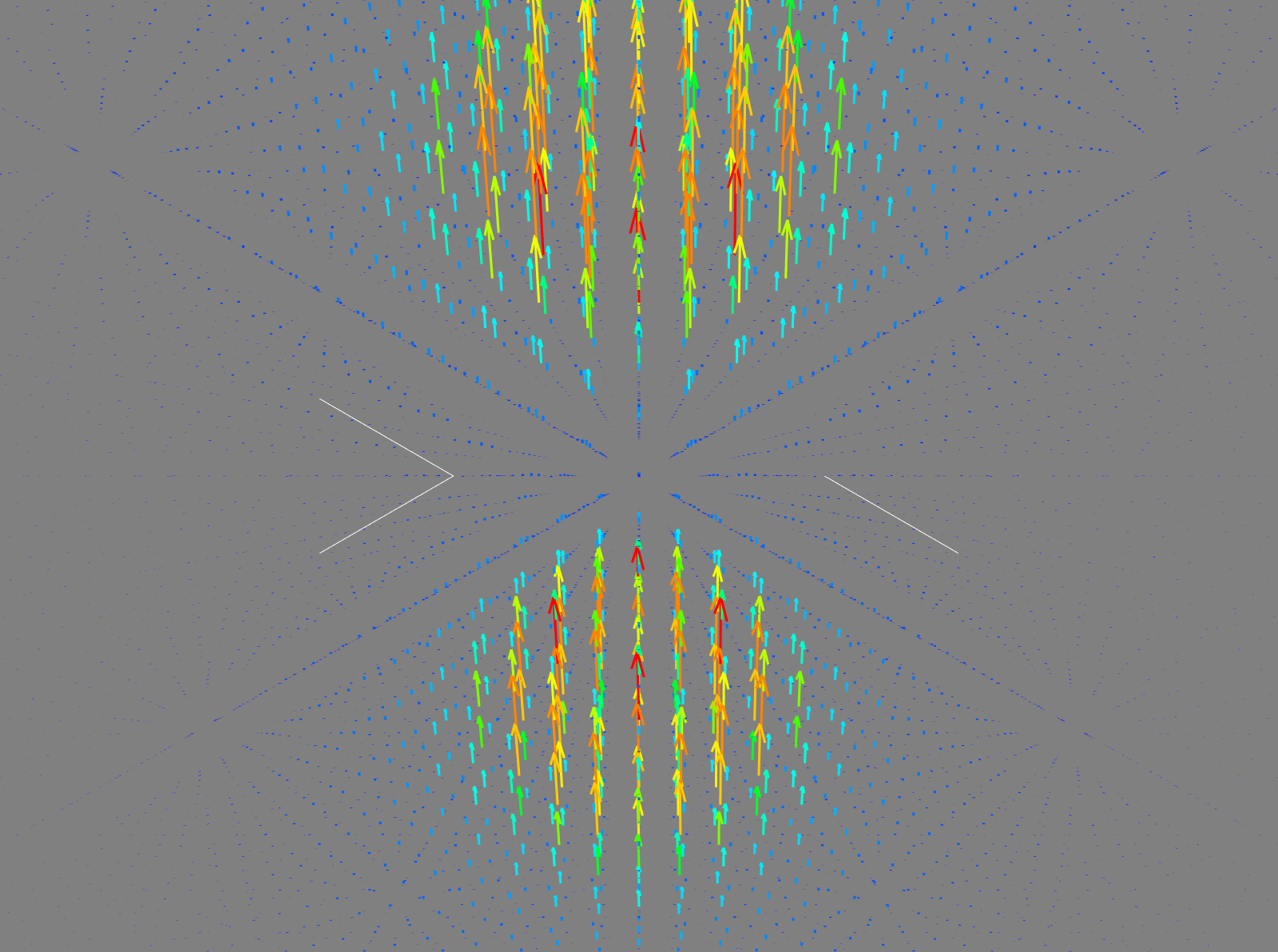} }}%
    \qquad
    \subfloat[\centering \ce{Fe2} spin density with spin-orbit interaction]{{\includegraphics[width=0.46\textwidth]{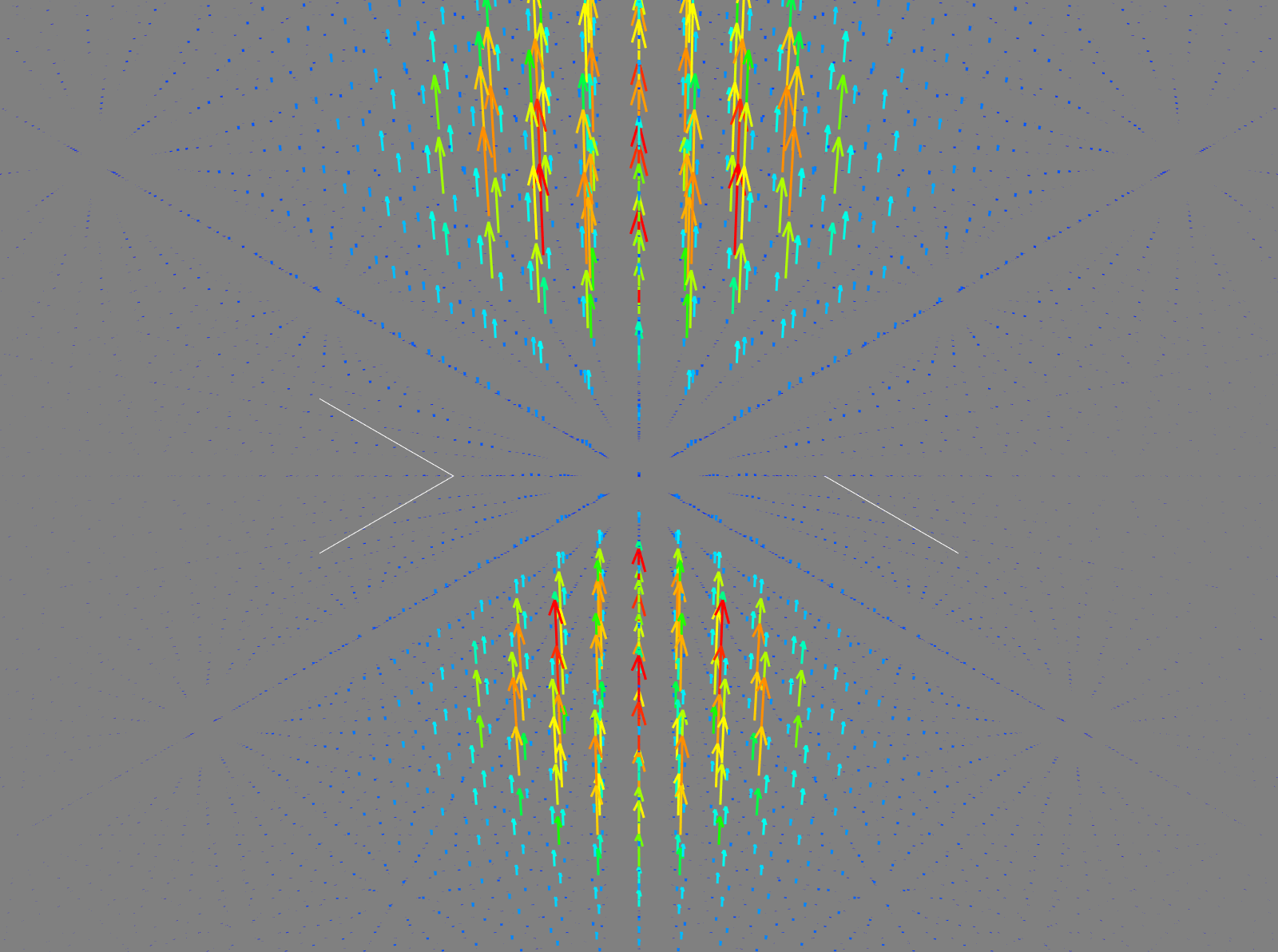} }}%
    \\
    \subfloat[\centering \ce{Fe6} spin density with spin-orbit interaction]{{\includegraphics[width=0.46\textwidth]{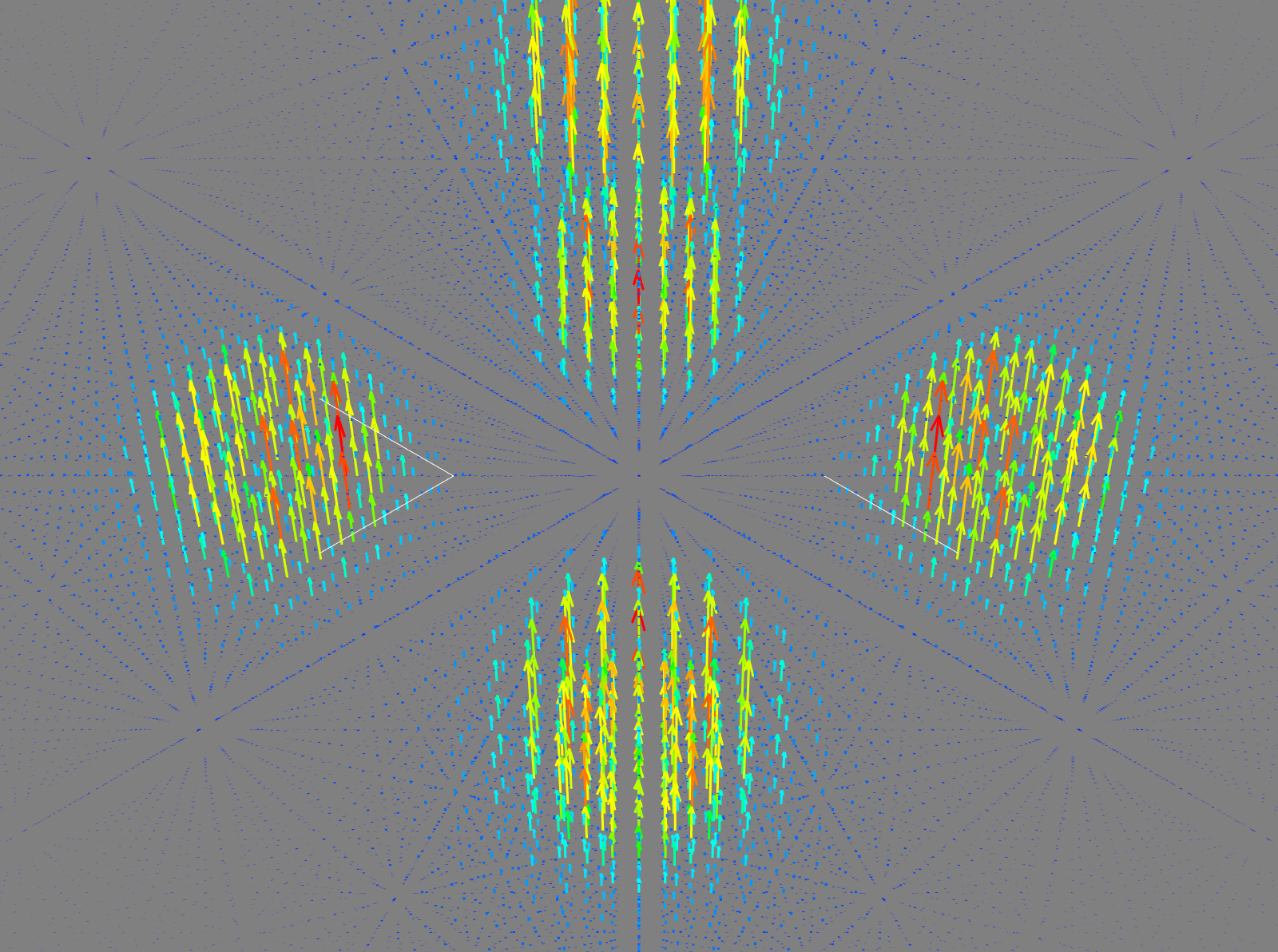} }}%
    \qquad
    \subfloat[\centering Non-collinear (\(\rm x\) and \(\rm y\)) components of the \ce{Fe6} spin density]{{\includegraphics[width=0.46\textwidth]{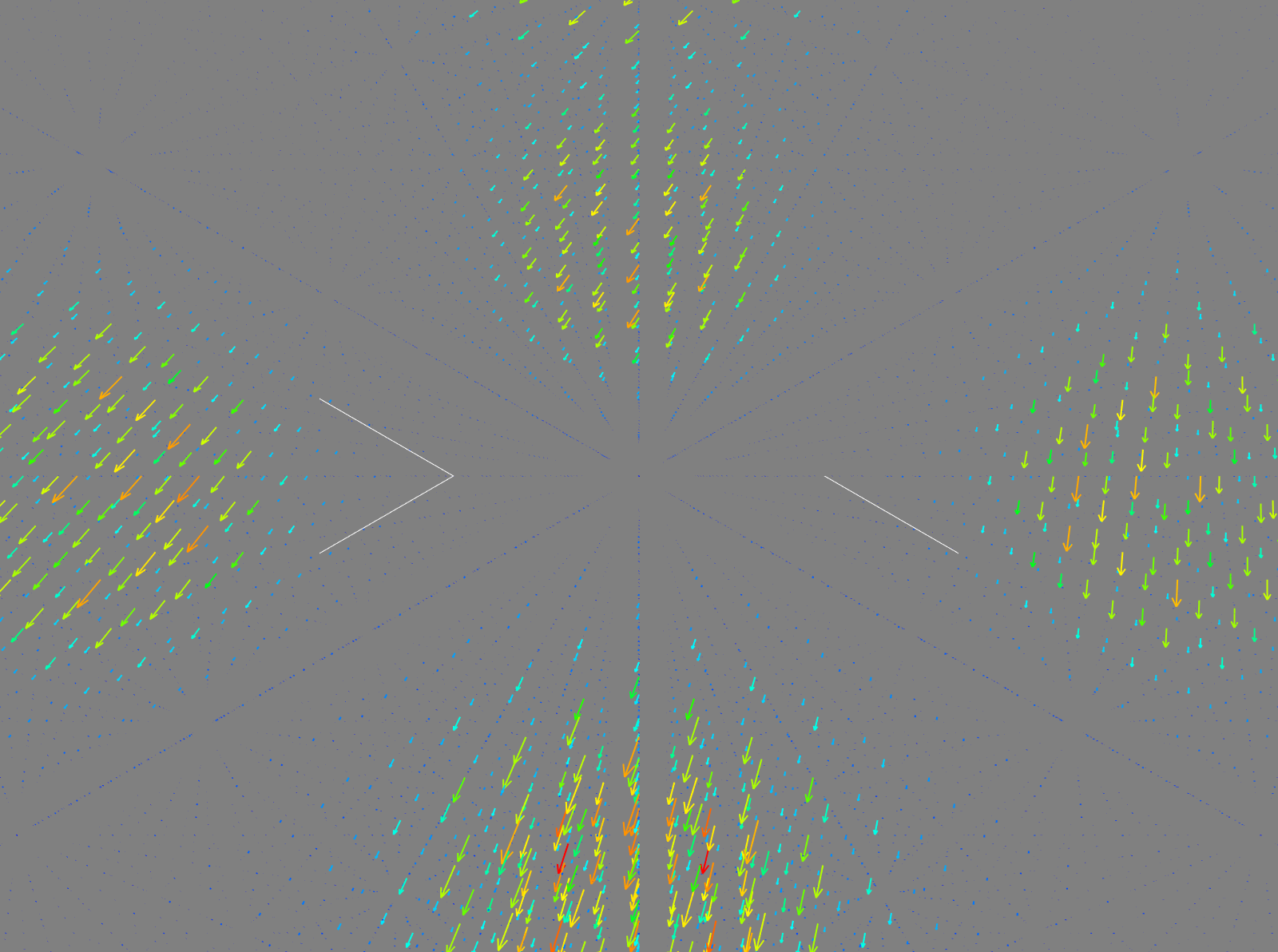} }}%
    \caption{
    Spin density profile of magnetic Fe clusters.
    In panel (a) and (b) we show the spin density profile for the \ce{Fe2} molecule without and with spin-orbit interaction, respectively.
    The degree of spin non-collinearity induced by spin-orbit is small and the two calculations produce very similar spin density profiles. This is not the case for \ce{Fe6} cluster.
    In panel (c) we consider the full spin-density profile \((m_{\rm x},m_{\rm y},m_{\rm z})\) of \ce{Fe6}.
    In panel (d) we remove the z-component and show \((m_{\rm x},m_{\rm y},0)\) to emphasize the importance of the spin non-collinearity for the description of this magnetic cluster.
    The two \ce{Fe} atoms in the dimer are at a distance of \(1.99~\text{\AA}\) and aligned along the z-axis.
    The \ce{Fe6} cluster forms a octahedral structure (6 vertices) with four atoms in the \(x-y\) plane forming a regular square with distance between atoms \(d\simeq 12.7~\text{\AA}\) and the two apex atoms aligned along the $z$ axis at a distance \(d\simeq 11.4~\text{\AA}\).}%
    \label{fig:nc-clusters}%
\end{figure*}

We tested our non-collinear spin implementation by computing the electronic ground state of different magnetic clusters.
In Fig.~(\ref{fig:nc-clusters}) we show the magnetization density field of the \ce{Fe2} magnetic molecule.
The distance between the \ce{Fe} atoms is \(1.99\,\text{\AA}\).
The calculation is performed in a cubic box of size \(15\,\text{\AA}\) to ensure that the molecule is isolated. 
We include the spin-orbit interaction using full-relativistic pseudopotentials.
The total magnetic moment of the \ce{Fe2} molecule is in agreement with previous results and calculations using different codes.
The magnetic moment along the z-axis is \(M_z = 6.12\mu_\text{B}\).
Calculations with and without spin-orbit interaction produce a negligible difference in spin texture, as can be seen from figs.~(\ref{fig:nc-clusters}(a)) and (\ref{fig:nc-clusters}(b)).
The arrows in the figure point along the direction of magnetization at each point in the spatial grid.
It is easy to observe that at each point the spin density is approximately oriented along the z-axis, producing a total net magnetic moment oriented along the z-axis.
The magnetic structure of the \ce{Fe6} cluster is more complex.
The optimized geometry of the cluster and its magnetic structure is consistent with previous works\cite{PhysRevB.94.014423}.
The magnetic moment of the system is approximately oriented along the z-axis with magnitude \(M_{\rm z} = 19.98\,\mu_\text{B}\);
in addition, it has small but not negligible \(x\) and \(y\) components (\(M_{\rm x} = 0.71\,\mu_\text{B}\), \(M_{\rm y} = 0.40\,\mu_\text{B}\)).
This is reflected in a magnetic texture characterized by a higher degree of non-collinearity compared to the \ce{Fe2} case.
In Fig.~(\ref{fig:nc-clusters}(c)) we show the complete spin density profile of the system.
This is compared in panel (d) of the same figure, which shows only the \(x\) and \(y\) components of the spin density obtained after removing the dominant \(z\) component.
Although much smaller in magnitude, this non-collinear contribution is clearly not negligible, as can be seen from Fig.~(\ref{fig:nc-clusters}(d)), and must be accounted for in a complete description of the magnetism of the molecule.

\subsection{Spin non-collinear calculation of magnetically ordered solids}

\begin{table}[ht]
    \centering
    \resizebox{\columnwidth}{!}{%
    \begin{tabular}{cccc}
    \hline
    Ferromagnet &     \multicolumn{2}{c}{Magnetic moment $\left[\mu_{\rm B}\right]$} \\
    & \textsc{INQ} & \textsc{QE}\\
    \hline
    bcc \ce{Fe}   &  \(2.2208\)  &  \(2.2210\) \\
    fcc \ce{Ni}   &  \(0.6379\)  &  \(0.6365\) \\
    fcc \ce{Co}   &  \(1.6362\)  &  \(1.6367\) \\
    \hline
    \end{tabular}
    }
    \caption{Comparison of the magnetic moment calculated by \textsc{INQ} and \textsc{Quantum Espresso (QE)} for different ferromagnetic solids.}
    \label{Tab:magmom}
\end{table}

In Table (\ref{Tab:magmom}) we show the computed magnetic moments of three metallic ferromagnets: bcc \ce{Fe}, fcc \ce{Ni} and fcc \ce{Co}.
The calculations were performed using the locally-collinear LSDA \ac{XC} functional and compared with analogous calculations from \textsc{Quantum Espresso} (\textsc{QE})\cite{Giannozzi_2017}.
The locally-collinear PBE functional produces only minor changes to the final magnetic moments in comparison to the results obtained using LSDA. We use a {\bf k}-points grid of size \(13\times 13\times 13\) and include \(20\) empty bands to reach convergence in \textsc{INQ}.
We also used a Fermi-Dirac smearing with a fictitious temperature of \(300~\mathrm{K}\) to compute the occupations of the electronic states.
We initialized the magnetic moments following the procedure discussed in Algorithm~(\ref{alg:mag_init}).
In comparison to the \ce{Fe6} cluster, here the degree of non-collinearity is much lower, with the magnetic moments fully oriented along the z-axis.
The comparison with the results from \textsc{QE} is quite accurate, and we did not observe sensible variations in the values of the magnetic moments and ground-state energies.

\subsection{Spin-orbit coupling induced energy splittings}

\begin{table}[ht]
    \caption{Energy levels splitting with fully relativistic pseudopotentials for different atoms. Comparison between \textsc{INQ} and \textsc{Quantum Espresso (QE)} results.}
    \centering
    \resizebox{\columnwidth}{!}{%
    \begin{tabular}{cccc}
    \hline
    Element & Orbital & \multicolumn{2}{c}{Level splitting [Ha]} \\
    & & \textsc{INQ} & \textsc{QE}\\
    \hline

    \ce{Xe} & 5p & \(0.0453\) & \(0.0464\) \\
    \ce{Au} & 5d & \(0.0045\) & \(0.0045\) \\
    \ce{Pb} & 5d & \(0.0002\) & \(0.0003\) \\
    \ce{Ag} & 5d & \(0.0030\) & \(0.0032\) \\
    \hline
    \end{tabular}
    }
    \label{Tab:SOCsplit}
\end{table}

One effect of spin-orbit interactions is the splitting of the originally degenerate energy levels of electronic orbitals.
In atoms, the splitting of the energy levels due to spin-orbit coupling is similar in size to the relativistic corrections to the kinetic energy and the Zitterbewegung effect \cite{Strange_1998}, therefore it cannot be neglected.
These corrections contribute to the fine structure of the atom.
In Table~(\ref{Tab:SOCsplit}) we look at the fine structure of the atomic energy levels induced by the fully relativistic pseudopotentials.
For \ce{Xe} we consider the intra-level splitting of the \(5p\) states due to spin-orbit interaction.
For the other elements, we instead consider the intra-level splitting of the lowest energy \(5d\) degenerate states.
Calculations were performed for the isolated atoms shown in Table~(\ref{Tab:SOCsplit}), using the local collinear approximation of LSDA for the \ac{XC} functional.
We use fully relativistic pseudopotentials from the \textit{PseudoDojo} library\cite{VANSETTEN201839}.
Our results appear in good agreement with analogous simulations performed using \textsc{QE}.
As already discussed, the pseudopotential approximation in the case of spin-orbit coupling consists of removing the non-spherical atom-centered contribution in the electric field generated by the \ac{KS} potential in Eq.~(\ref{Eq:VSO_ACA}).
This is often considered a good approximation for the electrons close to the nuclei that experience an electric field mainly determined by the electrostatic bare nuclear potential and the interaction with the core electrons.
In the next section, we analyze the magnetic response to an applied external magnetic field.

\subsection{Calculation of Zeeman splittings}

\begin{table}[ht]
    \caption{Energy level splitting in the \ce{H} atom \(1s\) levels as a function of the applied magnetic field.
    The results are compared with calculations from \textsc{VASP}\cite{KRESSE199615,PhysRevB.47.558}.}
    \centering
    \begin{tabular}{ccc}
    \hline
    \({b}^{\rm z}_\text{ext}\) [eV] & \multicolumn{2}{c}{\(\Delta E_{\rm Z}-\Delta E(B=0)\) [eV] } \\
    & \textsc{INQ} & \textsc{VASP}\\
    \hline
    0.1   & 0.2000    &0.2000\\
    0.3   & 0.6000    &0.6000\\
    0.5   & 1.0000    &1.0000\\
    0.7   & 1.4000    &1.4000\\
    1.0   & 2.0000    &2.0000\\
    \hline
    \end{tabular}
    \label{Tab:ZsplitH}
\end{table}

\begin{table}[ht]
    \caption{Energy level splitting in the \ce{Al} atom \(3s\) levels as a function of the applied magnetic field.
    The results are compared with calculations from \textsc{VASP}\cite{KRESSE199615,PhysRevB.47.558}.}
    \centering
    \begin{tabular}{ccc}
    \hline
     \({b}^{\rm z}_\text{ext}\) [eV] &\multicolumn{2}{c}{ \(\Delta E_\text{Z}-\Delta E(B=0)\) [eV]} \\
   & \textsc{INQ} & \textsc{VASP}\cite{}\\
    \hline
    0.1   & 0.2000    &0.1998\\
    0.3   & 0.6000    &0.5998\\
    0.5   & 1.0000    &0.9998\\
    0.7   & 1.4000    &1.3998\\
    1.0   & 2.0000    &1.9998\\
    \hline
    \end{tabular}
    \label{Tab:ZsplitAl}
\end{table}

In this section, we discuss the calculation of the Zeeman energy splitting in the case of a few simple atoms and compare the results with analogous calculations performed using the \textsc{VASP} code\cite{PhysRevLett.106.107202}.
We calculate the energy splitting under an applied external magnetic field of the \(\mathrm{1s}\) energy levels in \ce{H} (table~\ref{Tab:ZsplitH}) and the \(\mathrm{3s}\) levels of \ce{Al} (table \ref{Tab:ZsplitAl}).
For each case, we apply an external magnetic field of different intensities.
The simulation is performed in \textsc{INQ} using the non-collinear implementation, but the same results are obtained also using the standard collinear implementation (as expected for this simple case and in the presence of an external magnetic field).
The results agree well with the two codes and are easy to understand from the expression of the Zeeman coupling.
In this particular case, following Eq.~(\ref{eq:zeeman_potential}) we have the Zeeman potential
\begin{equation}
    \hat{\mathcal{V}}_{\rm z} = \mu_\text{B}\sigma_{\rm z}\cdot b_\text{ext}^{\rm z}=
    \mu_\text{B}\left [ {\begin{array}{cc}
    b_\text{ext}^z & 0\\
    0 & - b_\text{ext}^z\\
  \end{array} } \right ]\ .
\end{equation}
This term breaks the degeneracy in the spin space, producing different energies for the spin up- and down-states \(s = \pm 1\), \(E_{\rm z}^\text{s} = s b_\text{ext}^{\rm z}[eV]\).
Theoretically, the  Zeeman splitting energy is given by \(\Delta E_{\rm z} = 2b_\text{ext}^{\rm z}[eV]\), which is consistent with the values in the tables~(\ref{Tab:ZsplitH}) and (\ref{Tab:ZsplitAl}) for both \ce{H} and \ce{Al}.

\subsection{Calculation of Zeeman response}

\begin{figure*}%
    \centering
    \subfloat[\centering Comparison between \textsc{INQ} and \textsc{VASP} magnetic-moment response to an external magnetic field in BCC \ce{Na}.]{{\includegraphics[width=0.46\textwidth]{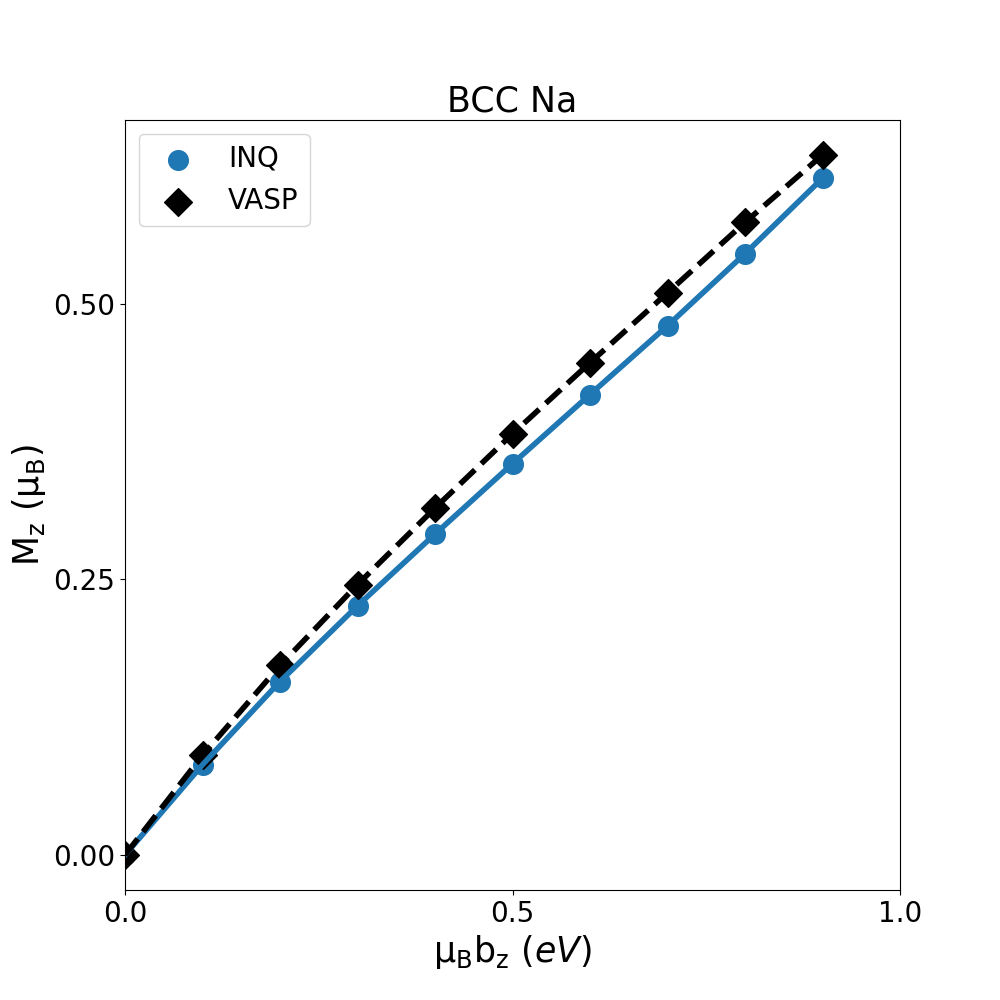} }}%
    \qquad
    \subfloat[\centering Comparison between \textsc{INQ} and \textsc{VASP} magnetic-moment response to an external magnetic field BCC \ce{Al}.]{{\includegraphics[width=0.46\textwidth]{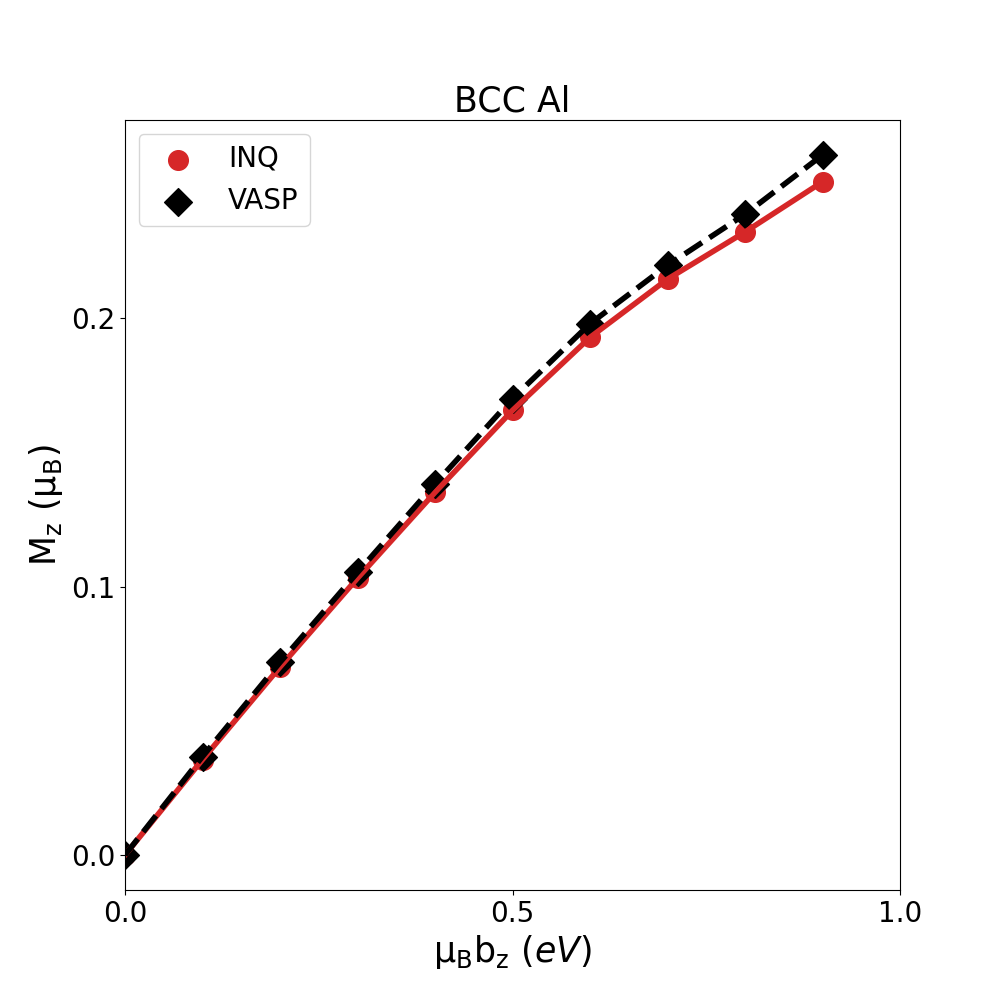} }}%
    \\
    \subfloat[\centering Comparison between \textsc{INQ} and \textsc{VASP}magnetic-moment response to an external magnetic field BCC \ce{Rb}.]{{\includegraphics[width=0.48\textwidth]{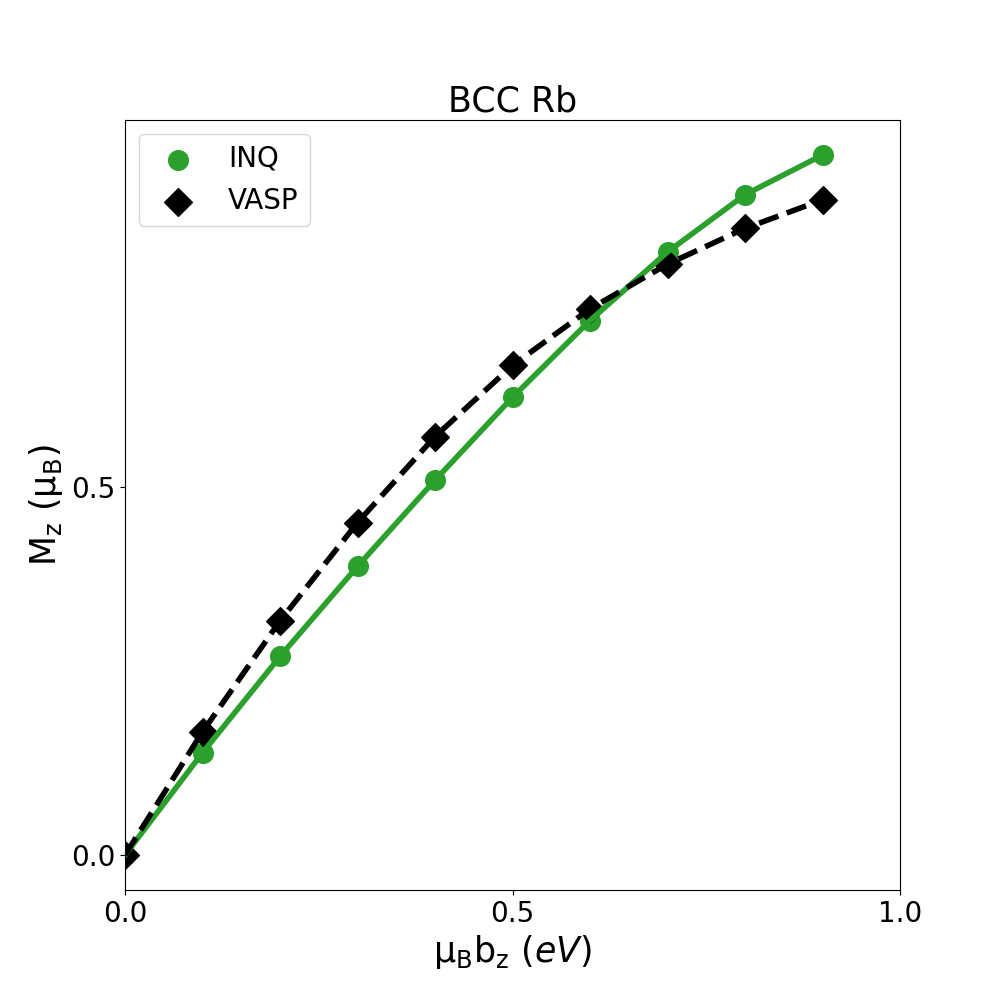} }}%
    \qquad
    \subfloat[\centering Comparing of the paramagnetic metals as a function of magnetic field strength for BCC \ce{Na}, \ce{Al} and \ce{Rb}.]{{\includegraphics[width=0.44\textwidth]{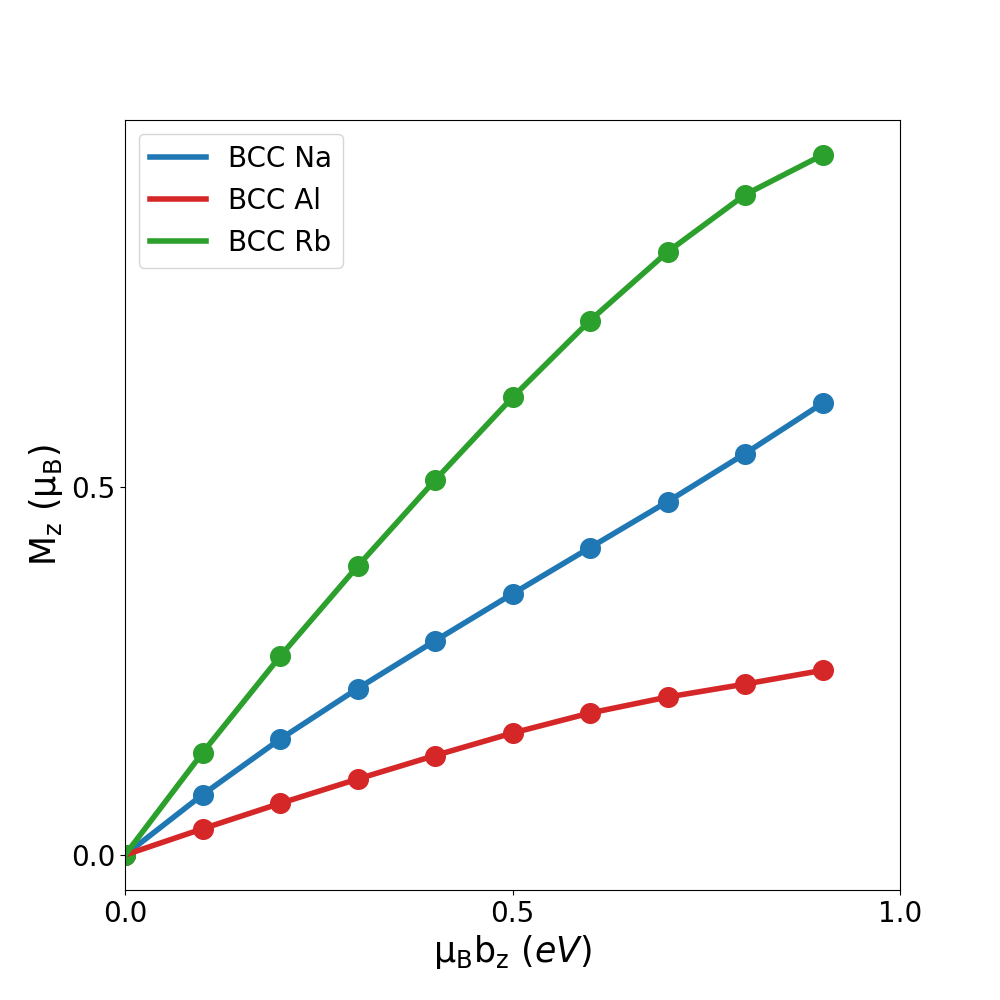} }}%
    \caption{Magnetic moment response under externally applied magnetic fields in paramagnetic metal solids. The materials are BCC \ce{Na}, BCC \ce{Al} and BCC \ce{Rb}.}
    \label{fig:paramag-extB}%
\end{figure*}

The first-order magnetic susceptibility of conduction electrons in metals consists of two contributions\cite{dAvezac2007}.
The first, \(\chi_\text{P}\), comes from the spin magnetic moment response under the application of an external magnetic field \cite{jensen1991rare}, the second is the orbital response to an applied external magnetic field\cite{Varsano2009,Andrade2010,Lebedeva2019-er}. Second-order magnetic response can be important in some cases \cite{2022Kaneko} too.
Here we consider the first contribution, the Pauli paramagnetism that arises from the Zeeman coupling between the electron spin and the externally applied magnetic field. As a consequence, these results should not be taken as an accurate description of the magnetic response in real materials, but only as benchmarks of our implementation of the Zeeman paramagnetic coupling.

In Fig.~\ref{fig:paramag-extB} we show the comparison in the magnetic-moment response under externally-applied magnetic fields of different strengths in the case of three different paramagnetic metals: {\rm BCC} \ce{Na} (Fig.~\ref{fig:paramag-extB}a), {\rm BCC} \ce{Al} (Fig.~\ref{fig:paramag-extB}b) and {\rm BCC} \ce{Rb} (Fig.~\ref{fig:paramag-extB}c).
The results are compared with calculations performed with \textsc{VASP}.
In both cases, we used a local-collinear LSDA functional and a \(8\times 8\times 8\) {\bf k}-points grid to compute the electronic ground state in the presence of the magnetic field. We use a Fermi-Dirac electronic smearing with a linewidth of approximately \(0.2\,eV\) in both calculations.
The external magnetic field, \({\bf b}_\text{ext}\), is spatially homogeneous and oriented along the {\rm z}-axis.
The results appear in good agreement in the case of \ce{Al} and \ce{Na}, while in the case of \ce{Rb} the differences between the two curves are more evident. In all the cases we observe that the difference between the two curves increases for larger external fields. This can be attributed to the different pseudopotentials employed in the two calculations, \ac{PAW} in \textsc{VASP} and norm-conserving in \textsc{INQ}. Differences in the magnetic moments computed with norm-conserving and \ac{PAW} potentials have been observed before \cite{PhysRevB.53.10685,10.1063/9.0000202,10.1063/1.4992138},  and the effect can become larger at high magnetic fields when semi-core states become more important.
In the case of \ce{Rb}, there is a steady increase in the magnetic moment as a function of the applied field, with a higher slope compared to the other two metals \ce{Al} and \ce{Na} (Fig.~\ref{fig:paramag-extB}d) which is associated with a higher magnetic susceptibility.

\subsection{Real time dynamics under applied magnetic fields}

We have established our simulations yield reliable results for the ground state.
We can now explore the results of our non-collinear real-time TDDFT simulations for spin dynamics.

\subsubsection{single atom dynamics}

\begin{figure*}%
    \centering
    \subfloat[\centering \ce{H} atom: magnetization under external homogeneous magnetic field oriented along three different directions.]{{\includegraphics[width=0.46\textwidth]{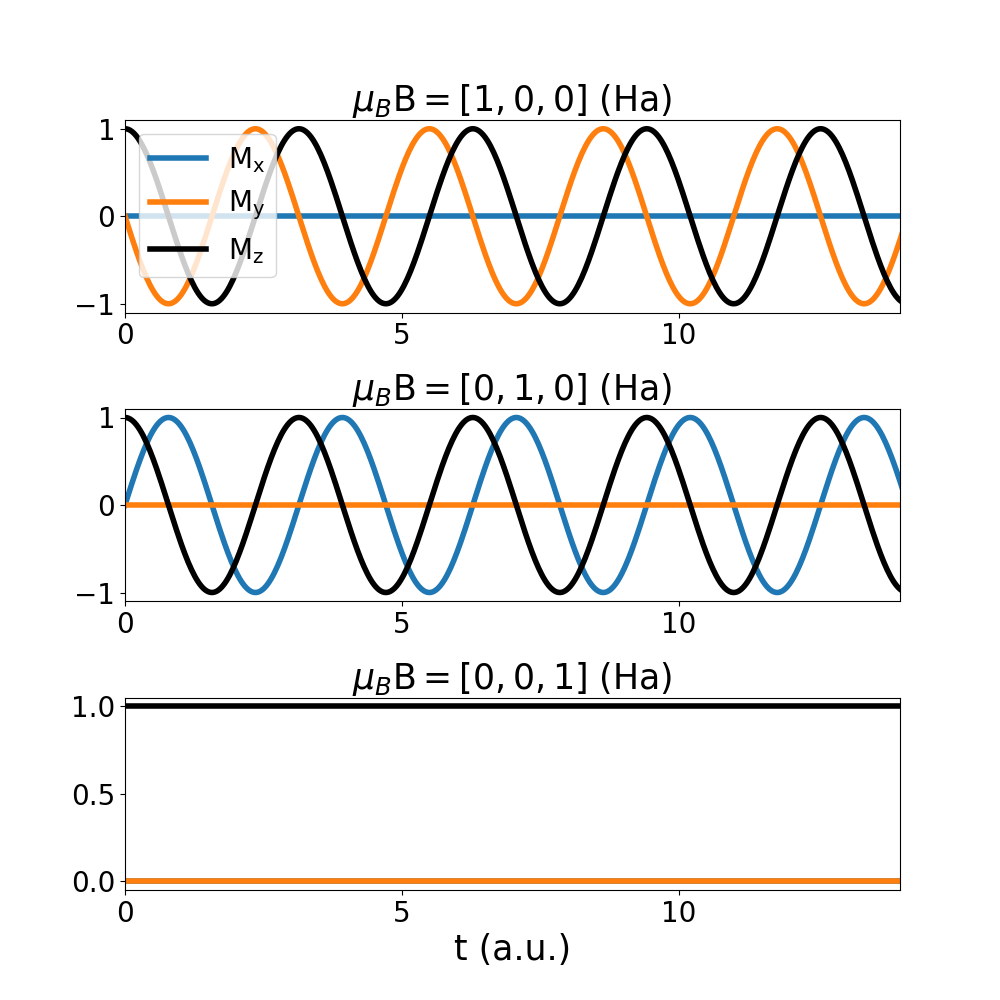} }}%
    \qquad
    \subfloat[\centering \ce{H} atom: magnetization under external homogeneous magnetic field of three different amplitudes.]{{\includegraphics[width=0.46\textwidth]{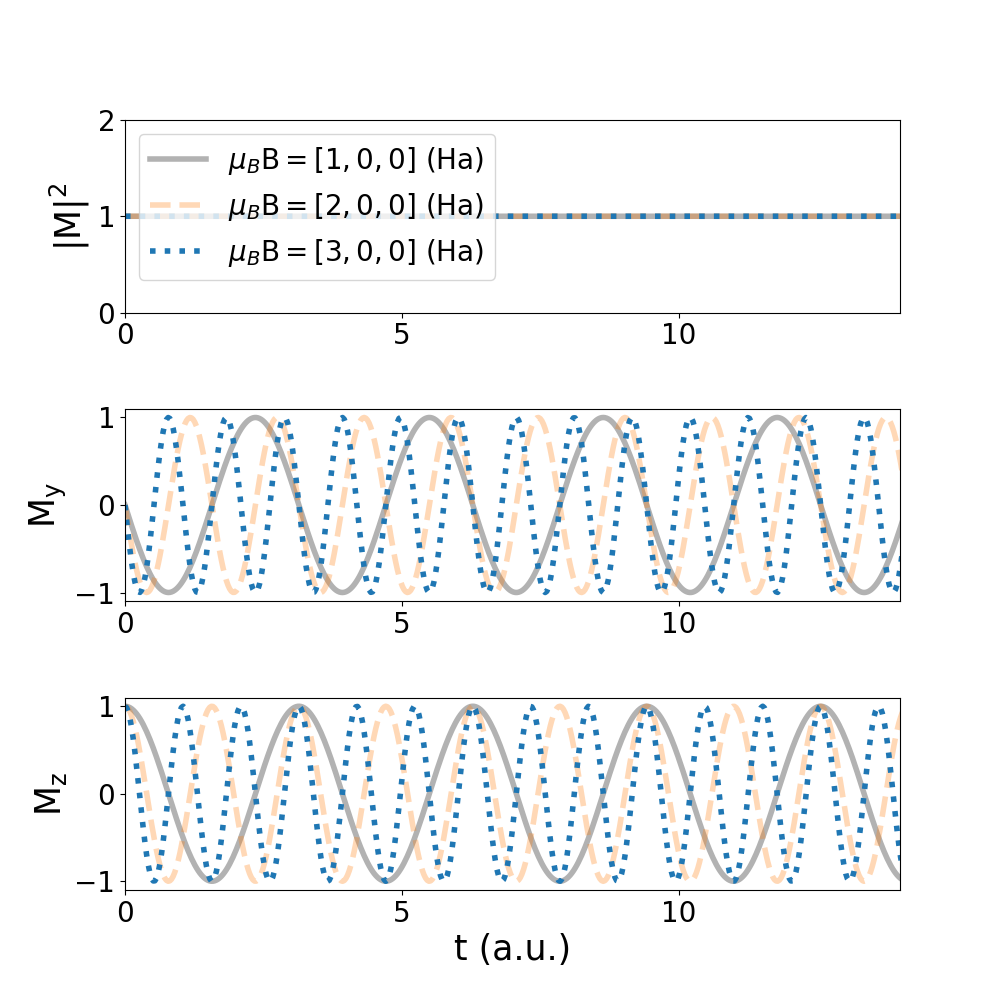} }}%
    \\
    \subfloat[\centering \ce{H} atom: fit of magnetization dynamics over sine function for different magnetic field strengths.]{{\includegraphics[width=0.48\textwidth]{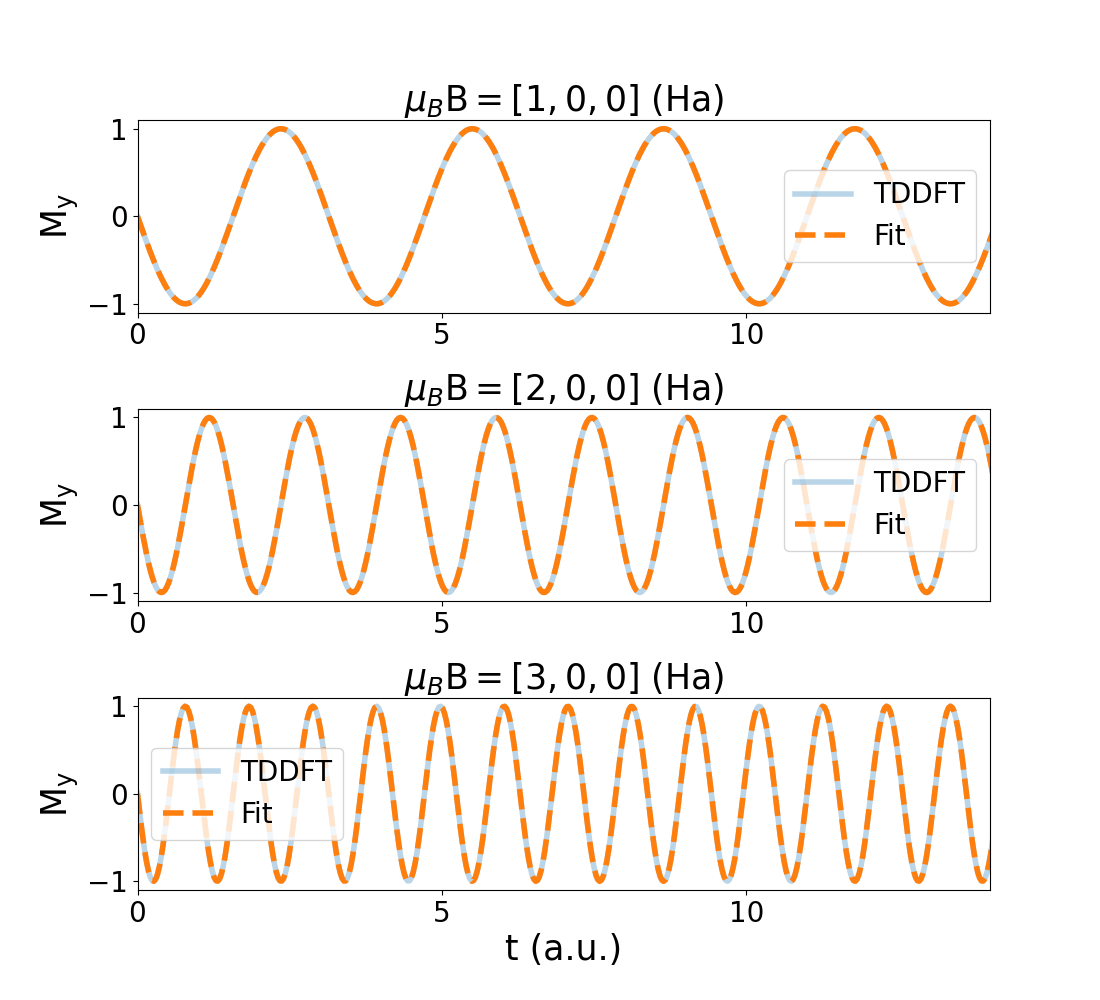} }}%
    \qquad
    \subfloat[\centering \ce{H} atom: \(\omega_{\rm B}\) as a function of magnetic field strength.]{{\includegraphics[width=0.44\textwidth]{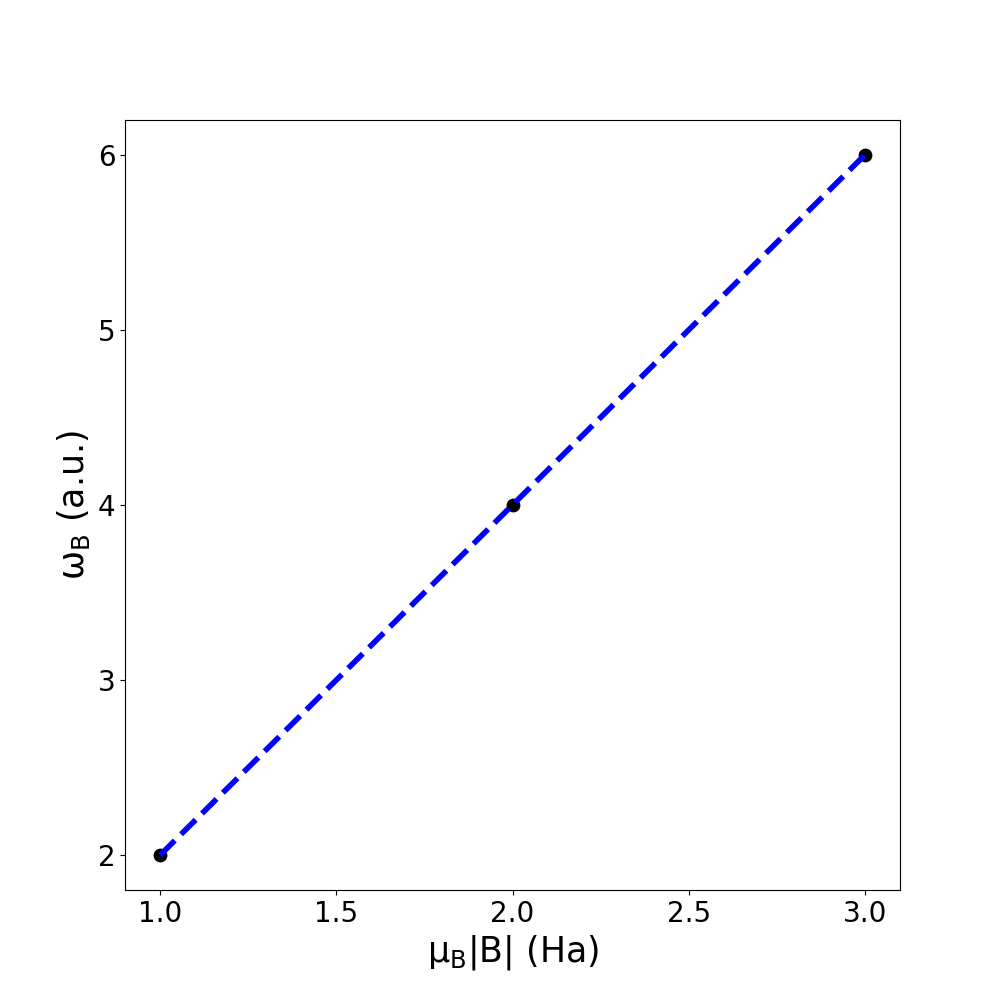} }}%
    \caption{
    Time-dependent evolution of \ce{H} atom under an external magnetic field.
    In panel (a) we compare the dynamics using fields with different polarization directions;
    in panel (b) we consider magnetic fields with different amplitudes;
    in panel (c) we fit the results against a sine function;
    in panel (d) we plot the oscillation frequency as a function of the field amplitude.
    The magnetic moments are always in units of \(\mu_{B}\). }%
    \label{fig:H-extB}%
\end{figure*}

We start with a simple case, the real-time TDDFT dynamics of a single \ce{H} atom under the influence of an external magnetic field. 
Without loss of generality we set the initial magnetic moment oriented along the {\rm z}-axis while changing the orientation of the magnetic field.
To analyze our simulation results we can derive the expected spin dynamics.
In the absence of spin-orbit interaction, the spin continuity equation, Eq.~(\ref{eq:spincont}), can be simplified further given that the effect of the spin currents is negligible in a single isolated atom.
The \ac{XC} magnetic field for a spin polarized hydrogen atom sums up to the applied external magnetic field (\(B = 1\) is in atomic units and corresponds to \(0.5~\mathrm{Ha}\) of Zeeman energy).
The equation reduces to Eq.~(\ref{eq:GMT}), but now the magnetization field \({\bf m}({\bf r},t)\) is completely uniform throughout the simulation box.
We can then write 
\begin{equation}\label{eq:MOmega}
    \frac{\mathrm{d}}{\mathrm{d}t}{\bf M}_\Omega = 2\mu_\text{B}{\bf B}\times{\bf M}_\Omega(t)\ .
\end{equation}
The \ac{XC} field cannot, in fact, exert a global torque on the magnetization vector. 

Fig.~\ref{fig:H-extB}a shows the dynamics of the magnetic moment when the external magnetic field \(\mathbf{B}\) is applied along three different directions.
As expected, this produces perfectly oscillatory dynamics with only the component along the field axis that remains static.
When the field is applied along the spin polarization direction \(z\), the atom has no dynamics.

In Fig.~\ref{fig:H-extB}b we keep the magnetic field direction fixed along the \({\rm x}\) axis and change its amplitude instead.
The magnetization module \(|\mathbf{M}_\Omega|\) remains constant over time and the dynamics is fully coherent, since there is no source of dissipation.
Meanwhile, the \({\rm y}\) and \({\rm z}\) components of \({\bf M}_\Omega\) show the oscillatory behavior, with a frequency that changes with the intensity.
As shown in Fig.~\ref{fig:H-extB}c we can perfectly fit a sinusoidal function \(A\sin(\omega_\text{B}t)\) to the magnetization (in the \({\rm z}\) direction) for each value of the intensity.
In Fig.~\ref{fig:H-extB}d we plot the fitted value of \(\omega_\text{B}\) for the different values of the magnetic field amplitude.
The linear relation between frequency and magnetic field is perfectly reproduced with a factor \(2\) consistent with Eq.~(\ref{eq:MOmega}).

\subsubsection{Bulk \ce{NiO}}

\begin{figure*}%
    \centering
    \subfloat[\centering simulation at \(t=0\)]{{\includegraphics[width=0.46\textwidth]{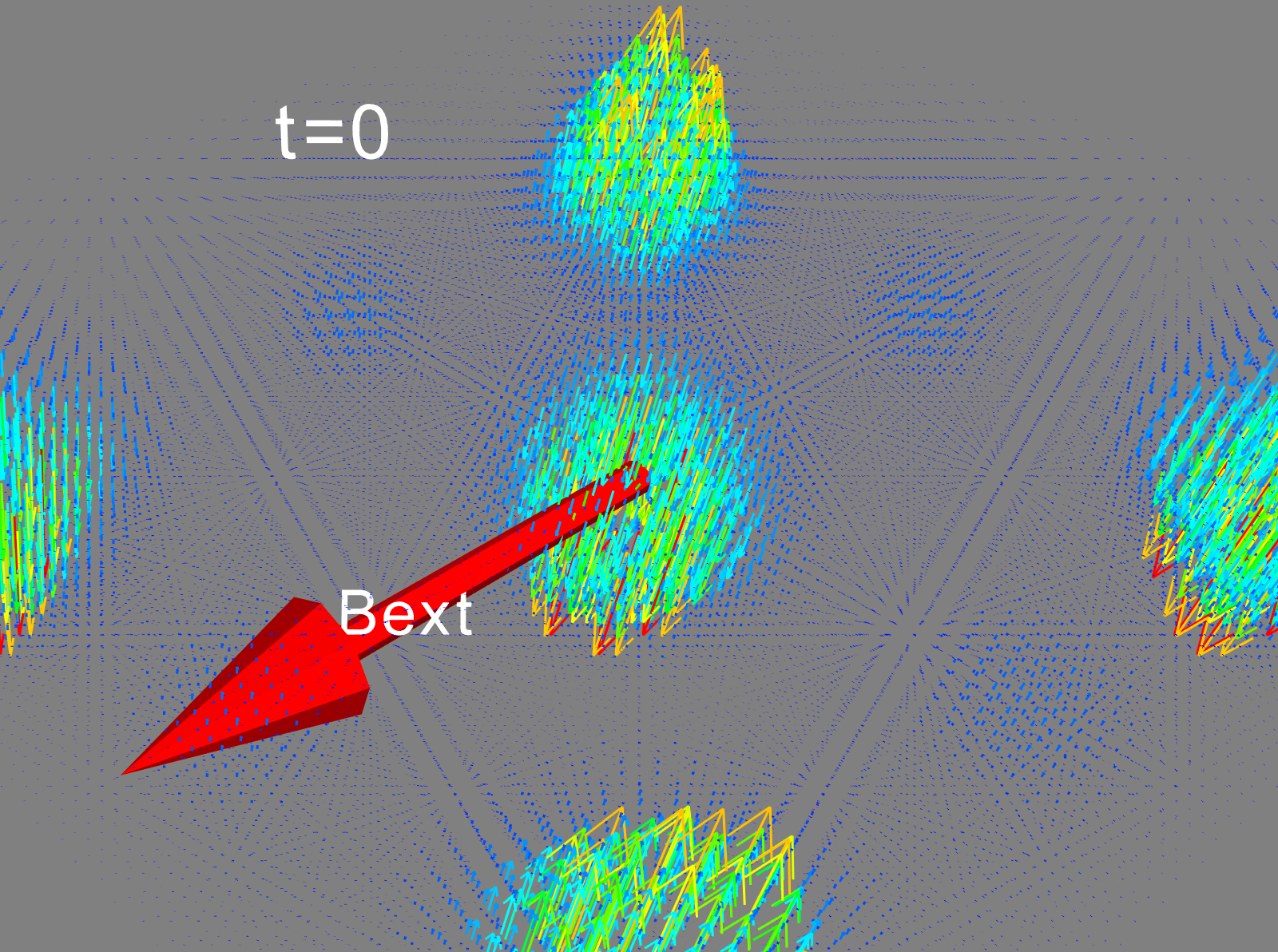} }}%
    \qquad
    \subfloat[\centering simulation at \(t=40\)]{{\includegraphics[width=0.46\textwidth]{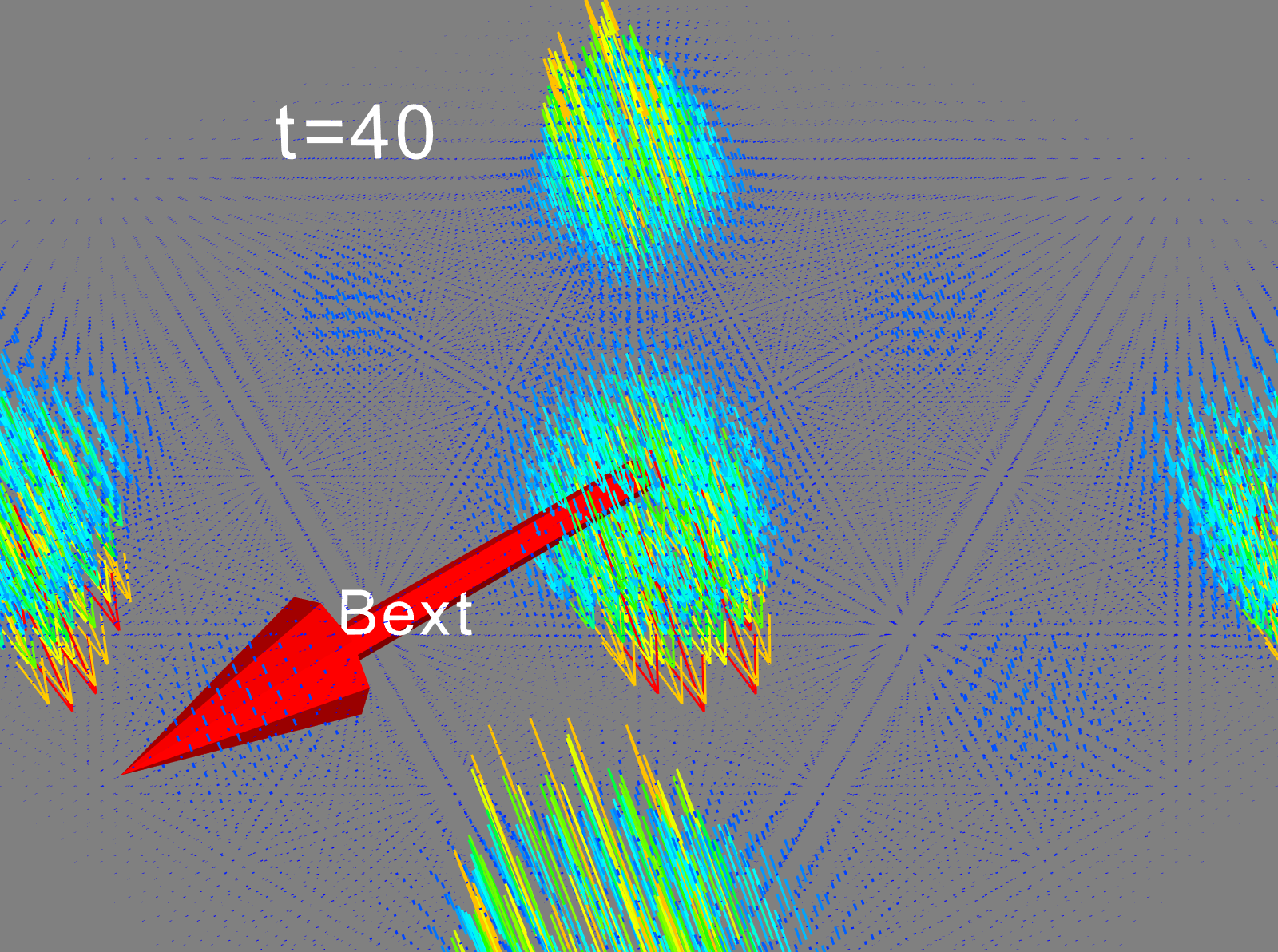} }}%
    \\
    \subfloat[\centering simulation at \(t=60\)]{{\includegraphics[width=0.46\textwidth]{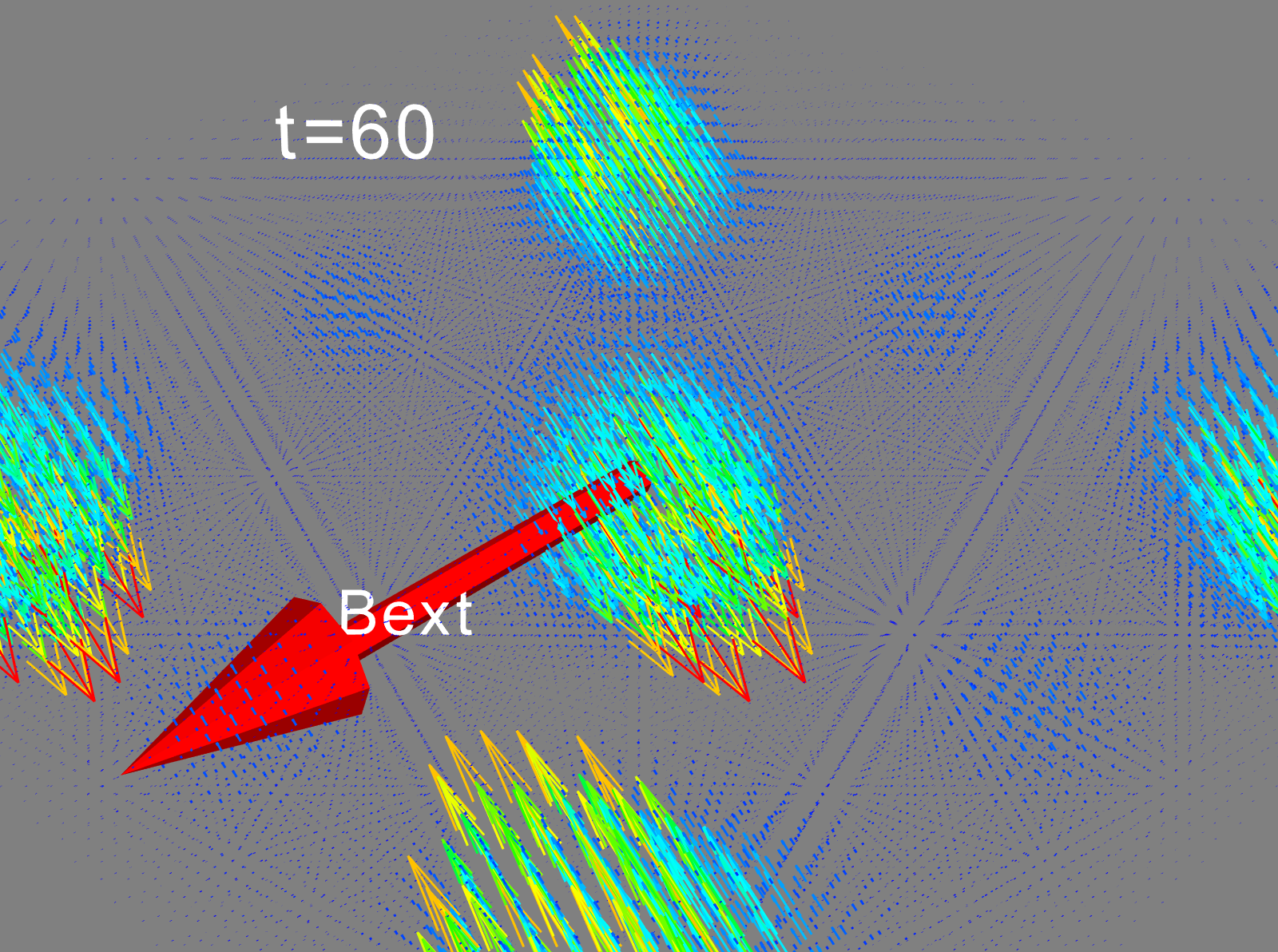} }}%
    \qquad
    \subfloat[\centering simulation at \(t=80\)]{{\includegraphics[width=0.46\textwidth]{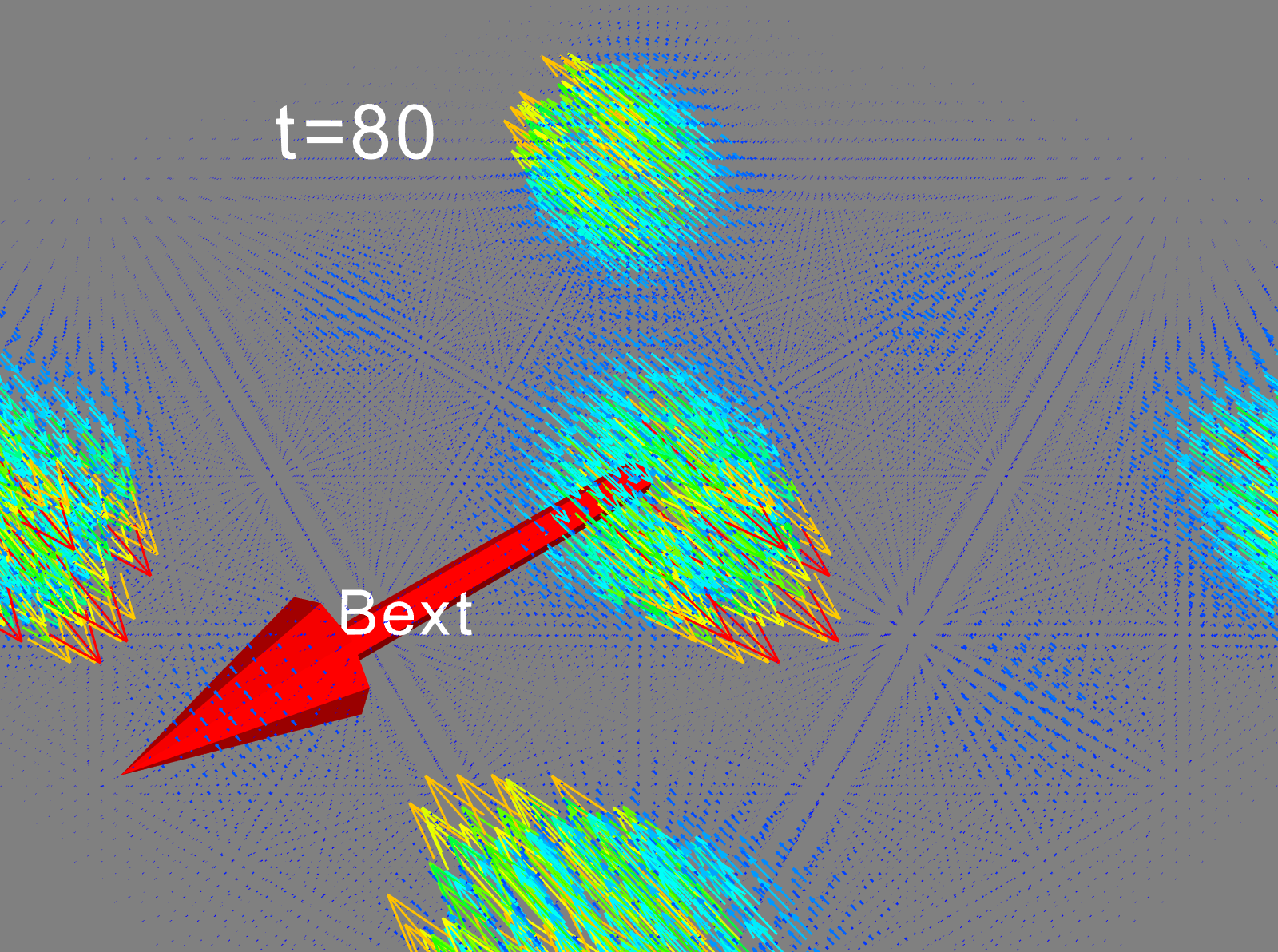} }}%
    \\
    \subfloat[\centering simulation at \(t=100\)]{{\includegraphics[width=0.46\textwidth]{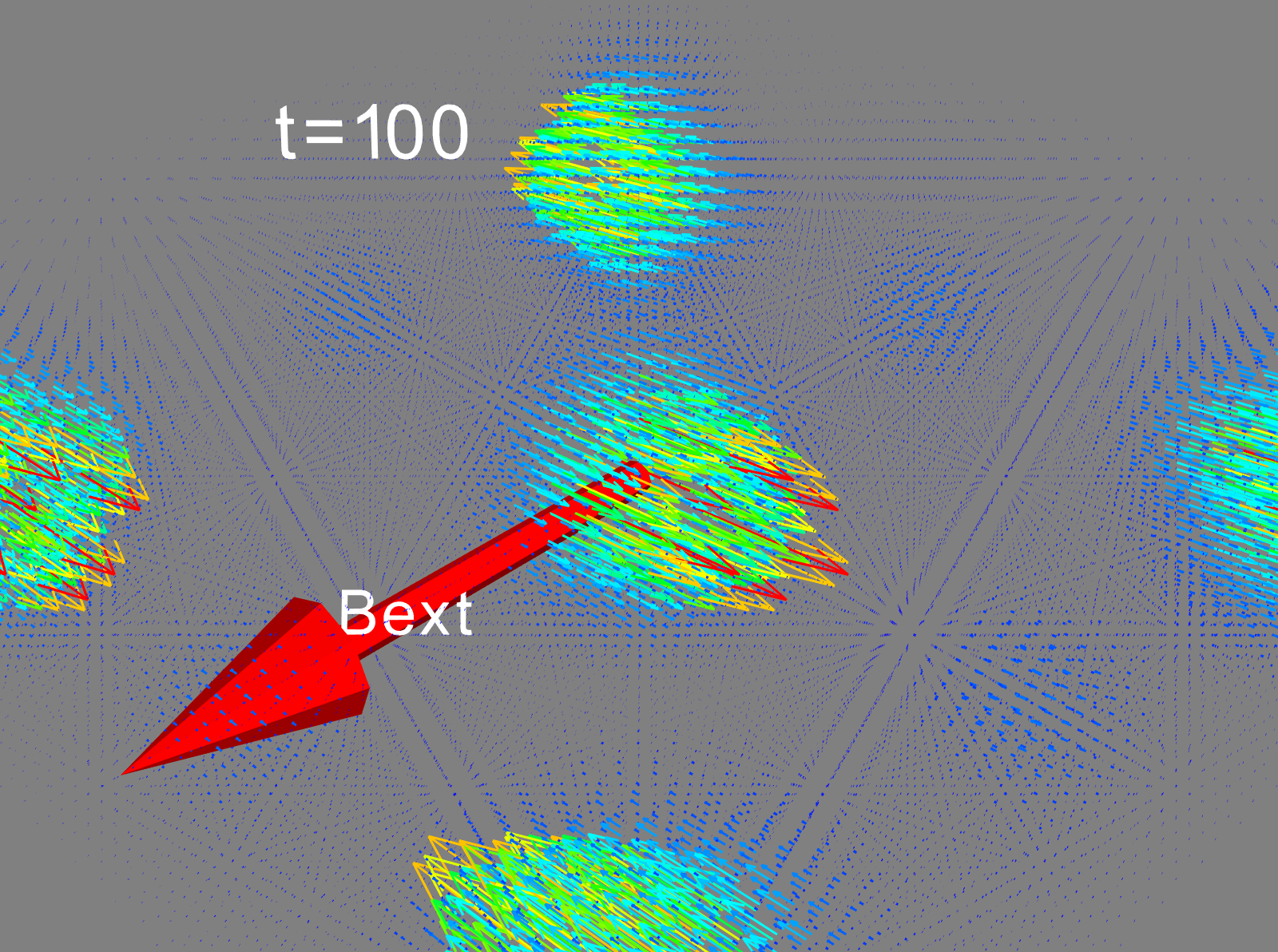} }}%
    \qquad
    \subfloat[\centering simulation at \(t=130\)]{{\includegraphics[width=0.46\textwidth]{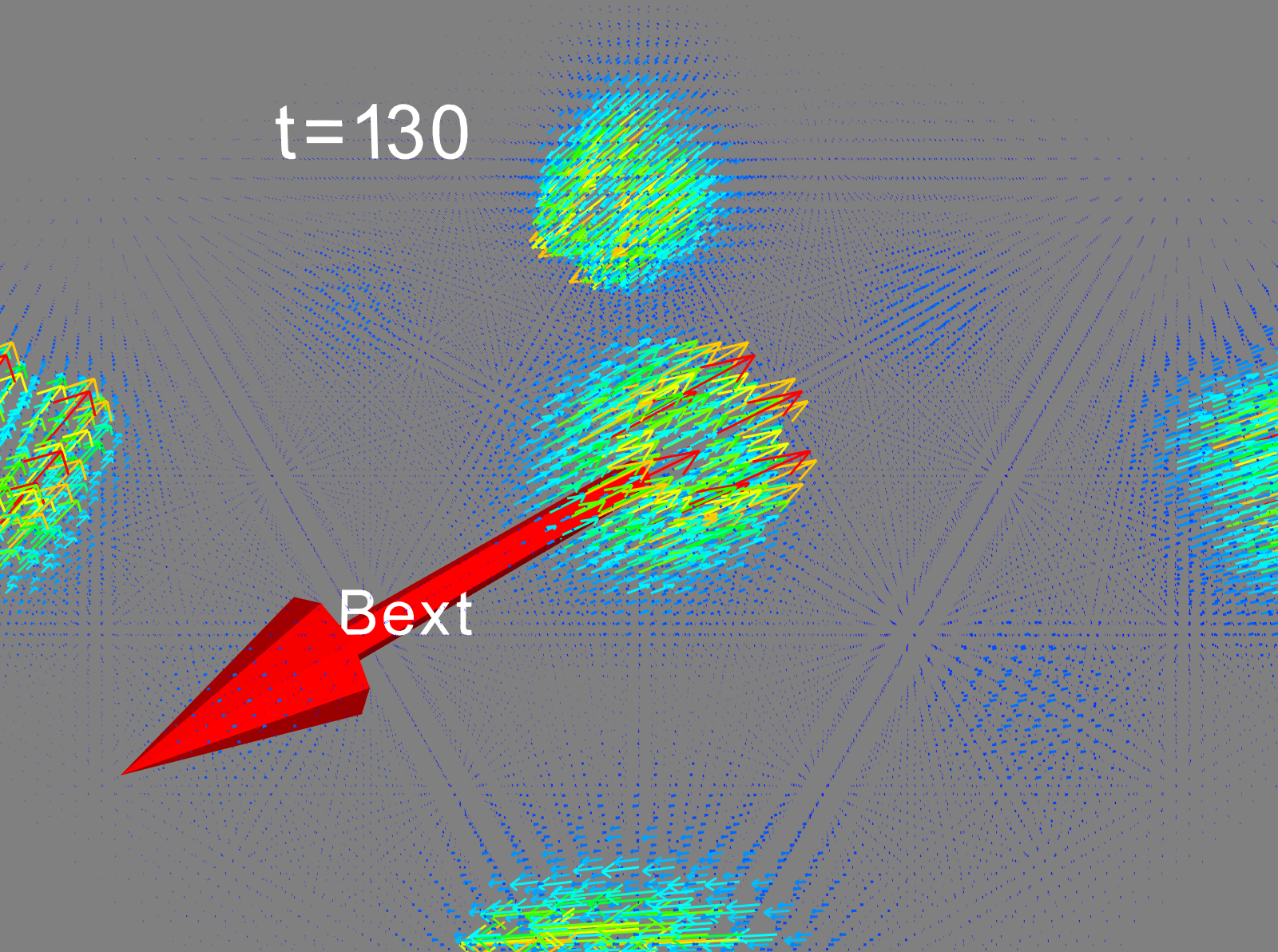} }}%
    \caption{
    Time dependent evolution of \ce{NiO} under an external magnetic field of \(2~\mathrm{a.u.}\) of amplitude (\(1~\mathrm{Ha}\) of Zeeman energy).
    The external magnetic field is applied along the \({\rm x}\)-axis and the magnetic moments are approximately aligned along the \({\rm z}\)-axis at \(t = 0\).
    In the different panels we show the magnetic moments precessing around the magnetic field direction over time.
    The time is expressed in atomic units, and the time step during the simulation is set to \(\Delta t=0.005~\mathrm{a.u.}\).
    }%
    \label{fig:nio-dyn1}%
\end{figure*}

We now look at the real-time dynamics of the anti-ferromagnet \ce{NiO} under applied external and homogeneous magnetic fields.
The results are shown in Fig.~(\ref{fig:nio-dyn1}).
The electronic ground state is characterized by two opposite local magnetic moments with approximate magnitude \(1.1~\mu_\text{B}\) and oriented along the z-axis localized around the two \ce{Ni} atoms (the \ce{O} atoms are non-magnetic).
The calculation converges on a \(2\times 2\times 2\) {\bf k}-grid that allows us to perform real-time simulations more efficiently.
The applied magnetic field shown in the figure is oriented along the {\rm x} axis and remains constant throughout the simulation.
The total magnetization of the system does not change over time;
the system remains in its anti-ferromagnetic state.
From Eq.~(\ref{eq:GMT}) we have that the global magnetic torque acting on the anti-ferromagnet is zero.
The effective equation for the two magnetic moments, \({\bf M}^\text{A/B}\) is
\begin{equation}
    \frac{\mathrm{d}}{\mathrm{d}t}{\bf M}^\text{A/B} = 2\mu_\text{B}{\bf B}\times{\bf M}^\text{A/B}(t)
\end{equation}
The total magnetization \({\bf M}_\Omega={\bf M}^\text{A}+{\bf M}^\text{B}\) integrated in the cell is approximately zero in the ground state, and therefore its derivative is also zero \(\dot{\bf M}_\Omega = 0\) and its value remains constant over time.
The applied magnetic field does not perturb the electronic density;
as a consequence, the dynamics is adiabatic and remains fully coherent, with magnetic moments not changing in magnitude over time.
In the six panels of Fig.~(\ref{fig:nio-dyn1}) we show snapshots of the spin density profile at different times during the real-time simulation.
The magnetization density texture clearly shows the magnetic moments rotating around the axis parallel to the direction of the applied field.
This simulation demonstrates the capability of \textsc{INQ} of performing real-time spin dynamics in magnetic solids.

\subsubsection{\ce{Fe6} and bcc \ce{Fe}}

\begin{figure*}%
    \centering
    \subfloat[\centering \ce{Fe6} - \(M_z\) dynamics under an external magnetic pulse.
    The pulse has polarization as shown in panel (c).
    Dashed black line is with spin orbit and the blue line without.]{{\includegraphics[width=0.46\textwidth]{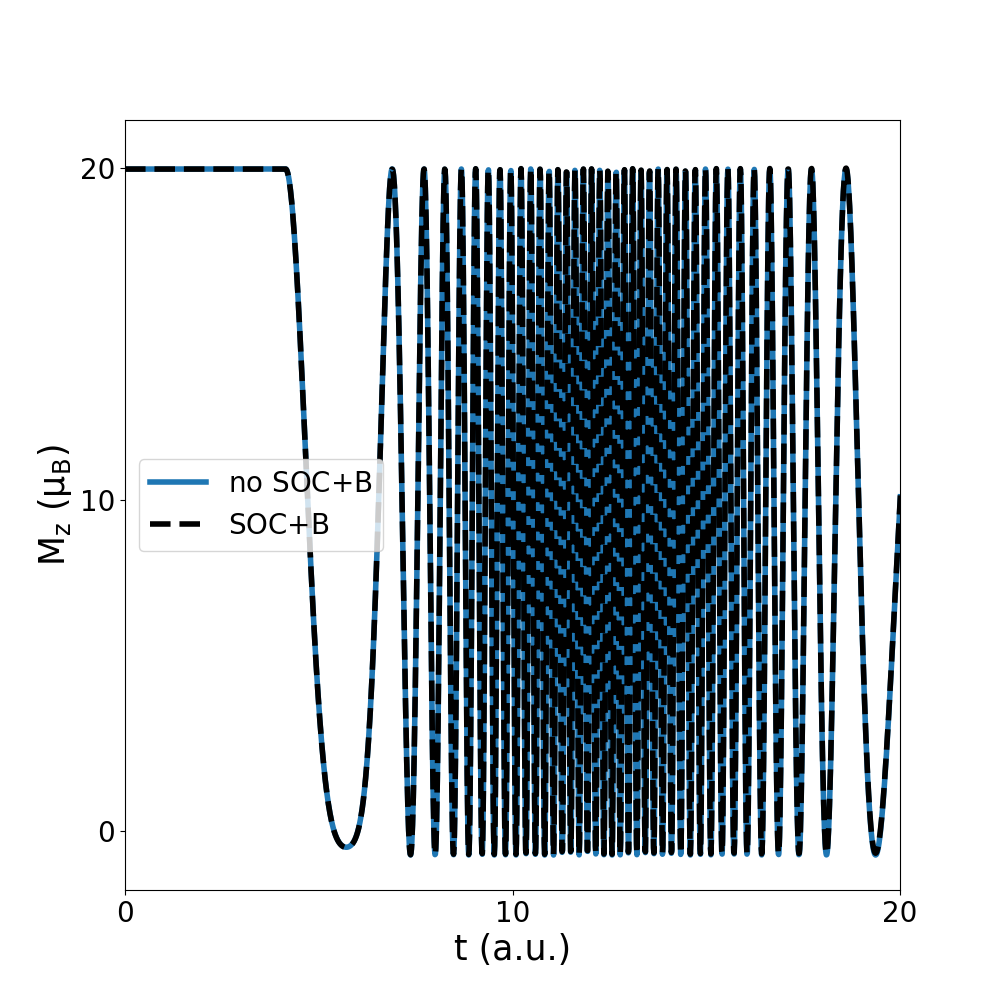} }}%
    \qquad
    \subfloat[\centering bcc \ce{Fe} - \(M_z\) dynamics under an external magnetic pulse. Same external pulse as in panel (a); blue line and dashed black line without and with spin-orbit interaction. ]{{\includegraphics[width=0.46\textwidth]{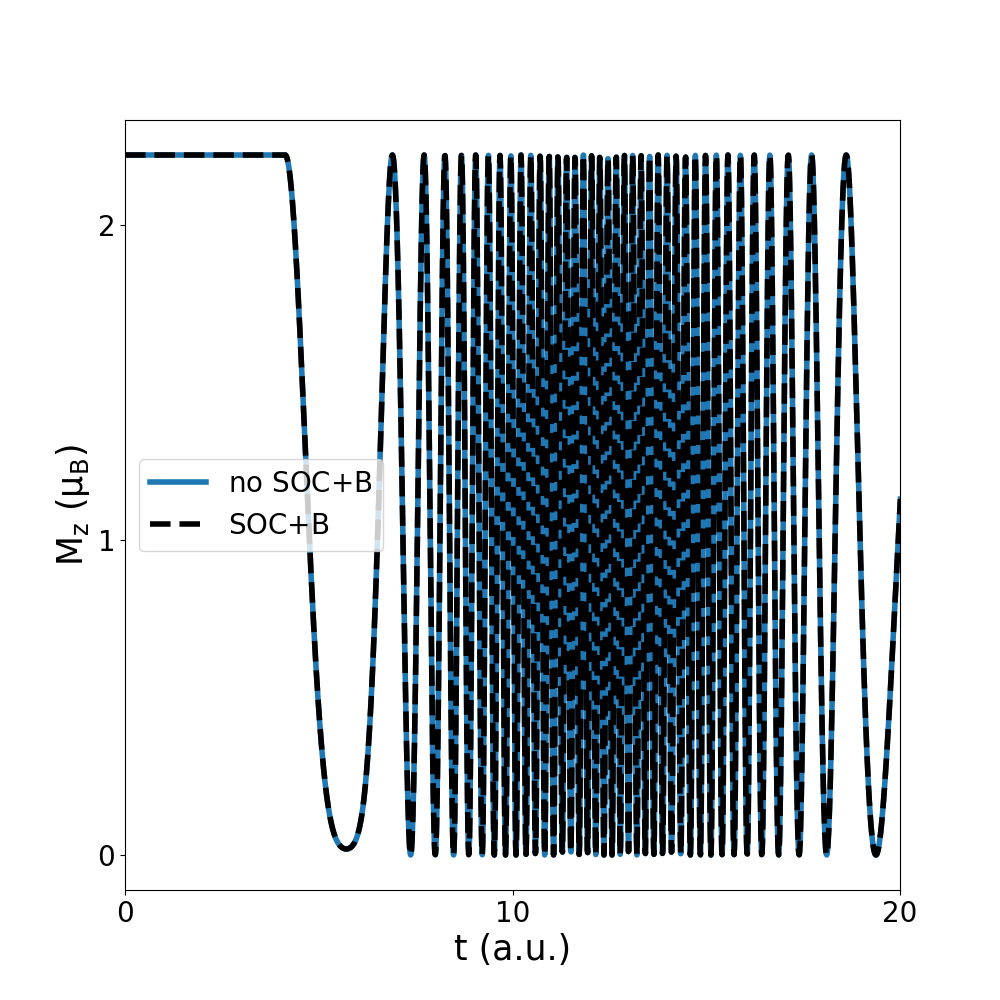} }}%
    \qquad
    \subfloat[\centering Applied external magnetic pulse with polarization vector $\boldsymbol{\epsilon}=\begin{psmallmatrix} 1 & 0 & 1 \end{psmallmatrix}$. ]{{\includegraphics[width=0.46\textwidth]{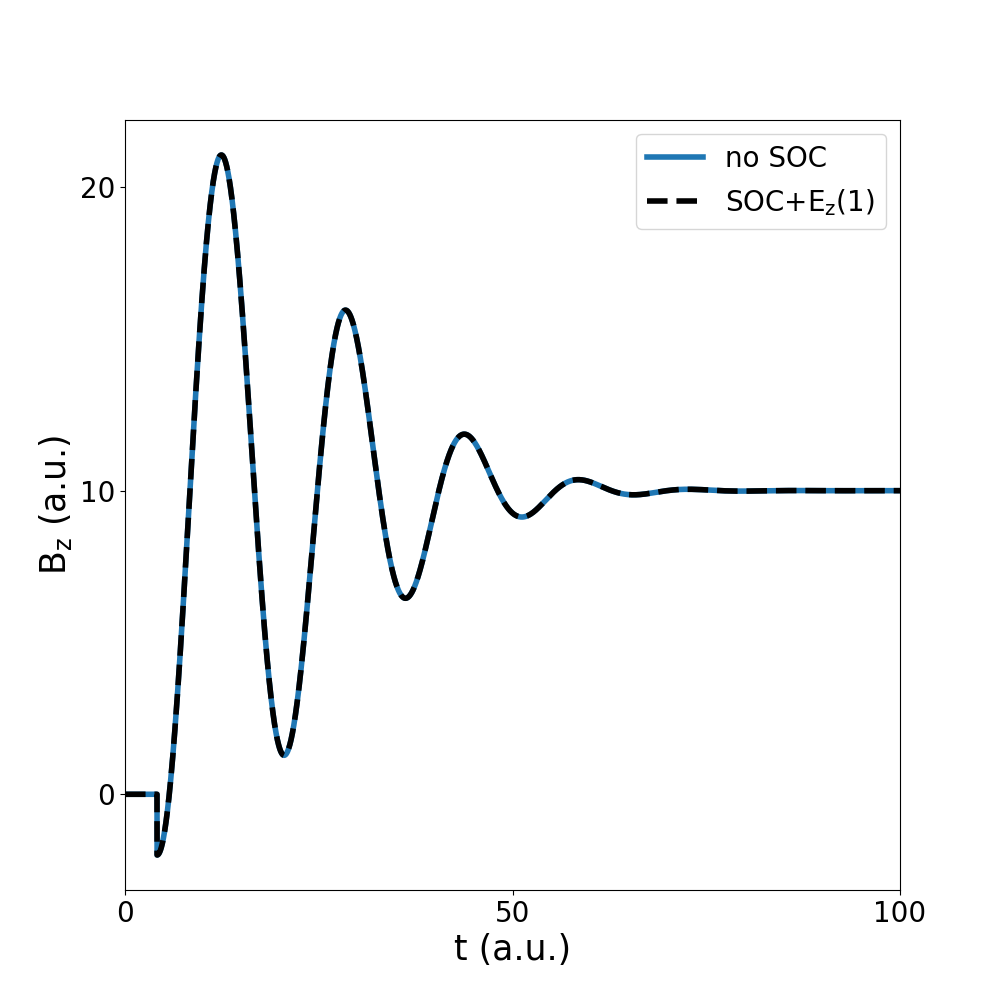} }}%
    \caption{Time-dependent evolution of \ce{Fe6} and bcc \ce{Fe} under an external time-dependent magnetic pulse.
    The functional form of the pulse is \({\bf B}(t)= A\boldsymbol{\epsilon} \big[\cos(\omega_0(t - t_0)+\phi)\exp\big(-\frac{(t-t_0)^2}{2\tau^2}\big) -1\big]\Theta(t-t_0) + \Delta_{B}\Theta(t-t_0)\), where \(\Theta\) is the Heaviside step function.
    In panels (a) and (b) we look at the induced spin dynamics in \ce{Fe6} and bcc \ce{Fe}.
    In panel (c) we show the time-dependent profile of the magnetic pulse.
    Note that in current implementation, the external magnetic field is only introduced through spin Zeeman coupling;
    therefore no difference of magnetization dynamics between with and without SOC.}%
    \label{fig:fe-dyn-Bext}%
\end{figure*}

We now consider the case of a time-dependent magnetic field.
In Fig.~(\ref{fig:fe-dyn-Bext}) we look at the evolution of magnetization in the \ce{Fe6} cluster;
this forms a octahedral structure with a spin density profile shown in panels (c) and (d) of Fig.~\ref{fig:nc-clusters}, and bcc \ce{Fe} under an applied external magnetic field that is uniform but that changes in time.
The two systems are initially, at time \(t = 0\), in their equilibrium ground-state configuration.
In both cases, the total magnetic moment is oriented along the {\rm z}-axis;
\ce{Fe6} has a total magnetic moment of approximately \(20\,\mu_\text{B}\) (\(3.33~\mu_\text{B}\) per atom), while bulk bcc \ce{Fe} has a computed magnetic moment in the unit cell of \(2.22~\mu_\text{B}\) per atom.
Both systems are excited using the same magnetic pulse, shown in Fig.~\ref{fig:fe-dyn-Bext}c.
The dynamics are extremely similar, except for the different amplitude determined by the initial magnetization of the two systems and the small negative value of the spin oscillations in the case of \ce{Fe6}.
The precession frequency of oscillations is the same since it is induced by the same externally applied field.
We do not observe any difference in the simulation with and without spin-orbit interaction.
In general, an orbital response to an external magnetic field is not zero due to orbital angular momentum contributions~\cite{Xu2024-lw}, which may affect spin due to spin-orbit interactions,  
but we are not considering this contribution here with only spin Zeeman interactions introduced in section 3.4.
The external magnetic field couples with the spin degrees of freedom via \(\hat{\mathcal{V}}_{\rm z}\);
a dynamical modification of the spin-orbit requires an electronic excitation close to the nuclei that modifies the orbital state of the system.
This excitation can be obtained by applying an optical pulse;
in contrast, a magnetic pulse cannot induce this type of excitation within the current implementation level.
In the next section, we discuss real-time simulations in the presence of optical pulses.

\subsection{Real-time dynamics under uniform electric fields}

The dynamics of magnetic systems under external electric pulses is qualitatively different and richer compared to the dynamics under magnetic pulses alone.
In the first place, the electric pulse influences the magnetization dynamics via the spin-orbit interaction.
The spin-orbit, in fact, provides a possible conversion mechanism between orbital and spin degrees of freedom.
At the same time, the electric pulse excites the orbital degrees of freedom of the magnetic system and causes a charge imbalance that can induce a modification of the spin degrees of freedom.
These effects have been observed experimentally \cite{PhysRevLett.76.4250, PhysRevLett.85.844, RevModPhys.82.2731} and numerically simulated using \ac{TDDFT} in the case of ferromagnetic metals and magnetic clusters \cite{PhysRevB.94.014423, PhysRevB.96.054411, PhysRevB.95.024412, Krieger_2017, PhysRevB.105.134425, Krieger_2015}.
On the other hand, ultrafast spin dynamics in semiconductors has been studied with first-principles density-matrix dynamics including applied pump pulses, electron-phonon, electron-electron interactions with spin-orbit couplings~\cite{Xu2021-eo,Xu2024-cb,Li2024-lu,Xu2024-lw}.

\subsubsection{\ce{Fe2} molecule}

\begin{figure}%
    \centering
    \subfloat[\centering \ce{Fe2}: \(M_{\rm z}\) dynamics under an external laser. ]{{\includegraphics[width=0.9\columnwidth]{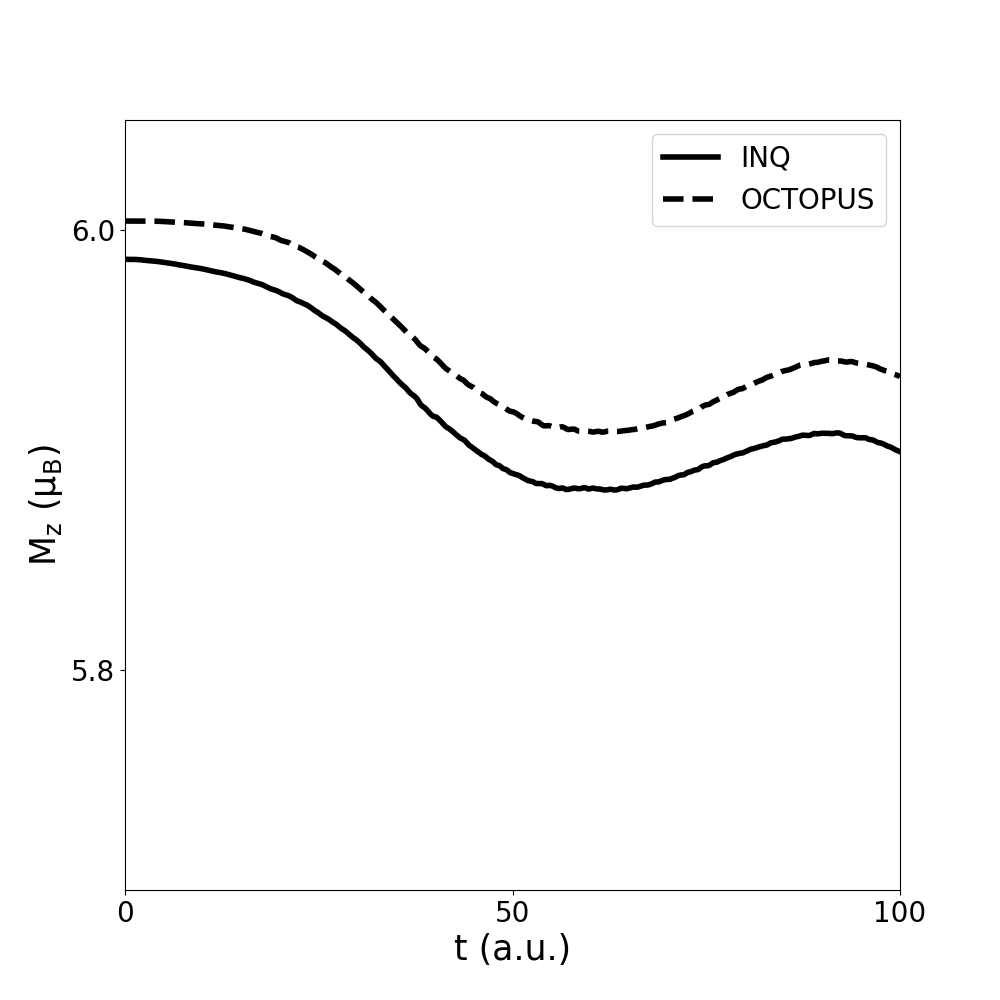} }}\\
    \subfloat[\centering Applied laser field polarized along the $z$ direction; $\boldsymbol{\epsilon}=\begin{psmallmatrix} 0 & 0 & 1 \end{psmallmatrix}$]{{\includegraphics[width=0.9\columnwidth]{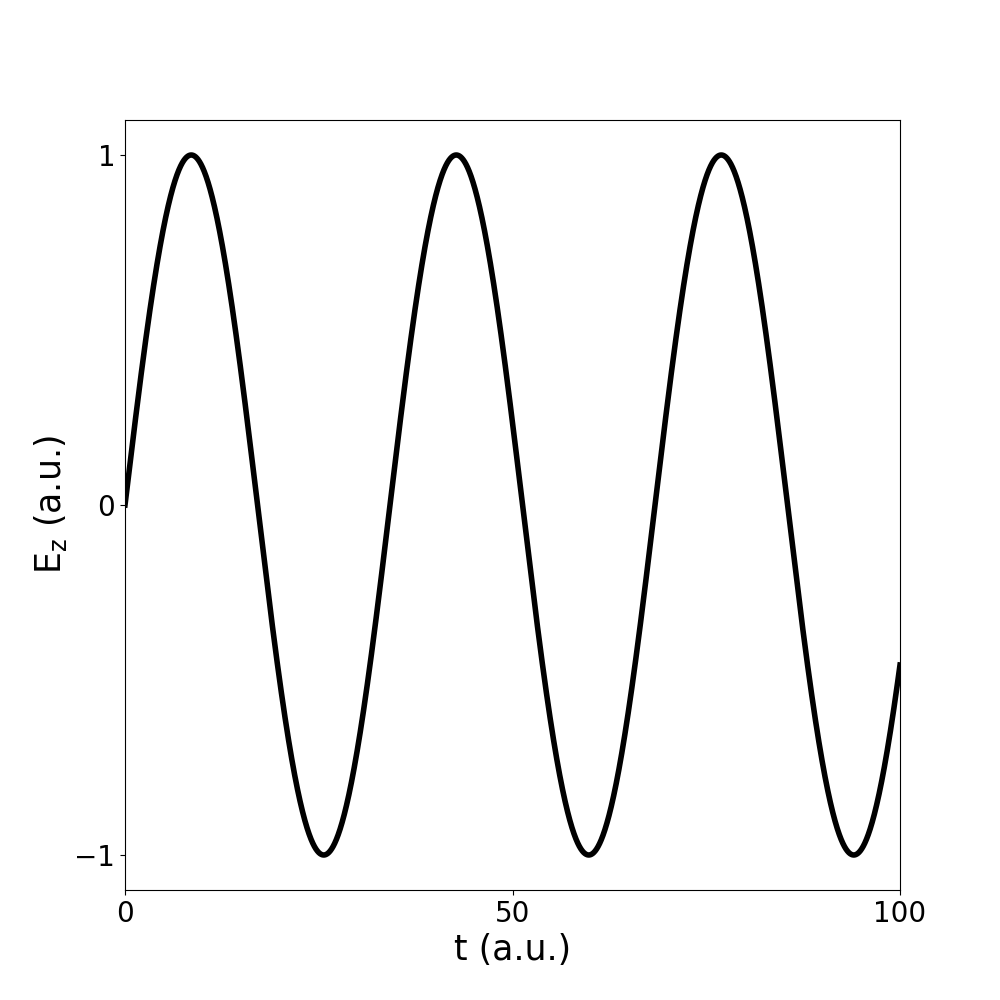} }}%
    \caption{
    Time dependent evolution of \ce{Fe2} under an external laser field.
    Comparison between \textsc{INQ} and \textsc{Octopus};
    in (a) panel the solid black line is obtained from the \textsc{INQ} simulation, the dashed line from the \textsc{Octopus} simulation.
    In the (b) panel we show the laser pulse used in the two simulations;
    the laser field has shape \({\bf E}(t) = A\boldsymbol{\epsilon}\sin(\omega t)\).
    The amplitude of the field is large (\(A = 1\) in atomic units) to make the precession more evident in the time-scale shown, the oscillation frequency is \(\omega = 5~\mathrm{eV}\).}%
    \label{fig:fe2-inq-oct}%
\end{figure}

\begin{figure}%
    \centering
    \subfloat[\centering \ce{Fe2}: \(M_{\rm z}\) component]{{\includegraphics[width=0.9\columnwidth]{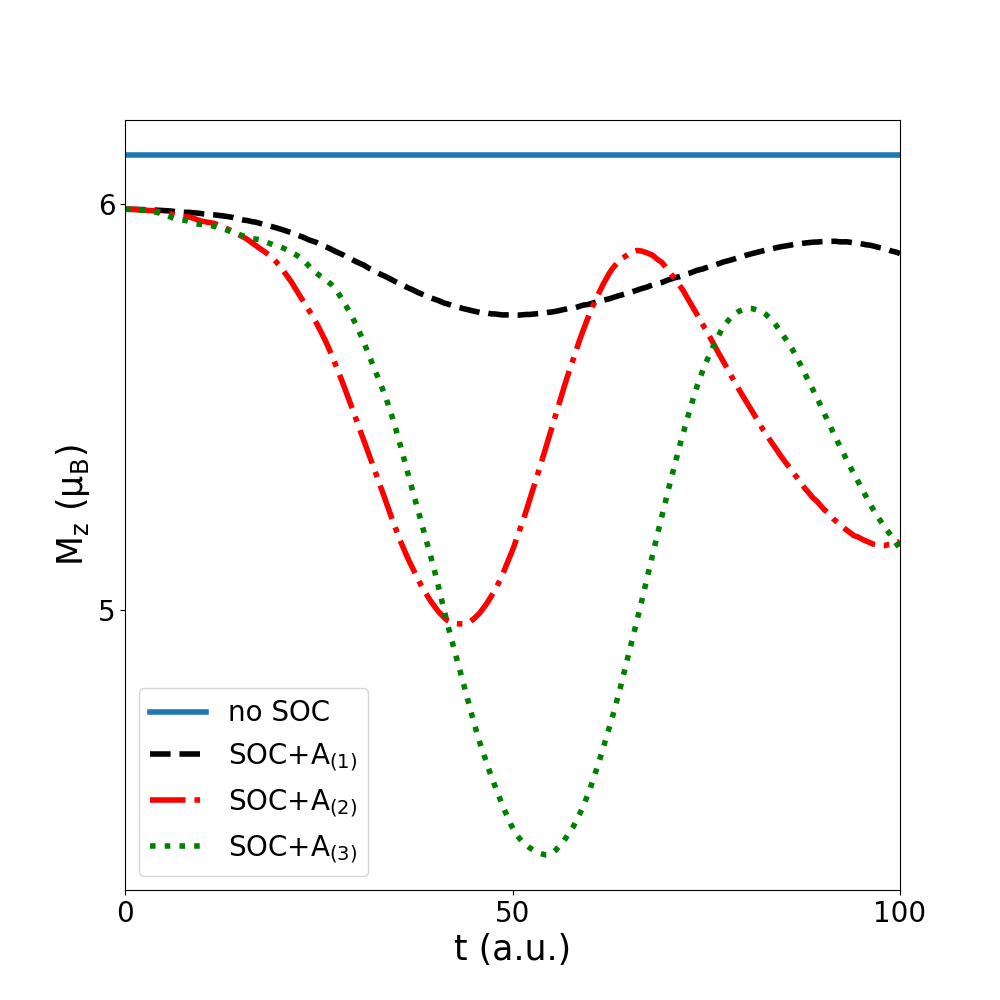} }}\\
    \subfloat[\centering \ce{Fe2}: applied external electric pulse with polarization \(\boldsymbol{\epsilon}=\begin{psmallmatrix} 0 & 0 & 1 \end{psmallmatrix}\) and different amplitudes \(A\). ]{{\includegraphics[width=0.9\columnwidth]{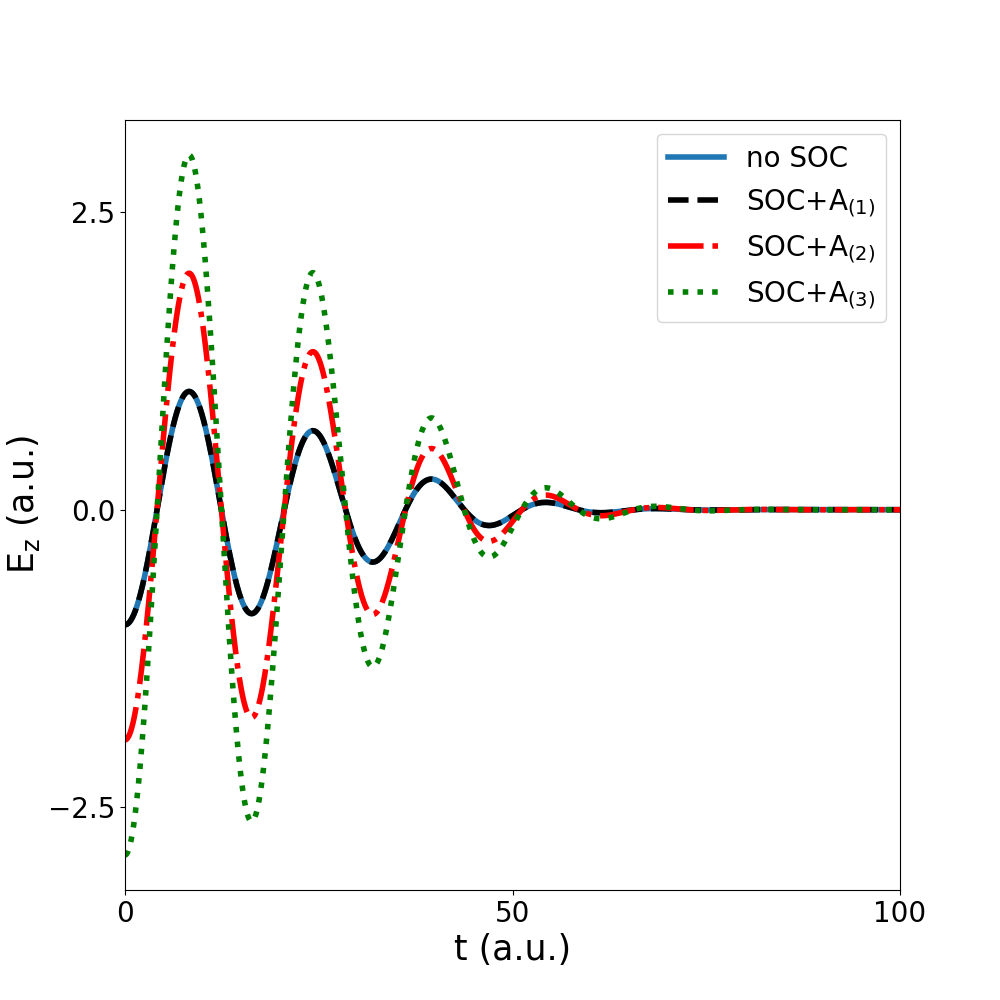} }}%
    \caption{
    Time-dependent evolution of \ce{Fe2} under three different external laser pulses.
    In panel (a) we look at the z-component of the magnetization of the dimer.
    The pulse profiles are shown in panel (b) with the same line colors and shapes.
    The magnetization change correlates with the pulse amplitude. The pulse temporal profile is given by \({\bf E}(t)=A\boldsymbol{\epsilon}\cos(\omega t+\phi)\exp\big(-\frac{(t-t_0)^2}{2\tau^2}\big)\). We use quite large amplitudes \(A=1, 2, 3\) in atomic units to start the magnetization dynamics; the other parameters are the same in the different simulations, \(\omega=10~\mathrm{eV}\), \(t_0 = 0.1~\mathrm{fs}\), \(\tau = 0.5~\mathrm{fs}\). }%
    \label{fig:fe2-laser}%
\end{figure}

First, we study the case of a \ce{Fe2} molecule under a monochromatic laser pulse using both \textsc{INQ} and the \textsc{Octopus} code~\cite{C5CP00351B, tancogne_dejean_octopus_2020}.
The \ce{Fe2} molecule is the same used in the simulation of Fig.~(\ref{fig:nc-clusters}).
The laser is represented by an oscillating electric field shown in Fig.~\ref{fig:fe2-inq-oct}b.
The magnetic field component of the laser is neglected.
In Fig.~\ref{fig:fe2-inq-oct}a we plot the evolution of the magnetization.
The initial value of the magnetization is slightly different in the ground states obtained from the two codes;
as a consequence, the two curves appear as shifted.
We observe that their temporal behavior is very similar and they agree with each other up to the initial difference in magnetization (of the order of \(0.017~\mu_\text{B}\)).

In Fig.~(\ref{fig:fe2-laser}) we consider the same magnetic molecule \ce{Fe2} while we perturb the system under laser pulses of different intensities.
The temporal profile of the total magnetization is given in Fig.~\ref{fig:fe2-laser}a, while the laser pulse profiles are given in panel Fig.~\ref{fig:fe2-laser}b.
There are four curves in the figure.
The solid blue line represents the expectation value of the spin, \(\expval*{\hat{S}_z}\), in the absence of the spin-orbit interaction.
In this case, we observe no change in the magnetization during the dynamics, independently of the electric field amplitude.
This is expected given that in the absence of spin-orbit interaction there is no direct coupling between the spin and the external laser field.
The black, red, and green curves are instead obtained in the presence of spin-orbit interaction, under three different amplitudes of the laser pulse.
The absolute change in magnetization increases proportionally to the applied pulse amplitude.
The pulses are linearly polarized along the \({\rm z}\)-axis in all the different cases.
The pulse with a larger amplitude will produce transitions to higher energy orbital states with higher probability compared to the other pulses.
This induces a fully coherent excitation with the system transitioning to a new electronic state with a different orbital and spin expectation value.
The dynamical modification in the spin expectation value does not reflect, in general, an opposite change in the orbital expectation value given that the total electronic angular momentum is not conserved within our formalism.

Such a dynamics can be better understood starting from the pseudopotential expression in Eq.~(\ref{eq:Vps}) and the spin orbit term in Eq.~(\ref{Eq:Vps-soc}).
As already mentioned, \(\hat{v}_\text{ps}\) can be expressed in a form that makes the spin-orbit operator more explicit \cite{Fernandez-Seivane_2007}
\begin{multline}
  \hat{v}_\text{ps}=\sum_{\rm a=1}^{N_{\rm a}}\Big\{V_\text{local}^{\rm a}(r) +\sum_{\ell mm'}\ket{\ell m;a}\big[\bar{V}_\ell^{\rm a}(r)\\
  + V_\ell^{a;\text{SO}}(r)\mel*{\ell m}{\hat{\bf L}_{\rm a}}{\ell m'}\cdot\hat{\bf S}\big]\bra{\ell m';a}\Big\}\,.
\end{multline}
The spin operator is no longer a constant of motion in the presence of these fully relativistic pseudopotentials.
The time derivative of the magnetization over the volume \(\Omega\) of the simulation box can be written using Eq.~(\ref{eq:spincont3}) as
%
\begin{multline} \label{eq:SOtorque}
  \frac{\mathrm{d}}{\mathrm{d}t}M^{\rm i}_\Omega =
  \sum_{n{\bf k}}f_{n{\bf k}}\sum_{ju}\epsilon_{iju}\sum_{\rm a=1}^{N_a}\sum_{\ell mm'}\bra{n{\bf k}}\big[\\\
  V_\ell^{a;\rm SO}({\bf r}-{\bf R}_a)\mel*{\ell m;a}{L_j}{\ell m';a}\ket{\ell m;a}\,\hat{S}^u\times\\\
  \times\bra{\ell m';a}\big]\ket{n{\bf k}}\\\,,
\end{multline}
where, in the absence of an external magnetic field, we removed the other contributions in the equation.
We observe that the internal double summation over the spherical harmonics in Eq.~(\ref{eq:SOtorque}) is a potential source of problems.
The summation is truncated in the calculation, neglecting the projection of the wave function on higher-order harmonics. This is ultimately a problem related to the pseudopotential approximation and to the fact that the orbital momentum operator is described on the basis of the available atomic projected harmonics.

\begin{figure}[!h]%
    \centering
    \subfloat[\centering \ce{Fe6}: \(M_{\rm z}\) component under laser pulses of panel (b). ]{{\includegraphics[width=0.46\textwidth]{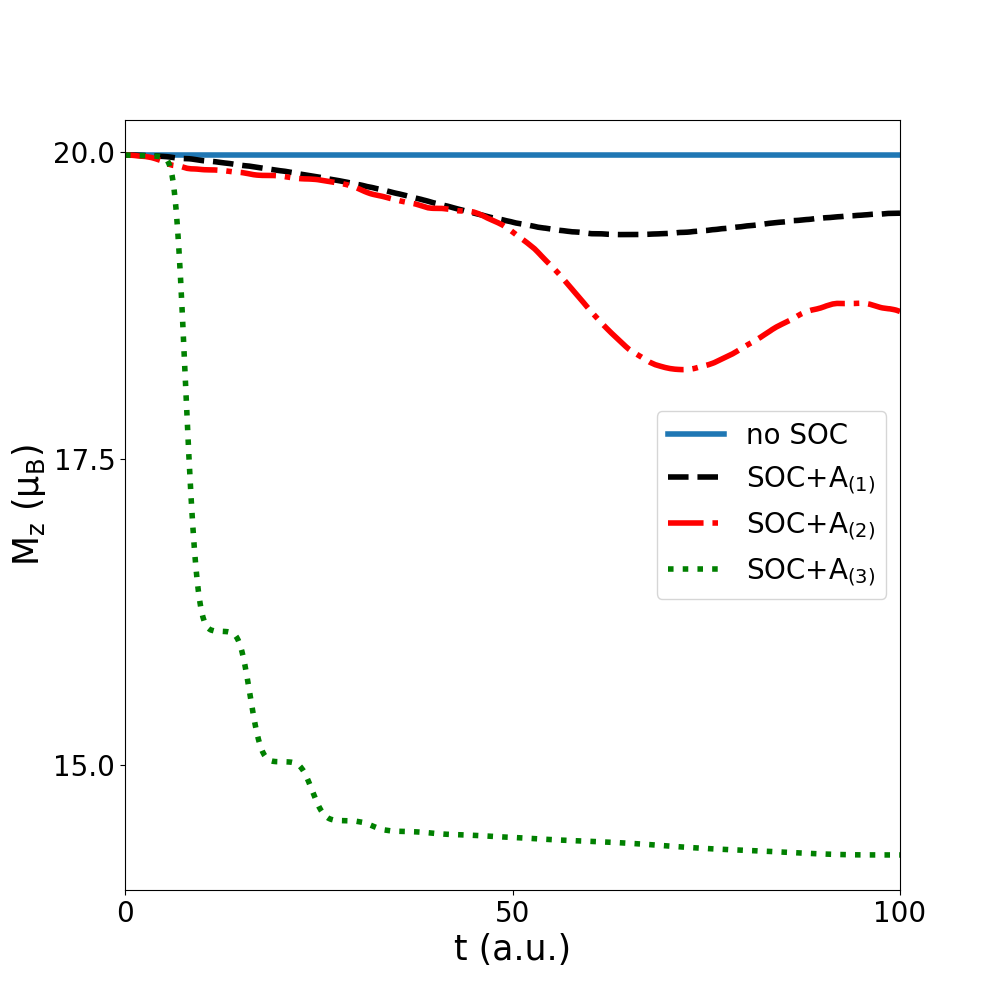} }}%
    \qquad
    \subfloat[\centering \ce{Fe6}: applied external electric pulses. The electric field polarization is \(\boldsymbol{\epsilon}=\begin{psmallmatrix} 0 & 0 & 1 \end{psmallmatrix}\) in all three cases. ]{{\includegraphics[width=0.46\textwidth]{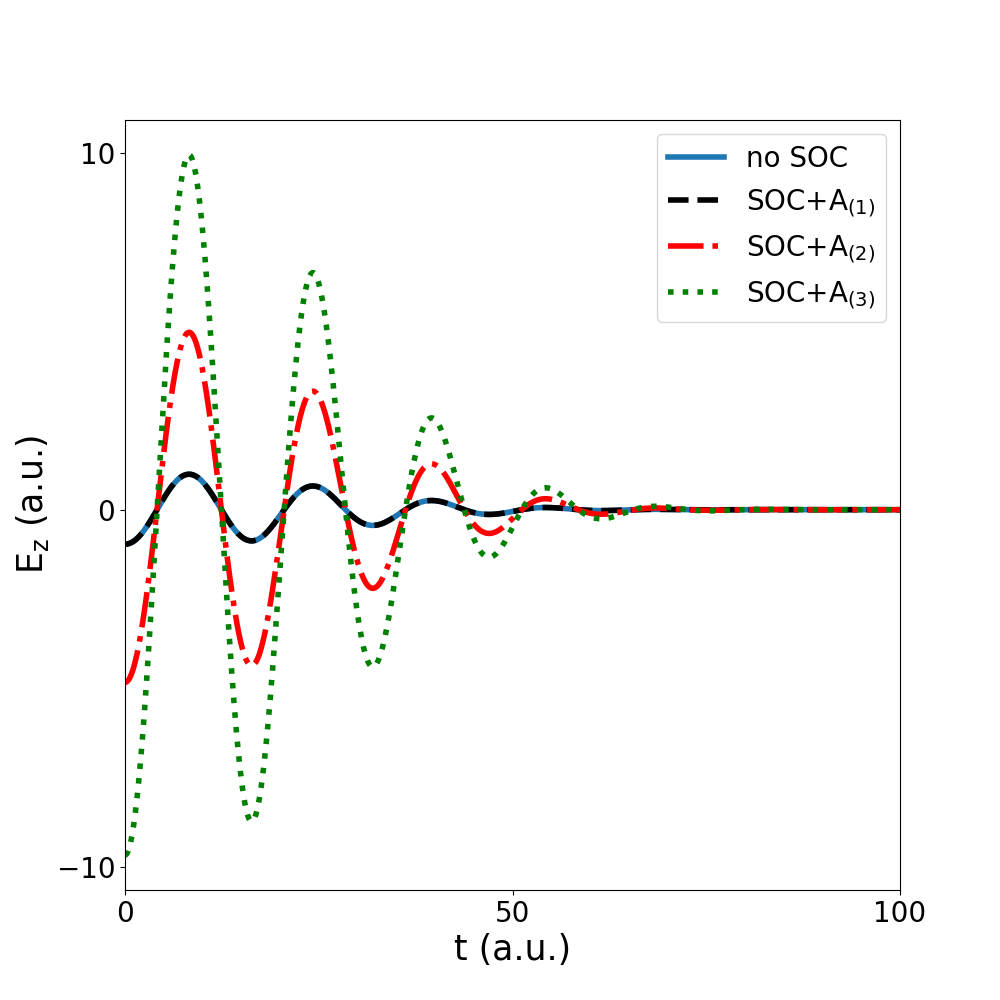} }}%
    \caption{
    Time dependent evolution of \ce{Fe6} under external laser pulses. The pulse temporal profile is given by \({\bf E}(t)=A\boldsymbol{\epsilon}\cos(\omega t+\phi)\exp\big(-\frac{(t-t_0)^2}{2\tau^2}\big)\).
    The lines in panels (a) and (b) correspond to different amplitudes \(A = 1, 5, 10\) in atomic units, labeled with \(A_{(1)}\), \(A_{(2)}\) and \(A_{(3)}\). %
    The other parameters are the same in the different simulations, \(\omega=10~\mathrm{eV}\), \(t_0 = 0.1~\mathrm{fs}\), \(\tau = 0.5~\mathrm{fs}\). }%
    \label{fig:fe6-laser}%
\end{figure}

\begin{figure}[!h]
    \centering
    {\includegraphics[width=0.46\textwidth]{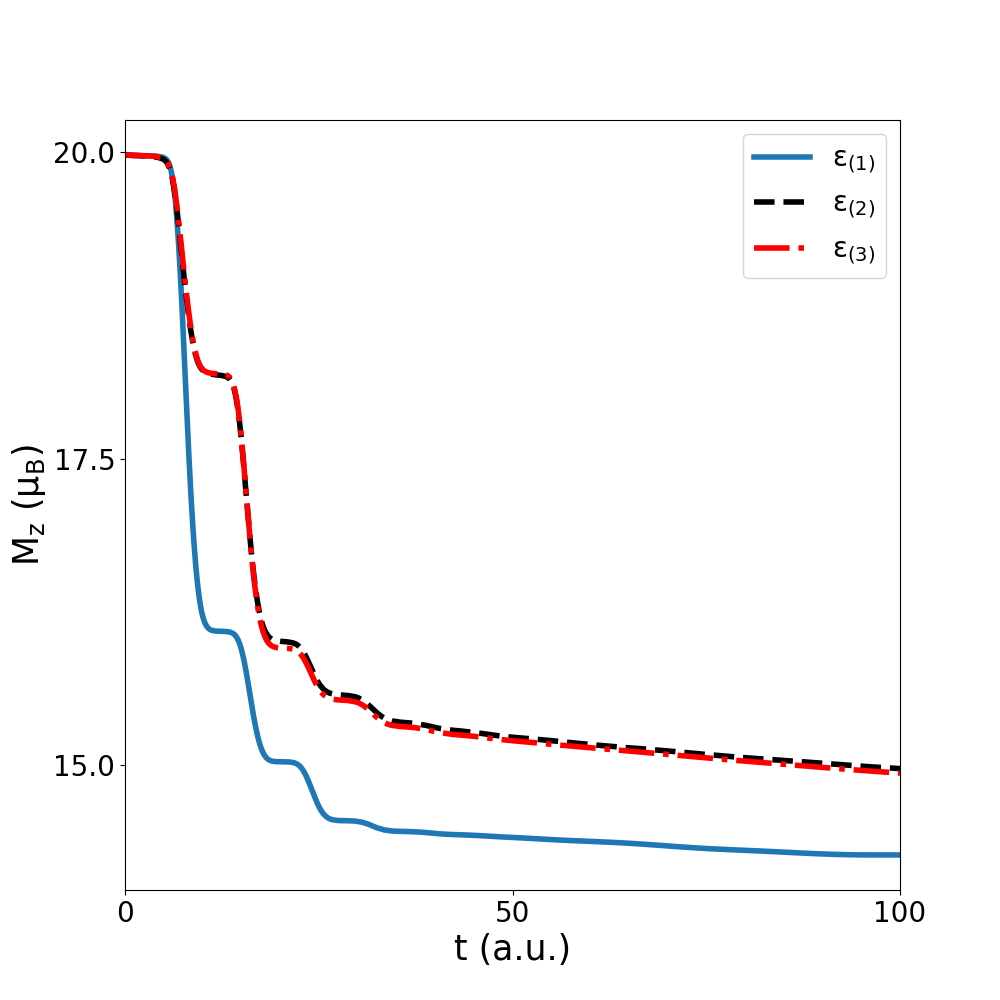} }
    \caption{Time-dependent magnetization of \ce{Fe6} under external laser pulses with different polarization directions: \(\boldsymbol{\epsilon}_{(1)} = \begin{psmallmatrix} 0 & 0 & 1 \end{psmallmatrix}\), \(\boldsymbol{\epsilon}_{(2)}=\begin{psmallmatrix} 1 & 1 & 0 \end{psmallmatrix}/\sqrt{2}\), and \(\boldsymbol{\epsilon}_{(3)}=\begin{psmallmatrix} 1 & 0 & 1 \end{psmallmatrix}/\sqrt{2}\).
    The pulse temporal profile is the same as in Fig.~(\ref{fig:fe6-laser}) with amplitude \(A = 10\) in atomic units left unchanged.
    }
    \label{fig:fe6-laser-2}
\end{figure}

\subsubsection{\ce{Fe6} cluster}

In Fig.~\ref{fig:fe6-laser} we consider again the real-time dynamics of the magnetic cluster \ce{Fe6}, under electric pulses.
In the absence of spin-orbit interaction, we observe no magnetization dynamics consistent with the \ce{Fe2} case. This corresponds to the solid blue line in the left panel.
The dashed black and red lines show a moderate demagnetization effect, which appears to be stronger for the pulse with larger amplitude.
The strongest magnetization loss is observed with the pulse with the largest amplitude (the green dotted line in both panels) and amounts to approximately \(27\%\) of the initial magnetic moment of the atomic cluster.
In all of the different cases, the dynamics of the magnetic moments are activated by the pulse and persist throughout its duration and even after the pulse disappears.

Fig.~\ref{fig:fe6-laser-2} shows the spin dynamics of the \ce{Fe6} cluster under electric pulses with different polarization directions.
We can see a dependence of the demagnetization effect on the polarization of the electric pulse.
The demagnetization is larger for the pulse polarized along the \(z\) axis.
The other two pulses do not produce sensible differences in the demagnetization signal.
In general, we expect such a demagnetization process to be highly dependent on the specific nature of the pulse excitation, frequency, pulse amplitude, and polarization.
The polarization direction influences the orbital states excited by the pulse; in turn, this affects the change in magnetization via the spin-orbit interaction. We emphasize that with these simulations we aim to demonstrate that \textsc{INQ} is capable of performing real-time TDDFT for ultrafast magnetism. However, we should keep in mind that the strong amplitude of the electric pulses and the approximations used for the exchange-correlation potentials make these ultrafast dynamics simulations not completely realistic. A better choice for the exchange-correlation potential, in particular, is crucial for a correct description of magnetism and spin dynamics. Other functionals like meta-GGA\cite{PhysRevMaterials.2.063801}, source-free functionals\cite{doi:10.1021/acs.jctc.7b01049,PhysRevB.111.094417}, or hybrid\cite{PhysRevB.79.245129,10.1063/1.3006419,PhysRevB.69.085115,PhysRevB.105.L100401} are in principle more suitable to account for these effects compared to LDA. The locally collinear approximation is also problematic in this context given that it sets automatically to zero the XC-torque in the spin continuity equation (\ref{eq:spincont}), neglecting potentially important many body torque effects. The adiabatic approximation completely neglects memory effects, which prevents a correct description of dissipation processes \cite{PhysRevLett.95.086401,PhysRevLett.77.2037,10.1063/1.2406069}. \textsc{INQ} is designed to easily include additional functionals thanks to its modularity, and we will discuss these new implementations in future works.

\section{Numerical Scalability}

\begin{figure}%
    \centering
    \subfloat[\centering Time per \ac{TDDFT} step as a function of the number of GPUs]{{\includegraphics[width=0.85\columnwidth]{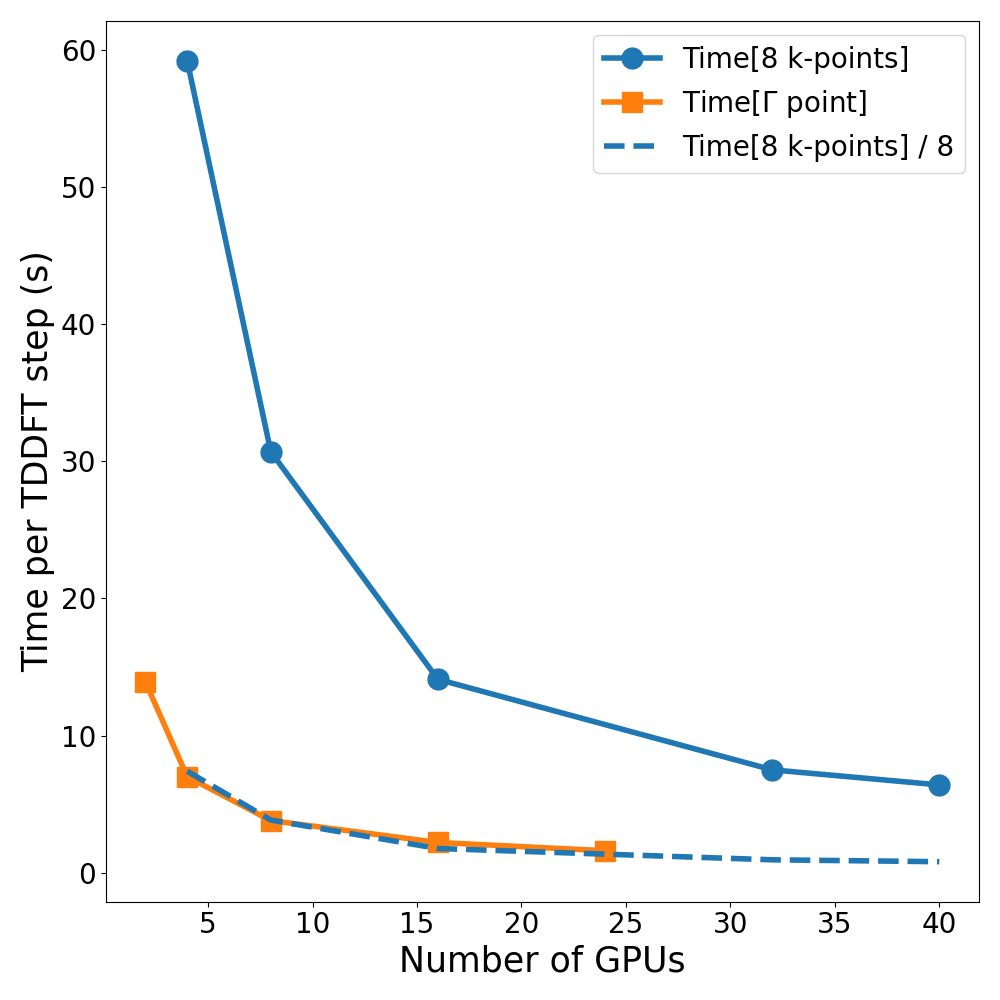} }}\\
    \subfloat[\centering Speed up as a function of the number of GPUs ]{{\includegraphics[width=0.85\columnwidth]{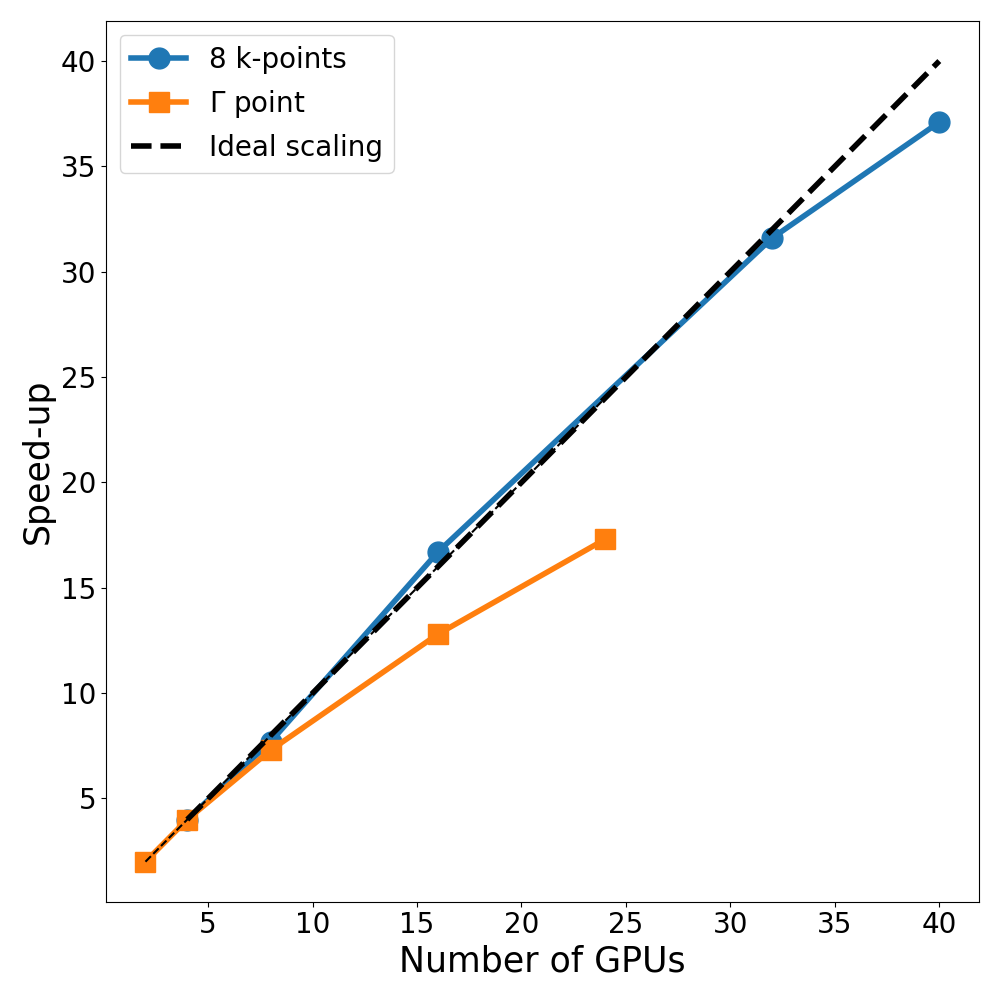} }}%
    \caption{
    Panel (a) shows a strong scaling plot of the wall time per \ac{TDDFT} simulation step for \ce{Fe} BCC supercell. Panel (b) shows the speedup of the simulations as a function of the number of GPUs. The simulations were run in LLNL's Tuolumne supercomputer.}%
    \label{fig:GPU-scal}%
\end{figure}

\textsc{INQ} achieves distributed memory parallelism (parallelism between different nodes in a supercomputer) using the \ac{MPI} system. \textsc{INQ} uses one \ac{MPI} task per GPU, in LLNL's Tuolumne supercomputer, where we perform the simulations shown in Fig.~\ref{fig:GPU-scal}. Each node has 4 GPUs and each GPU is controlled by one \ac{MPI} task. \textsc{INQ} is designed to maximize its degree of parallelization, i.e. each \ac{DFT} or \ac{TDDFT} simulation is parallelized over {\bf k}-points, electronic bands, spin for spin-polarized calculations, and basis set coefficients. This helps reducing the memory bottleneck in the case of large systems simulations.\newline
In Fig.~\ref{fig:GPU-scal} we perform  \ac{TDDFT} simulations for a $4\times 4\times 4$ \ce{Fe} BCC supercell. The electronic ground state is computed in spin non-collinear form, where we compare two calculations one with a single {\bf k}-point ($\Gamma$) and another with $8$ {\bf k}-points. No symmetry-operations are used to reduce the number of {\bf k}-points in the simulation. In panel (a) of the figure, we show the time per single \ac{TDDFT} step as a function of the number of GPUs used. The dependence on the number of {\bf k}-points here is exactly linear; the orange dotted line corresponds to the simulation at the $\Gamma$ point and their simulation time is $1/8$ of the simulation time for the system of $8$ {\bf k}. In panel (b), we show the speed-up as a function of the number of GPUs. The speed-up is approximately linear with the GPU number and very close to its ideal value when the number of GPUs is small compared to the number of parallelized degrees of freedom. When the number of GPUs used becomes too large, the speed-up starts to become less than the ideal scaling. As expected, in the $\Gamma$ point calculation case the deviation from the ideal curve happens at a lower number of GPUs compared to the 8 {\bf k}-points one.

\section{Conclusions} \label{sec:concl}

In summary, we have presented our implementation of non-collinear \ac{DFT} and \ac{TDDFT} in the \textsc{INQ} code.
We have tested the effect of the \ac{XC} magnetic field in the locally-collinear approximation, the spin Zeeman coupling with external magnetic fields, and the spin-orbit interaction in different systems.
We compared the results with other codes available, finding good agreement.
We have tested the dynamics of different magnetic systems under applied external fields, both electric and magnetic.
In the presence of external magnetic fields, we recover the expected spin-precession dynamics.
The dynamics under applied laser pulses are richer and predict optically driven demagnetization effects consistent with previous studies \cite{PhysRevB.94.014423, Krieger_2017, PhysRevB.105.134425, Krieger_2015}.

This implementation sets up the computational platform for future developments, for example, the calculation of orbital angular momentum dynamics or orbital magnetization, Ehrenfest dynamics~\cite{Andrade2009} for spin-phonon relaxation in magnetic systems, improved XC functionals for the description of dynamical magnetic properties beyond the adiabatic and the locally collinear approximations, as well as advanced relativistic treatment beyond the pseudopotential approximation.
A potential application area is pump-probe ultrafast magnetization dynamics, observed in experiments using techniques like Time-resolved magneto-optical Kerr effect (TR-MOKE) \cite{PhysRevB.86.125139, D0TC01322F, Ebert_1996, Kunes_2004} and Ultrafast X-ray Magnetic Circular Dichroism (XMCD) \cite{Mason2003,Lee2011}. Thanks to its high scalability, \textsc{INQ} can be potentially used to study the formation of spin currents induced by ultrafast demagnetization processes \cite{Choi2014} and predict their magnitude and time scales. The demagnetization times in experiments are strongly affected by disorder, laser fluence \cite{PhysRevLett.125.127201,29511738}, temperature \cite{PhysRevX.2.021006} and sample purity \cite{doi:10.1073/pnas.1201371109}. The inclusion of such effects in our \ac{TDDFT} implementation is possible, and this could help designing new experiments and control the demagnetization process. A second application area is the calculation of magnon excitations and early spin waves dynamical formation \cite{PhysRevB.98.054429,PhysRevLett.110.097201,Iacocca2019}. This is very interesting considering that the formation of damped spin waves following ultrafast excitations have been observed in different experiments \cite{Mizukami_2010}. Finally, \textsc{INQ} can be used to computationally screen materials like ferromagnets, anti-ferromagnets and low dimensional magnets for different properties e.g. low damping, spin orbit torque efficiencies, switching current densities and magnetic anisotropies, providing guidance to the design of new magnetic materials. In conclusion \textsc{INQ} represents a valuable tool for the study of spin dynamics in real materials from first principles.

The Gitlab repositories used to perform ground-state and real-time simulations can be found here \cite{inq_gitlab}.

\begin{acknowledgement}
We acknowledge support from the Computational Materials Sciences Program funded by the US Department of Energy, Office of Science, Basic Energy Sciences, Materials Sciences and Engineering Division for the materials application and the code development.
W.F. acknowledges the support from the NSF through the University of Wisconsin Materials Research Science and Engineering Center (DMR-2309000) for the code benchmark and  debugging. 
Part of this work was performed under the auspices of the U.S. Department of Energy by Lawrence Livermore National Laboratory under Contract No. DE-AC52-07NA27344.
\end{acknowledgement}

\bibliography{biblio}

\providecommand{\latin}[1]{#1}
\makeatletter
\providecommand{\doi}
  {\begingroup\let\do\@makeother\dospecials
  \catcode`\{=1 \catcode`\}=2 \doi@aux}
\providecommand{\doi@aux}[1]{\endgroup\texttt{#1}}
\makeatother
\providecommand*\mcitethebibliography{\thebibliography}
\csname @ifundefined\endcsname{endmcitethebibliography}  {\let\endmcitethebibliography\endthebibliography}{}
\begin{mcitethebibliography}{158}
\providecommand*\natexlab[1]{#1}
\providecommand*\mciteSetBstSublistMode[1]{}
\providecommand*\mciteSetBstMaxWidthForm[2]{}
\providecommand*\mciteBstWouldAddEndPuncttrue
  {\def\EndOfBibitem{\unskip.}}
\providecommand*\mciteBstWouldAddEndPunctfalse
  {\let\EndOfBibitem\relax}
\providecommand*\mciteSetBstMidEndSepPunct[3]{}
\providecommand*\mciteSetBstSublistLabelBeginEnd[3]{}
\providecommand*\EndOfBibitem{}
\mciteSetBstSublistMode{f}
\mciteSetBstMaxWidthForm{subitem}{(\alph{mcitesubitemcount})}
\mciteSetBstSublistLabelBeginEnd
  {\mcitemaxwidthsubitemform\space}
  {\relax}
  {\relax}

\bibitem[Sanvito(2019)]{Sanvito2019}
Sanvito,~S. In \emph{Handbook of Materials Modeling : Methods: Theory and Modeling}; Andreoni,~W., Yip,~S., Eds.; Springer International Publishing: Cham, 2019; pp 1--4\relax
\mciteBstWouldAddEndPuncttrue
\mciteSetBstMidEndSepPunct{\mcitedefaultmidpunct}
{\mcitedefaultendpunct}{\mcitedefaultseppunct}\relax
\EndOfBibitem
\bibitem[Bihlmayer(2018)]{Bihlmayer2018}
Bihlmayer,~G. In \emph{Handbook of Materials Modeling : Methods: Theory and Modeling}; Andreoni,~W., Yip,~S., Eds.; Springer International Publishing: Cham, 2018; pp 1--23\relax
\mciteBstWouldAddEndPuncttrue
\mciteSetBstMidEndSepPunct{\mcitedefaultmidpunct}
{\mcitedefaultendpunct}{\mcitedefaultseppunct}\relax
\EndOfBibitem
\bibitem[Brooks \latin{et~al.}(2001)Brooks, Richter, and Sandratskii]{BROOKS20012059}
Brooks,~M.; Richter,~M.; Sandratskii,~L. In \emph{Encyclopedia of Materials: Science and Technology}; Buschow,~K.~J., Cahn,~R.~W., Flemings,~M.~C., Ilschner,~B., Kramer,~E.~J., Mahajan,~S., Veyssière,~P., Eds.; Elsevier: Oxford, 2001; pp 2059--2070\relax
\mciteBstWouldAddEndPuncttrue
\mciteSetBstMidEndSepPunct{\mcitedefaultmidpunct}
{\mcitedefaultendpunct}{\mcitedefaultseppunct}\relax
\EndOfBibitem
\bibitem[Antropov \latin{et~al.}(1995)Antropov, Katsnelson, van Schilfgaarde, and Harmon]{PhysRevLett.75.729}
Antropov,~V.~P.; Katsnelson,~M.~I.; van Schilfgaarde,~M.; Harmon,~B.~N. $\mathit{Ab}\mathit{}\mathit{Initio}$ Spin Dynamics in Magnets. \emph{Phys. Rev. Lett.} \textbf{1995}, \emph{75}, 729--732\relax
\mciteBstWouldAddEndPuncttrue
\mciteSetBstMidEndSepPunct{\mcitedefaultmidpunct}
{\mcitedefaultendpunct}{\mcitedefaultseppunct}\relax
\EndOfBibitem
\bibitem[Antropov \latin{et~al.}(1996)Antropov, Katsnelson, Harmon, van Schilfgaarde, and Kusnezov]{PhysRevB.54.1019}
Antropov,~V.~P.; Katsnelson,~M.~I.; Harmon,~B.~N.; van Schilfgaarde,~M.; Kusnezov,~D. Spin dynamics in magnets: Equation of motion and finite temperature effects. \emph{Phys. Rev. B} \textbf{1996}, \emph{54}, 1019--1035\relax
\mciteBstWouldAddEndPuncttrue
\mciteSetBstMidEndSepPunct{\mcitedefaultmidpunct}
{\mcitedefaultendpunct}{\mcitedefaultseppunct}\relax
\EndOfBibitem
\bibitem[Onida \latin{et~al.}(2002)Onida, Reining, and Rubio]{Onida2002}
Onida,~G.; Reining,~L.; Rubio,~A. Electronic excitations: density-functional versus many-body Green’s-function approaches. \emph{Rev. Mod. Phys.} \textbf{2002}, \emph{74}, 601–659\relax
\mciteBstWouldAddEndPuncttrue
\mciteSetBstMidEndSepPunct{\mcitedefaultmidpunct}
{\mcitedefaultendpunct}{\mcitedefaultseppunct}\relax
\EndOfBibitem
\bibitem[Marques and Gross(2004)Marques, and Gross]{Marques2004}
Marques,~M.; Gross,~E. TIME-DEPENDENT DENSITY FUNCTIONAL THEORY. \emph{Ann. Rev. Phys. Chem.} \textbf{2004}, \emph{55}, 427–455\relax
\mciteBstWouldAddEndPuncttrue
\mciteSetBstMidEndSepPunct{\mcitedefaultmidpunct}
{\mcitedefaultendpunct}{\mcitedefaultseppunct}\relax
\EndOfBibitem
\bibitem[Casida and Huix-Rotllant(2012)Casida, and Huix-Rotllant]{Casida2012}
Casida,~M.; Huix-Rotllant,~M. Progress in Time-Dependent Density-Functional Theory. \emph{Ann. Rev. Phys. Chem.} \textbf{2012}, \emph{63}, 287–323\relax
\mciteBstWouldAddEndPuncttrue
\mciteSetBstMidEndSepPunct{\mcitedefaultmidpunct}
{\mcitedefaultendpunct}{\mcitedefaultseppunct}\relax
\EndOfBibitem
\bibitem[Jornet-Somoza \latin{et~al.}(2015)Jornet-Somoza, Alberdi-Rodriguez, Milne, Andrade, Marques, Nogueira, Oliveira, Stewart, and Rubio]{JornetSomoza2015}
Jornet-Somoza,~J.; Alberdi-Rodriguez,~J.; Milne,~B.~F.; Andrade,~X.; Marques,~M. A.~L.; Nogueira,~F.; Oliveira,~M. J.~T.; Stewart,~J. J.~P.; Rubio,~A. Insights into colour-tuning of chlorophyll optical response in green plants. \emph{Phys. Chem. Chem. Phys.} \textbf{2015}, \emph{17}, 26599–26606\relax
\mciteBstWouldAddEndPuncttrue
\mciteSetBstMidEndSepPunct{\mcitedefaultmidpunct}
{\mcitedefaultendpunct}{\mcitedefaultseppunct}\relax
\EndOfBibitem
\bibitem[Correa(2018)]{Correa2018}
Correa,~A.~A. Calculating electronic stopping power in materials from first principles. \emph{Comput. Mat. Sci.} \textbf{2018}, \emph{150}, 291–303\relax
\mciteBstWouldAddEndPuncttrue
\mciteSetBstMidEndSepPunct{\mcitedefaultmidpunct}
{\mcitedefaultendpunct}{\mcitedefaultseppunct}\relax
\EndOfBibitem
\bibitem[Byun \latin{et~al.}(2020)Byun, Sun, and Ullrich]{Byun2020}
Byun,~Y.-M.; Sun,~J.; Ullrich,~C.~A. Time-dependent density-functional theory for periodic solids: assessment of excitonic exchange–correlation kernels. \emph{Electronic Structure} \textbf{2020}, \emph{2}, 023002\relax
\mciteBstWouldAddEndPuncttrue
\mciteSetBstMidEndSepPunct{\mcitedefaultmidpunct}
{\mcitedefaultendpunct}{\mcitedefaultseppunct}\relax
\EndOfBibitem
\bibitem[Kononov \latin{et~al.}(2022)Kononov, Lee, dos Santos, Robinson, Yao, Yao, Andrade, Baczewski, Constantinescu, Correa, Kanai, Modine, and Schleife]{Kononov2022}
Kononov,~A.; Lee,~C.-W.; dos Santos,~T.~P.; Robinson,~B.; Yao,~Y.; Yao,~Y.; Andrade,~X.; Baczewski,~A.~D.; Constantinescu,~E.; Correa,~A.~A.; Kanai,~Y.; Modine,~N.; Schleife,~A. Electron dynamics in extended systems within real-time time-dependent density-functional theory. \emph{MRS Commun.} \textbf{2022}, \emph{12}, 1002–1014\relax
\mciteBstWouldAddEndPuncttrue
\mciteSetBstMidEndSepPunct{\mcitedefaultmidpunct}
{\mcitedefaultendpunct}{\mcitedefaultseppunct}\relax
\EndOfBibitem
\bibitem[Xu \latin{et~al.}(2024)Xu, Carney, Zhou, Shepard, and Kanai]{Xu2024}
Xu,~J.; Carney,~T.; Zhou,~R.; Shepard,~C.; Kanai,~Y. Real-Time Time-Dependent Density Functional Theory for Simulating Nonequilibrium Electron Dynamics. \emph{J. Am. Chem. Soc.} \textbf{2024}, \emph{146}, 5011--5029\relax
\mciteBstWouldAddEndPuncttrue
\mciteSetBstMidEndSepPunct{\mcitedefaultmidpunct}
{\mcitedefaultendpunct}{\mcitedefaultseppunct}\relax
\EndOfBibitem
\bibitem[\ifmmode \check{Z}\else \v{Z}\fi{}uti\ifmmode~\acute{c}\else \'{c}\fi{} \latin{et~al.}(2004)\ifmmode \check{Z}\else \v{Z}\fi{}uti\ifmmode~\acute{c}\else \'{c}\fi{}, Fabian, and Das~Sarma]{Igor2004}
\ifmmode \check{Z}\else \v{Z}\fi{}uti\ifmmode~\acute{c}\else \'{c}\fi{},~I.; Fabian,~J.; Das~Sarma,~S. Spintronics: Fundamentals and applications. \emph{Rev. Mod. Phys.} \textbf{2004}, \emph{76}, 323--410\relax
\mciteBstWouldAddEndPuncttrue
\mciteSetBstMidEndSepPunct{\mcitedefaultmidpunct}
{\mcitedefaultendpunct}{\mcitedefaultseppunct}\relax
\EndOfBibitem
\bibitem[Gu \latin{et~al.}(2024)Gu, Zheng, Jia, Shi, Zhao, Zeng, and Zhang]{Gu2024}
Gu,~Y.; Zheng,~Z.; Jia,~L.; Shi,~S.; Zhao,~T.; Zeng,~T.; Zhang,~Q.~a. Ferroelectric Control of Spin-Orbitronics. \emph{Advanced Functional Materials} \textbf{2024}, \emph{34}, 2406444\relax
\mciteBstWouldAddEndPuncttrue
\mciteSetBstMidEndSepPunct{\mcitedefaultmidpunct}
{\mcitedefaultendpunct}{\mcitedefaultseppunct}\relax
\EndOfBibitem
\bibitem[Jo \latin{et~al.}(2024)Jo, Go, Choi, and Lee]{Jo2024}
Jo,~D.; Go,~D.; Choi,~G.-M.; Lee,~H.-W. Spintronics meets orbitronics: Emergence of orbital angular momentum in solids. \emph{npj Spintronics} \textbf{2024}, \emph{2}, 19\relax
\mciteBstWouldAddEndPuncttrue
\mciteSetBstMidEndSepPunct{\mcitedefaultmidpunct}
{\mcitedefaultendpunct}{\mcitedefaultseppunct}\relax
\EndOfBibitem
\bibitem[Rezende(2020)]{magnonics}
Rezende,~S. \emph{Fundamentals of Magnonics}; Springer Cham, 2020\relax
\mciteBstWouldAddEndPuncttrue
\mciteSetBstMidEndSepPunct{\mcitedefaultmidpunct}
{\mcitedefaultendpunct}{\mcitedefaultseppunct}\relax
\EndOfBibitem
\bibitem[Yuan \latin{et~al.}(2022)Yuan, Cao, Kamra, Duine, and Yan]{YUAN20221}
Yuan,~H.; Cao,~Y.; Kamra,~A.; Duine,~R.~A.; Yan,~P. Quantum magnonics: When magnon spintronics meets quantum information science. \emph{Physics Reports} \textbf{2022}, \emph{965}, 1--74\relax
\mciteBstWouldAddEndPuncttrue
\mciteSetBstMidEndSepPunct{\mcitedefaultmidpunct}
{\mcitedefaultendpunct}{\mcitedefaultseppunct}\relax
\EndOfBibitem
\bibitem[Zhang(2023)]{ZHANG2023100044}
Zhang,~X. A review of common materials for hybrid quantum magnonics. \emph{Materials Today Electronics} \textbf{2023}, \emph{5}, 100044\relax
\mciteBstWouldAddEndPuncttrue
\mciteSetBstMidEndSepPunct{\mcitedefaultmidpunct}
{\mcitedefaultendpunct}{\mcitedefaultseppunct}\relax
\EndOfBibitem
\bibitem[Vandersypen and Eriksson(2019)Vandersypen, and Eriksson]{10.1063/PT.3.4270}
Vandersypen,~L. M.~K.; Eriksson,~M.~A. Quantum computing with semiconductor spins. \emph{Physics Today} \textbf{2019}, \emph{72}, 38--45\relax
\mciteBstWouldAddEndPuncttrue
\mciteSetBstMidEndSepPunct{\mcitedefaultmidpunct}
{\mcitedefaultendpunct}{\mcitedefaultseppunct}\relax
\EndOfBibitem
\bibitem[Nagaosa and Tokura(2013)Nagaosa, and Tokura]{Nagaosa2013}
Nagaosa,~N.; Tokura,~Y. Topological properties and dynamics of magnetic skyrmions. \emph{Nature Nanotech.} \textbf{2013}, \emph{8}, 899--911\relax
\mciteBstWouldAddEndPuncttrue
\mciteSetBstMidEndSepPunct{\mcitedefaultmidpunct}
{\mcitedefaultendpunct}{\mcitedefaultseppunct}\relax
\EndOfBibitem
\bibitem[G{\"o}bel \latin{et~al.}(2021)G{\"o}bel, Mertig, and Tretiakov]{GOBEL20211}
G{\"o}bel,~B.; Mertig,~I.; Tretiakov,~O.~A. Beyond skyrmions: Review and perspectives of alternative magnetic quasiparticles. \emph{Physics Reports} \textbf{2021}, \emph{895}, 1--28\relax
\mciteBstWouldAddEndPuncttrue
\mciteSetBstMidEndSepPunct{\mcitedefaultmidpunct}
{\mcitedefaultendpunct}{\mcitedefaultseppunct}\relax
\EndOfBibitem
\bibitem[Fert \latin{et~al.}(2017)Fert, Reyren, and Cros]{Fert2017}
Fert,~A.; Reyren,~N.; Cros,~V. Magnetic skyrmions: advances in physics and potential applications. \emph{Nat. Rev. Mater.} \textbf{2017}, \emph{2}, 17031\relax
\mciteBstWouldAddEndPuncttrue
\mciteSetBstMidEndSepPunct{\mcitedefaultmidpunct}
{\mcitedefaultendpunct}{\mcitedefaultseppunct}\relax
\EndOfBibitem
\bibitem[F{\"a}hnle \latin{et~al.}(2018)F{\"a}hnle, Haag, Illg, Mueller, Weng, Tsatsoulis, Huang, Briones, Teeny, Zhang, and Kuhn]{UFM2018}
F{\"a}hnle,~M.; Haag,~M.; Illg,~C.; Mueller,~B.; Weng,~W.; Tsatsoulis,~T.; Huang,~H.; Briones,~J.; Teeny,~N.; Zhang,~L.; Kuhn,~T. Review of Ultrafast Demagnetization After Femtosecond Laser Pulses: A Complex Interaction of Light with Quantum Matter. \emph{Am. J. Mod. Phys.} \textbf{2018}, \emph{7}, 68--74\relax
\mciteBstWouldAddEndPuncttrue
\mciteSetBstMidEndSepPunct{\mcitedefaultmidpunct}
{\mcitedefaultendpunct}{\mcitedefaultseppunct}\relax
\EndOfBibitem
\bibitem[Bloom \latin{et~al.}(2024)Bloom, Paltiel, Naaman, and Waldeck]{Bloom2024}
Bloom,~B.; Paltiel,~Y.; Naaman,~R.; Waldeck,~D. Chiral Induced Spin Selectivity. \emph{Chem. Rev.} \textbf{2024}, \emph{124}, 1950--1991\relax
\mciteBstWouldAddEndPuncttrue
\mciteSetBstMidEndSepPunct{\mcitedefaultmidpunct}
{\mcitedefaultendpunct}{\mcitedefaultseppunct}\relax
\EndOfBibitem
\bibitem[Luo \latin{et~al.}(2023)Luo, Lin, Zhang, Chen, Blackert, Xu, Yakobson, and Zhu]{doi:10.1126/science.adi9601}
Luo,~J.; Lin,~T.; Zhang,~J.; Chen,~X.; Blackert,~E.~R.; Xu,~R.; Yakobson,~B.~I.; Zhu,~H. Large effective magnetic fields from chiral phonons in rare-earth halides. \emph{Science} \textbf{2023}, \emph{382}, 698--702\relax
\mciteBstWouldAddEndPuncttrue
\mciteSetBstMidEndSepPunct{\mcitedefaultmidpunct}
{\mcitedefaultendpunct}{\mcitedefaultseppunct}\relax
\EndOfBibitem
\bibitem[Johansson(2024)]{Annika2024}
Johansson,~A. Theory of spin and orbital Edelstein effects. \emph{Journal of Physics: Condensed Matter} \textbf{2024}, \emph{36}, 423002\relax
\mciteBstWouldAddEndPuncttrue
\mciteSetBstMidEndSepPunct{\mcitedefaultmidpunct}
{\mcitedefaultendpunct}{\mcitedefaultseppunct}\relax
\EndOfBibitem
\bibitem[Beaurepaire \latin{et~al.}(1996)Beaurepaire, Merle, Daunois, and Bigot]{PhysRevLett.76.4250}
Beaurepaire,~E.; Merle,~J.-C.; Daunois,~A.; Bigot,~J.-Y. Ultrafast Spin Dynamics in Ferromagnetic Nickel. \emph{Phys. Rev. Lett.} \textbf{1996}, \emph{76}, 4250--4253\relax
\mciteBstWouldAddEndPuncttrue
\mciteSetBstMidEndSepPunct{\mcitedefaultmidpunct}
{\mcitedefaultendpunct}{\mcitedefaultseppunct}\relax
\EndOfBibitem
\bibitem[Yang \latin{et~al.}(2017)Yang, Wilson, Gorchon, Lambert, Salahuddin, and Bokor]{doi:10.1126/sciadv.1603117}
Yang,~Y.; Wilson,~R.~B.; Gorchon,~J.; Lambert,~C.-H.; Salahuddin,~S.; Bokor,~J. Ultrafast magnetization reversal by picosecond electrical pulses. \emph{Science Advances} \textbf{2017}, \emph{3}, e1603117\relax
\mciteBstWouldAddEndPuncttrue
\mciteSetBstMidEndSepPunct{\mcitedefaultmidpunct}
{\mcitedefaultendpunct}{\mcitedefaultseppunct}\relax
\EndOfBibitem
\bibitem[Kimel \latin{et~al.}(2005)Kimel, Kirilyuk, Usachev, Pisarev, Balbashov, and Rasing]{kimel2005}
Kimel,~A.; Kirilyuk,~A.; Usachev,~P.; Pisarev,~R.; Balbashov,~A.; Rasing,~T. Ultrafast non-thermal control of magnetization by instantaneous photomagnetic pulses. \emph{Nature} \textbf{2005}, \emph{435}, 655--7\relax
\mciteBstWouldAddEndPuncttrue
\mciteSetBstMidEndSepPunct{\mcitedefaultmidpunct}
{\mcitedefaultendpunct}{\mcitedefaultseppunct}\relax
\EndOfBibitem
\bibitem[Ju \latin{et~al.}(2004)Ju, Hohlfeld, Bergman, van~de Veerdonk, Mryasov, Kim, Wu, Weller, and Koopmans]{PhysRevLett.93.197403}
Ju,~G.; Hohlfeld,~J.; Bergman,~B.; van~de Veerdonk,~R. J.~M.; Mryasov,~O.~N.; Kim,~J.-Y.; Wu,~X.; Weller,~D.; Koopmans,~B. Ultrafast Generation of Ferromagnetic Order via a Laser-Induced Phase Transformation in FeRh Thin Films. \emph{Phys. Rev. Lett.} \textbf{2004}, \emph{93}, 197403\relax
\mciteBstWouldAddEndPuncttrue
\mciteSetBstMidEndSepPunct{\mcitedefaultmidpunct}
{\mcitedefaultendpunct}{\mcitedefaultseppunct}\relax
\EndOfBibitem
\bibitem[Jakowski \latin{et~al.}(2025)Jakowski, Lu, Briggs, Lingerfelt, Sumpter, Ganesh, and Bernholc]{RMG}
Jakowski,~J.; Lu,~W.; Briggs,~E.; Lingerfelt,~D.; Sumpter,~B.~G.; Ganesh,~P.; Bernholc,~J. Simulation of 24,000 Electron Dynamics: Real-Time Time-Dependent Density Functional Theory (TDDFT) with the Real-Space Multigrids (RMG). \emph{Journal of Chemical Theory and Computation} \textbf{2025}, \emph{21}, 1322--1339\relax
\mciteBstWouldAddEndPuncttrue
\mciteSetBstMidEndSepPunct{\mcitedefaultmidpunct}
{\mcitedefaultendpunct}{\mcitedefaultseppunct}\relax
\EndOfBibitem
\bibitem[Lopata and Govind(2011)Lopata, and Govind]{NWChem}
Lopata,~K.; Govind,~N. Modeling Fast Electron Dynamics with Real-Time Time-Dependent Density Functional Theory: Application to Small Molecules and Chromophores. \emph{Journal of Chemical Theory and Computation} \textbf{2011}, \emph{7}, 1344--1355\relax
\mciteBstWouldAddEndPuncttrue
\mciteSetBstMidEndSepPunct{\mcitedefaultmidpunct}
{\mcitedefaultendpunct}{\mcitedefaultseppunct}\relax
\EndOfBibitem
\bibitem[Baczewski \latin{et~al.}(2014)Baczewski, Shulenburger, Desjarlais, and Magyar]{Baczewski2014}
Baczewski,~A.~D.; Shulenburger,~L.; Desjarlais,~M.~P.; Magyar,~R.~J. \emph{Numerical Implementation of Time-Dependent Density Functional Theory for Extended Systems in Extreme Environments}; 2014\relax
\mciteBstWouldAddEndPuncttrue
\mciteSetBstMidEndSepPunct{\mcitedefaultmidpunct}
{\mcitedefaultendpunct}{\mcitedefaultseppunct}\relax
\EndOfBibitem
\bibitem[Draeger \latin{et~al.}(2017)Draeger, Andrade, Gunnels, Bhatele, Schleife, and Correa]{Draeger2017}
Draeger,~E.~W.; Andrade,~X.; Gunnels,~J.~A.; Bhatele,~A.; Schleife,~A.; Correa,~A.~A. Massively parallel first-principles simulation of electron dynamics in materials. \emph{Journal of Parallel and Distributed Computing} \textbf{2017}, \emph{106}, 205–214\relax
\mciteBstWouldAddEndPuncttrue
\mciteSetBstMidEndSepPunct{\mcitedefaultmidpunct}
{\mcitedefaultendpunct}{\mcitedefaultseppunct}\relax
\EndOfBibitem
\bibitem[Noda \latin{et~al.}(2019)Noda, Sato, Hirokawa, Uemoto, Takeuchi, Yamada, Yamada, Shinohara, Yamaguchi, Iida, Floss, Otobe, Lee, Ishimura, Boku, Bertsch, Nobusada, and Yabana]{Noda2019}
Noda,~M. \latin{et~al.}  SALMON: Scalable Ab-initio Light–Matter simulator for Optics and Nanoscience. \emph{Computer Physics Communications} \textbf{2019}, \emph{235}, 356–365\relax
\mciteBstWouldAddEndPuncttrue
\mciteSetBstMidEndSepPunct{\mcitedefaultmidpunct}
{\mcitedefaultendpunct}{\mcitedefaultseppunct}\relax
\EndOfBibitem
\bibitem[K\"{u}hne \latin{et~al.}(2020)K\"{u}hne, Iannuzzi, Del~Ben, Rybkin, Seewald, Stein, Laino, Khaliullin, Sch\"{u}tt, Schiffmann, Golze, Wilhelm, Chulkov, Bani-Hashemian, Weber, Borštnik, Taillefumier, Jakobovits, Lazzaro, Pabst, M\"{u}ller, Schade, Guidon, Andermatt, Holmberg, Schenter, Hehn, Bussy, Belleflamme, Tabacchi, Gl\"{o}ß, Lass, Bethune, Mundy, Plessl, Watkins, VandeVondele, Krack, and Hutter]{Khne2020}
K\"{u}hne,~T.~D. \latin{et~al.}  CP2K: An electronic structure and molecular dynamics software package - Quickstep: Efficient and accurate electronic structure calculations. \emph{The Journal of Chemical Physics} \textbf{2020}, \emph{152}\relax
\mciteBstWouldAddEndPuncttrue
\mciteSetBstMidEndSepPunct{\mcitedefaultmidpunct}
{\mcitedefaultendpunct}{\mcitedefaultseppunct}\relax
\EndOfBibitem
\bibitem[Mortensen \latin{et~al.}(2024)Mortensen, Larsen, Kuisma, Ivanov, Taghizadeh, Peterson, Haldar, Dohn, Sch\"{a}fer, Jónsson, Hermes, Nilsson, Kastlunger, Levi, Jónsson, H\"{a}kkinen, Fojt, Kangsabanik, Sødequist, Lehtom\"{a}ki, Heske, Enkovaara, Winther, Dulak, Melander, Ovesen, Louhivuori, Walter, Gjerding, Lopez-Acevedo, Erhart, Warmbier, W\"{u}rdemann, Kaappa, Latini, Boland, Bligaard, Skovhus, Susi, Maxson, Rossi, Chen, Schmerwitz, Schiøtz, Olsen, Jacobsen, and Thygesen]{Mortensen2024}
Mortensen,~J.~J. \latin{et~al.}  GPAW: An open Python package for electronic structure calculations. \emph{The Journal of Chemical Physics} \textbf{2024}, \emph{160}\relax
\mciteBstWouldAddEndPuncttrue
\mciteSetBstMidEndSepPunct{\mcitedefaultmidpunct}
{\mcitedefaultendpunct}{\mcitedefaultseppunct}\relax
\EndOfBibitem
\bibitem[Tancogne-Dejean \latin{et~al.}(2020)Tancogne-Dejean, Oliveira, Andrade, Appel, Borca, Le~Breton, Buchholz, Castro, Corni, Correa, De~Giovannini, Delgado, Bich, Flick, Gil, Gomez, Helbig, Huebener, Jestaedt, Jornet-Somoza, Larsen, Lebedeva, Martin, Marques, Ohlmann, Pipolo, Rampp, Rozzi, Strubbe, Sato, Schaefer, Theophilou, Welden, and Rubio]{tancogne_dejean_octopus_2020}
Tancogne-Dejean,~N. \latin{et~al.}  \textsc{Octopus}, a computational framework for exploring light-driven phenomena and quantum dynamics in extended and finite systems. \emph{The Journal of Chemical Physics} \textbf{2020}, \emph{152}, 124119\relax
\mciteBstWouldAddEndPuncttrue
\mciteSetBstMidEndSepPunct{\mcitedefaultmidpunct}
{\mcitedefaultendpunct}{\mcitedefaultseppunct}\relax
\EndOfBibitem
\bibitem[Andrade \latin{et~al.}(2015)Andrade, Strubbe, De~Giovannini, Larsen, Oliveira, Alberdi-Rodriguez, Varas, Theophilou, Helbig, Verstraete, Stella, Nogueira, Aspuru-Guzik, Castro, Marques, and Rubio]{C5CP00351B}
Andrade,~X. \latin{et~al.}  Real-space grids and the \textsc{Octopus} code as tools for the development of new simulation approaches for electronic systems. \emph{Phys. Chem. Chem. Phys.} \textbf{2015}, \emph{17}, 31371--31396\relax
\mciteBstWouldAddEndPuncttrue
\mciteSetBstMidEndSepPunct{\mcitedefaultmidpunct}
{\mcitedefaultendpunct}{\mcitedefaultseppunct}\relax
\EndOfBibitem
\bibitem[Dewhurst \latin{et~al.}()Dewhurst, Sangeeta, Nordstr{\"o}m, Cricchio, Gran{\"a}s, and Gross]{Elk_code}
Dewhurst,~K.; Sangeeta,~S.; Nordstr{\"o}m,~L.; Cricchio,~F.; Gran{\"a}s,~O.; Gross,~H. The \textsc{Elk} code. \url{http://elk.sourceforge.net/}\relax
\mciteBstWouldAddEndPuncttrue
\mciteSetBstMidEndSepPunct{\mcitedefaultmidpunct}
{\mcitedefaultendpunct}{\mcitedefaultseppunct}\relax
\EndOfBibitem
\bibitem[Andrade \latin{et~al.}(2021)Andrade, Pemmaraju, Kartsev, Xiao, Lindenberg, Rajpurohit, Tan, Ogitsu, and Correa]{INQ2021}
Andrade,~X.; Pemmaraju,~C.; Kartsev,~A.; Xiao,~J.; Lindenberg,~A.; Rajpurohit,~S.; Tan,~L.; Ogitsu,~T.; Correa,~A. Inq, a Modern GPU-Accelerated Computational Framework for (Time-Dependent) Density Functional Theory. \emph{J. Chem. Theory Comput.} \textbf{2021}, \emph{17}, 7447--7467\relax
\mciteBstWouldAddEndPuncttrue
\mciteSetBstMidEndSepPunct{\mcitedefaultmidpunct}
{\mcitedefaultendpunct}{\mcitedefaultseppunct}\relax
\EndOfBibitem
\bibitem[Kohn \latin{et~al.}(1996)Kohn, Becke, and Parr]{Kohn1996}
Kohn,~W.; Becke,~A.; Parr,~R. Density Functional Theory of Electronic Structure. \emph{J. Phys. Chem.} \textbf{1996}, \emph{100}, 12974--12980\relax
\mciteBstWouldAddEndPuncttrue
\mciteSetBstMidEndSepPunct{\mcitedefaultmidpunct}
{\mcitedefaultendpunct}{\mcitedefaultseppunct}\relax
\EndOfBibitem
\bibitem[Kohn and Sham(1965)Kohn, and Sham]{PhysRev.140.A1133}
Kohn,~W.; Sham,~L.~J. Self-Consistent Equations Including Exchange and Correlation Effects. \emph{Phys. Rev.} \textbf{1965}, \emph{140}, A1133--A1138\relax
\mciteBstWouldAddEndPuncttrue
\mciteSetBstMidEndSepPunct{\mcitedefaultmidpunct}
{\mcitedefaultendpunct}{\mcitedefaultseppunct}\relax
\EndOfBibitem
\bibitem[von Barth and Hedin(1972)von Barth, and Hedin]{vonBarth_1972}
von Barth,~U.; Hedin,~L. A local exchange-correlation potential for the spin polarized case. i. \emph{Journal of Physics C: Solid State Physics} \textbf{1972}, \emph{5}, 1629\relax
\mciteBstWouldAddEndPuncttrue
\mciteSetBstMidEndSepPunct{\mcitedefaultmidpunct}
{\mcitedefaultendpunct}{\mcitedefaultseppunct}\relax
\EndOfBibitem
\bibitem[Jacob and Reiher(2012)Jacob, and Reiher]{https://doi.org/10.1002/qua.24309}
Jacob,~C.~R.; Reiher,~M. Spin in density-functional theory. \emph{International Journal of Quantum Chemistry} \textbf{2012}, \emph{112}, 3661--3684\relax
\mciteBstWouldAddEndPuncttrue
\mciteSetBstMidEndSepPunct{\mcitedefaultmidpunct}
{\mcitedefaultendpunct}{\mcitedefaultseppunct}\relax
\EndOfBibitem
\bibitem[Runge and Gross(1984)Runge, and Gross]{PhysRevLett.52.997}
Runge,~E.; Gross,~E. K.~U. Density-Functional Theory for Time-Dependent Systems. \emph{Phys. Rev. Lett.} \textbf{1984}, \emph{52}, 997--1000\relax
\mciteBstWouldAddEndPuncttrue
\mciteSetBstMidEndSepPunct{\mcitedefaultmidpunct}
{\mcitedefaultendpunct}{\mcitedefaultseppunct}\relax
\EndOfBibitem
\bibitem[van Leeuwen(1999)]{PhysRevLett.82.3863}
van Leeuwen,~R. Mapping from Densities to Potentials in Time-Dependent Density-Functional Theory. \emph{Phys. Rev. Lett.} \textbf{1999}, \emph{82}, 3863--3866\relax
\mciteBstWouldAddEndPuncttrue
\mciteSetBstMidEndSepPunct{\mcitedefaultmidpunct}
{\mcitedefaultendpunct}{\mcitedefaultseppunct}\relax
\EndOfBibitem
\bibitem[Ullrich(2011)]{10.1093/acprof:oso/9780199563029.001.0001}
Ullrich,~C.~A. \emph{Time-Dependent Density-Functional Theory: Concepts and Applications}; Oxford University Press, 2011\relax
\mciteBstWouldAddEndPuncttrue
\mciteSetBstMidEndSepPunct{\mcitedefaultmidpunct}
{\mcitedefaultendpunct}{\mcitedefaultseppunct}\relax
\EndOfBibitem
\bibitem[Capelle \latin{et~al.}(2001)Capelle, Vignale, and Gy\"orffy]{PhysRevLett.87.206403}
Capelle,~K.; Vignale,~G.; Gy\"orffy,~B.~L. Spin Currents and Spin Dynamics in Time-Dependent Density-Functional Theory. \emph{Phys. Rev. Lett.} \textbf{2001}, \emph{87}, 206403\relax
\mciteBstWouldAddEndPuncttrue
\mciteSetBstMidEndSepPunct{\mcitedefaultmidpunct}
{\mcitedefaultendpunct}{\mcitedefaultseppunct}\relax
\EndOfBibitem
\bibitem[Hill \latin{et~al.}(2023)Hill, Shotton, and Ullrich]{PhysRevB.107.115134}
Hill,~D.; Shotton,~J.; Ullrich,~C.~A. Magnetization dynamics with time-dependent spin-density functional theory: Significance of exchange-correlation torques. \emph{Phys. Rev. B} \textbf{2023}, \emph{107}, 115134\relax
\mciteBstWouldAddEndPuncttrue
\mciteSetBstMidEndSepPunct{\mcitedefaultmidpunct}
{\mcitedefaultendpunct}{\mcitedefaultseppunct}\relax
\EndOfBibitem
\bibitem[Gross and Kohn(1985)Gross, and Kohn]{PhysRevLett.55.2850}
Gross,~E. K.~U.; Kohn,~W. Local density-functional theory of frequency-dependent linear response. \emph{Phys. Rev. Lett.} \textbf{1985}, \emph{55}, 2850--2852\relax
\mciteBstWouldAddEndPuncttrue
\mciteSetBstMidEndSepPunct{\mcitedefaultmidpunct}
{\mcitedefaultendpunct}{\mcitedefaultseppunct}\relax
\EndOfBibitem
\bibitem[Iwamoto and Gross(1987)Iwamoto, and Gross]{PhysRevB.35.3003}
Iwamoto,~N.; Gross,~E. K.~U. Correlation effects on the third-frequency-moment sum rule of electron liquids. \emph{Phys. Rev. B} \textbf{1987}, \emph{35}, 3003--3004\relax
\mciteBstWouldAddEndPuncttrue
\mciteSetBstMidEndSepPunct{\mcitedefaultmidpunct}
{\mcitedefaultendpunct}{\mcitedefaultseppunct}\relax
\EndOfBibitem
\bibitem[Constantin and Pitarke(2007)Constantin, and Pitarke]{PhysRevB.75.245127}
Constantin,~L.~A.; Pitarke,~J.~M. Simple dynamic exchange-correlation kernel of a uniform electron gas. \emph{Phys. Rev. B} \textbf{2007}, \emph{75}, 245127\relax
\mciteBstWouldAddEndPuncttrue
\mciteSetBstMidEndSepPunct{\mcitedefaultmidpunct}
{\mcitedefaultendpunct}{\mcitedefaultseppunct}\relax
\EndOfBibitem
\bibitem[Richardson and Ashcroft(1994)Richardson, and Ashcroft]{PhysRevB.50.8170}
Richardson,~C.~F.; Ashcroft,~N.~W. Dynamical local-field factors and effective interactions in the three-dimensional electron liquid. \emph{Phys. Rev. B} \textbf{1994}, \emph{50}, 8170--8181\relax
\mciteBstWouldAddEndPuncttrue
\mciteSetBstMidEndSepPunct{\mcitedefaultmidpunct}
{\mcitedefaultendpunct}{\mcitedefaultseppunct}\relax
\EndOfBibitem
\bibitem[Corradini \latin{et~al.}(1998)Corradini, Del~Sole, Onida, and Palummo]{PhysRevB.57.14569}
Corradini,~M.; Del~Sole,~R.; Onida,~G.; Palummo,~M. Analytical expressions for the local-field factor $G(q)$ and the exchange-correlation kernel ${K}_{\mathrm{xc}}(r)$ of the homogeneous electron gas. \emph{Phys. Rev. B} \textbf{1998}, \emph{57}, 14569--14571\relax
\mciteBstWouldAddEndPuncttrue
\mciteSetBstMidEndSepPunct{\mcitedefaultmidpunct}
{\mcitedefaultendpunct}{\mcitedefaultseppunct}\relax
\EndOfBibitem
\bibitem[Dabrowski(1986)]{PhysRevB.34.4989}
Dabrowski,~B. Dynamical local-field factor in the response function of an electron gas. \emph{Phys. Rev. B} \textbf{1986}, \emph{34}, 4989--4995\relax
\mciteBstWouldAddEndPuncttrue
\mciteSetBstMidEndSepPunct{\mcitedefaultmidpunct}
{\mcitedefaultendpunct}{\mcitedefaultseppunct}\relax
\EndOfBibitem
\bibitem[Kaplan \latin{et~al.}(2022)Kaplan, Nepal, Ruzsinszky, Ballone, and Perdew]{PhysRevB.105.035123}
Kaplan,~A.~D.; Nepal,~N.~K.; Ruzsinszky,~A.; Ballone,~P.; Perdew,~J.~P. First-principles wave-vector- and frequency-dependent exchange-correlation kernel for jellium at all densities. \emph{Phys. Rev. B} \textbf{2022}, \emph{105}, 035123\relax
\mciteBstWouldAddEndPuncttrue
\mciteSetBstMidEndSepPunct{\mcitedefaultmidpunct}
{\mcitedefaultendpunct}{\mcitedefaultseppunct}\relax
\EndOfBibitem
\bibitem[Panholzer \latin{et~al.}(2018)Panholzer, Gatti, and Reining]{PhysRevLett.120.166402}
Panholzer,~M.; Gatti,~M.; Reining,~L. Nonlocal and Nonadiabatic Effects in the Charge-Density Response of Solids: A Time-Dependent Density-Functional Approach. \emph{Phys. Rev. Lett.} \textbf{2018}, \emph{120}, 166402\relax
\mciteBstWouldAddEndPuncttrue
\mciteSetBstMidEndSepPunct{\mcitedefaultmidpunct}
{\mcitedefaultendpunct}{\mcitedefaultseppunct}\relax
\EndOfBibitem
\bibitem[Dobson \latin{et~al.}(1997)Dobson, B\"unner, and Gross]{PhysRevLett.79.1905}
Dobson,~J.~F.; B\"unner,~M.~J.; Gross,~E. K.~U. Time-Dependent Density Functional Theory beyond Linear Response: An Exchange-Correlation Potential with Memory. \emph{Phys. Rev. Lett.} \textbf{1997}, \emph{79}, 1905--1908\relax
\mciteBstWouldAddEndPuncttrue
\mciteSetBstMidEndSepPunct{\mcitedefaultmidpunct}
{\mcitedefaultendpunct}{\mcitedefaultseppunct}\relax
\EndOfBibitem
\bibitem[Vignale \latin{et~al.}(1997)Vignale, Ullrich, and Conti]{PhysRevLett.79.4878}
Vignale,~G.; Ullrich,~C.~A.; Conti,~S. Time-Dependent Density Functional Theory Beyond the Adiabatic Local Density Approximation. \emph{Phys. Rev. Lett.} \textbf{1997}, \emph{79}, 4878--4881\relax
\mciteBstWouldAddEndPuncttrue
\mciteSetBstMidEndSepPunct{\mcitedefaultmidpunct}
{\mcitedefaultendpunct}{\mcitedefaultseppunct}\relax
\EndOfBibitem
\bibitem[Qian and Vignale(2002)Qian, and Vignale]{PhysRevB.65.235121}
Qian,~Z.; Vignale,~G. Dynamical exchange-correlation potentials for an electron liquid. \emph{Phys. Rev. B} \textbf{2002}, \emph{65}, 235121\relax
\mciteBstWouldAddEndPuncttrue
\mciteSetBstMidEndSepPunct{\mcitedefaultmidpunct}
{\mcitedefaultendpunct}{\mcitedefaultseppunct}\relax
\EndOfBibitem
\bibitem[Lacombe and Maitra(2023)Lacombe, and Maitra]{Lacombe2023}
Lacombe,~L.; Maitra,~N. Non-Adiabatic Approximations in Time-Dependent Density Functional Theory: Progress and Prospects. \emph{Npj Computational Materials} \textbf{2023}, \emph{9}, 124--15\relax
\mciteBstWouldAddEndPuncttrue
\mciteSetBstMidEndSepPunct{\mcitedefaultmidpunct}
{\mcitedefaultendpunct}{\mcitedefaultseppunct}\relax
\EndOfBibitem
\bibitem[Wilken and Bauer(2006)Wilken, and Bauer]{PhysRevLett.97.203001}
Wilken,~F.; Bauer,~D. Adiabatic Approximation of the Correlation Function in the Density-Functional Treatment of Ionization Processes. \emph{Phys. Rev. Lett.} \textbf{2006}, \emph{97}, 203001\relax
\mciteBstWouldAddEndPuncttrue
\mciteSetBstMidEndSepPunct{\mcitedefaultmidpunct}
{\mcitedefaultendpunct}{\mcitedefaultseppunct}\relax
\EndOfBibitem
\bibitem[Perdew \latin{et~al.}(1996)Perdew, Burke, and Ernzerhof]{PhysRevLett.77.3865}
Perdew,~J.~P.; Burke,~K.; Ernzerhof,~M. Generalized Gradient Approximation Made Simple. \emph{Phys. Rev. Lett.} \textbf{1996}, \emph{77}, 3865--3868\relax
\mciteBstWouldAddEndPuncttrue
\mciteSetBstMidEndSepPunct{\mcitedefaultmidpunct}
{\mcitedefaultendpunct}{\mcitedefaultseppunct}\relax
\EndOfBibitem
\bibitem[Tozer and Handy(2000)Tozer, and Handy]{A910321J}
Tozer,~D.~J.; Handy,~N.~C. On the determination of excitation energies using density functional theory. \emph{Phys. Chem. Chem. Phys.} \textbf{2000}, \emph{2}, 2117--2121\relax
\mciteBstWouldAddEndPuncttrue
\mciteSetBstMidEndSepPunct{\mcitedefaultmidpunct}
{\mcitedefaultendpunct}{\mcitedefaultseppunct}\relax
\EndOfBibitem
\bibitem[Raghunathan and Nest(2011)Raghunathan, and Nest]{Raghunathan2011}
Raghunathan,~S.; Nest,~M. Critical Examination of Explicitly Time-Dependent Density Functional Theory for Coherent Control of Dipole Switching. \emph{J Chem Theory Comput} \textbf{2011}, \emph{7}, 2492--2497\relax
\mciteBstWouldAddEndPuncttrue
\mciteSetBstMidEndSepPunct{\mcitedefaultmidpunct}
{\mcitedefaultendpunct}{\mcitedefaultseppunct}\relax
\EndOfBibitem
\bibitem[Kubler \latin{et~al.}(1988)Kubler, Hock, Sticht, and Williams]{Kubler1988}
Kubler,~J.; Hock,~K.-H.; Sticht,~J.; Williams,~A. Density functional theory of non-collinear magnetism. \emph{J. Phys. F: Met. Phys.} \textbf{1988}, \emph{18}, 469\relax
\mciteBstWouldAddEndPuncttrue
\mciteSetBstMidEndSepPunct{\mcitedefaultmidpunct}
{\mcitedefaultendpunct}{\mcitedefaultseppunct}\relax
\EndOfBibitem
\bibitem[Sandratskii(1998)]{Sandratskii01011998}
Sandratskii,~L.~M. Noncollinear magnetism in itinerant-electron systems: Theory and applications. \emph{Advances in Physics} \textbf{1998}, \emph{47}, 91--160\relax
\mciteBstWouldAddEndPuncttrue
\mciteSetBstMidEndSepPunct{\mcitedefaultmidpunct}
{\mcitedefaultendpunct}{\mcitedefaultseppunct}\relax
\EndOfBibitem
\bibitem[Pu \latin{et~al.}(2023)Pu, Li, Zhang, Jiang, Gao, Xiao, Sun, Zhang, and Shao]{PhysRevResearch.5.013036}
Pu,~Z.; Li,~H.; Zhang,~N.; Jiang,~H.; Gao,~Y.; Xiao,~Y.; Sun,~Q.; Zhang,~Y.; Shao,~S. Noncollinear density functional theory. \emph{Phys. Rev. Res.} \textbf{2023}, \emph{5}, 013036\relax
\mciteBstWouldAddEndPuncttrue
\mciteSetBstMidEndSepPunct{\mcitedefaultmidpunct}
{\mcitedefaultendpunct}{\mcitedefaultseppunct}\relax
\EndOfBibitem
\bibitem[Sharma \latin{et~al.}(2007)Sharma, Dewhurst, Ambrosch-Draxl, Kurth, Helbig, Pittalis, Shallcross, Nordstr\"om, and Gross]{PhysRevLett.98.196405}
Sharma,~S.; Dewhurst,~J.~K.; Ambrosch-Draxl,~C.; Kurth,~S.; Helbig,~N.; Pittalis,~S.; Shallcross,~S.; Nordstr\"om,~L.; Gross,~E. K.~U. First-Principles Approach to Noncollinear Magnetism: Towards Spin Dynamics. \emph{Phys. Rev. Lett.} \textbf{2007}, \emph{98}, 196405\relax
\mciteBstWouldAddEndPuncttrue
\mciteSetBstMidEndSepPunct{\mcitedefaultmidpunct}
{\mcitedefaultendpunct}{\mcitedefaultseppunct}\relax
\EndOfBibitem
\bibitem[Katsnelson and Antropov(2003)Katsnelson, and Antropov]{PhysRevB.67.140406}
Katsnelson,~M.~I.; Antropov,~V.~P. Spin angular gradient approximation in the density functional theory. \emph{Phys. Rev. B} \textbf{2003}, \emph{67}, 140406\relax
\mciteBstWouldAddEndPuncttrue
\mciteSetBstMidEndSepPunct{\mcitedefaultmidpunct}
{\mcitedefaultendpunct}{\mcitedefaultseppunct}\relax
\EndOfBibitem
\bibitem[Scalmani and Frisch(2012)Scalmani, and Frisch]{Scalmani2012}
Scalmani,~G.; Frisch,~M. A New Approach to Noncollinear Spin Density Functional Theory beyond the Local Density Approximation. \emph{J Chem Theory Comput.} \textbf{2012}, \emph{8}, 2193--6\relax
\mciteBstWouldAddEndPuncttrue
\mciteSetBstMidEndSepPunct{\mcitedefaultmidpunct}
{\mcitedefaultendpunct}{\mcitedefaultseppunct}\relax
\EndOfBibitem
\bibitem[Eich and Gross(2013)Eich, and Gross]{PhysRevLett.111.156401}
Eich,~F.~G.; Gross,~E. K.~U. Transverse Spin-Gradient Functional for Noncollinear Spin-Density-Functional Theory. \emph{Phys. Rev. Lett.} \textbf{2013}, \emph{111}, 156401\relax
\mciteBstWouldAddEndPuncttrue
\mciteSetBstMidEndSepPunct{\mcitedefaultmidpunct}
{\mcitedefaultendpunct}{\mcitedefaultseppunct}\relax
\EndOfBibitem
\bibitem[Eich \latin{et~al.}(2013)Eich, Pittalis, and Vignale]{PhysRevB.88.245102}
Eich,~F.~G.; Pittalis,~S.; Vignale,~G. Transverse and longitudinal gradients of the spin magnetization in spin-density-functional theory. \emph{Phys. Rev. B} \textbf{2013}, \emph{88}, 245102\relax
\mciteBstWouldAddEndPuncttrue
\mciteSetBstMidEndSepPunct{\mcitedefaultmidpunct}
{\mcitedefaultendpunct}{\mcitedefaultseppunct}\relax
\EndOfBibitem
\bibitem[Pittalis \latin{et~al.}(2017)Pittalis, Vignale, and Eich]{PhysRevB.96.035141}
Pittalis,~S.; Vignale,~G.; Eich,~F.~G. $\text{U}(1)\ifmmode\times\else\texttimes\fi{}\mathrm{SU}(2)$ gauge invariance made simple for density functional approximations. \emph{Phys. Rev. B} \textbf{2017}, \emph{96}, 035141\relax
\mciteBstWouldAddEndPuncttrue
\mciteSetBstMidEndSepPunct{\mcitedefaultmidpunct}
{\mcitedefaultendpunct}{\mcitedefaultseppunct}\relax
\EndOfBibitem
\bibitem[Tancogne-Dejean \latin{et~al.}(2023)Tancogne-Dejean, Rubio, and Ullrich]{PhysRevB.107.165111}
Tancogne-Dejean,~N.; Rubio,~A.; Ullrich,~C.~A. Constructing semilocal approximations for noncollinear spin density functional theory featuring exchange-correlation torques. \emph{Phys. Rev. B} \textbf{2023}, \emph{107}, 165111\relax
\mciteBstWouldAddEndPuncttrue
\mciteSetBstMidEndSepPunct{\mcitedefaultmidpunct}
{\mcitedefaultendpunct}{\mcitedefaultseppunct}\relax
\EndOfBibitem
\bibitem[Komorovsky \latin{et~al.}(2019)Komorovsky, Cherry, and Repisky]{10.1063/1.5121713}
Komorovsky,~S.; Cherry,~P.~J.; Repisky,~M. Four-component relativistic time-dependent density-functional theory using a stable noncollinear DFT ansatz applicable to both closed- and open-shell systems. \emph{The Journal of Chemical Physics} \textbf{2019}, \emph{151}, 184111\relax
\mciteBstWouldAddEndPuncttrue
\mciteSetBstMidEndSepPunct{\mcitedefaultmidpunct}
{\mcitedefaultendpunct}{\mcitedefaultseppunct}\relax
\EndOfBibitem
\bibitem[Liu and Xiao(2018)Liu, and Xiao]{C8CS00175H}
Liu,~W.; Xiao,~Y. Relativistic time-dependent density functional theories. \emph{Chem. Soc. Rev.} \textbf{2018}, \emph{47}, 4481--4509\relax
\mciteBstWouldAddEndPuncttrue
\mciteSetBstMidEndSepPunct{\mcitedefaultmidpunct}
{\mcitedefaultendpunct}{\mcitedefaultseppunct}\relax
\EndOfBibitem
\bibitem[Dyall and F{\ae}gri(2007)Dyall, and F{\ae}gri]{Dyall2007IntroductionTR}
Dyall,~K.~G.; F{\ae}gri,~K.~J. Introduction to Relativistic Quantum Chemistry. 2007\relax
\mciteBstWouldAddEndPuncttrue
\mciteSetBstMidEndSepPunct{\mcitedefaultmidpunct}
{\mcitedefaultendpunct}{\mcitedefaultseppunct}\relax
\EndOfBibitem
\bibitem[Visscher \latin{et~al.}(1994)Visscher, Visser, Aerts, Merenga, and Nieuwpoort]{VISSCHER1994120}
Visscher,~L.; Visser,~O.; Aerts,~P.; Merenga,~H.; Nieuwpoort,~W. Relativistic quantum chemistry: the MOLFDIR program package. \emph{Computer Physics Communications} \textbf{1994}, \emph{81}, 120--144\relax
\mciteBstWouldAddEndPuncttrue
\mciteSetBstMidEndSepPunct{\mcitedefaultmidpunct}
{\mcitedefaultendpunct}{\mcitedefaultseppunct}\relax
\EndOfBibitem
\bibitem[Saue(2011)]{https://doi.org/10.1002/cphc.201100682}
Saue,~T. Relativistic Hamiltonians for Chemistry: A Primer. \emph{ChemPhysChem} \textbf{2011}, \emph{12}, 3077--3094\relax
\mciteBstWouldAddEndPuncttrue
\mciteSetBstMidEndSepPunct{\mcitedefaultmidpunct}
{\mcitedefaultendpunct}{\mcitedefaultseppunct}\relax
\EndOfBibitem
\bibitem[Sakurai(1987)]{sakurai1987advanced}
Sakurai,~J. \emph{Advanced Quantum Mechanics}; Addison-Wesley Series in Advanced Physics; Addison-Wesley, 1987\relax
\mciteBstWouldAddEndPuncttrue
\mciteSetBstMidEndSepPunct{\mcitedefaultmidpunct}
{\mcitedefaultendpunct}{\mcitedefaultseppunct}\relax
\EndOfBibitem
\bibitem[van Lenthe \latin{et~al.}(1994)van Lenthe, Baerends, and Snijders]{10.1063/1.467943}
van Lenthe,~E.; Baerends,~E.~J.; Snijders,~J.~G. Relativistic total energy using regular approximations. \emph{The Journal of Chemical Physics} \textbf{1994}, \emph{101}, 9783--9792\relax
\mciteBstWouldAddEndPuncttrue
\mciteSetBstMidEndSepPunct{\mcitedefaultmidpunct}
{\mcitedefaultendpunct}{\mcitedefaultseppunct}\relax
\EndOfBibitem
\bibitem[Douglas and Kroll(1974)Douglas, and Kroll]{DOUGLAS197489}
Douglas,~M.; Kroll,~N.~M. Quantum electrodynamical corrections to the fine structure of helium. \emph{Annals of Physics} \textbf{1974}, \emph{82}, 89--155\relax
\mciteBstWouldAddEndPuncttrue
\mciteSetBstMidEndSepPunct{\mcitedefaultmidpunct}
{\mcitedefaultendpunct}{\mcitedefaultseppunct}\relax
\EndOfBibitem
\bibitem[Reiher(2012)]{https://doi.org/10.1002/wcms.67}
Reiher,~M. Relativistic Douglas–Kroll–Hess theory. \emph{WIREs Computational Molecular Science} \textbf{2012}, \emph{2}, 139--149\relax
\mciteBstWouldAddEndPuncttrue
\mciteSetBstMidEndSepPunct{\mcitedefaultmidpunct}
{\mcitedefaultendpunct}{\mcitedefaultseppunct}\relax
\EndOfBibitem
\bibitem[Peng and Reiher(2012)Peng, and Reiher]{Peng2012}
Peng,~D.; Reiher,~M. Exact decoupling of the relativistic Fock operator. \emph{Theoretical Chemistry Accounts} \textbf{2012}, \emph{131}, 1081\relax
\mciteBstWouldAddEndPuncttrue
\mciteSetBstMidEndSepPunct{\mcitedefaultmidpunct}
{\mcitedefaultendpunct}{\mcitedefaultseppunct}\relax
\EndOfBibitem
\bibitem[Peng \latin{et~al.}(2013)Peng, Middendorf, Weigend, and Reiher]{10.1063/1.4803693}
Peng,~D.; Middendorf,~N.; Weigend,~F.; Reiher,~M. An efficient implementation of two-component relativistic exact-decoupling methods for large molecules. \emph{The Journal of Chemical Physics} \textbf{2013}, \emph{138}, 184105\relax
\mciteBstWouldAddEndPuncttrue
\mciteSetBstMidEndSepPunct{\mcitedefaultmidpunct}
{\mcitedefaultendpunct}{\mcitedefaultseppunct}\relax
\EndOfBibitem
\bibitem[Zheng \latin{et~al.}(2025)Zheng, Upadhyay, Wang, Shayit, Liu, Sun, and Li]{zheng2025chiralitydrivenmagnetizationemergesrelativistic}
Zheng,~X.; Upadhyay,~S.; Wang,~T.; Shayit,~A.; Liu,~J.; Sun,~D.; Li,~X. Chirality-Driven Magnetization Emerges from Relativistic Four-Current Dynamics. 2025; \url{https://arxiv.org/abs/2504.03781}\relax
\mciteBstWouldAddEndPuncttrue
\mciteSetBstMidEndSepPunct{\mcitedefaultmidpunct}
{\mcitedefaultendpunct}{\mcitedefaultseppunct}\relax
\EndOfBibitem
\bibitem[Strange(1998)]{Strange_1998}
Strange,~P. \emph{Relativistic Quantum Mechanics: With Applications in Condensed Matter and Atomic Physics}; Cambridge University Press, 1998\relax
\mciteBstWouldAddEndPuncttrue
\mciteSetBstMidEndSepPunct{\mcitedefaultmidpunct}
{\mcitedefaultendpunct}{\mcitedefaultseppunct}\relax
\EndOfBibitem
\bibitem[Fr\"ohlich and Studer(1993)Fr\"ohlich, and Studer]{RevModPhys.65.733}
Fr\"ohlich,~J.; Studer,~U.~M. Gauge invariance and current algebra in nonrelativistic many-body theory. \emph{Rev. Mod. Phys.} \textbf{1993}, \emph{65}, 733--802\relax
\mciteBstWouldAddEndPuncttrue
\mciteSetBstMidEndSepPunct{\mcitedefaultmidpunct}
{\mcitedefaultendpunct}{\mcitedefaultseppunct}\relax
\EndOfBibitem
\bibitem[Theurich and Hill(2001)Theurich, and Hill]{PhysRevB.64.073106}
Theurich,~G.; Hill,~N.~A. Self-consistent treatment of spin-orbit coupling in solids using relativistic fully separable ab initio pseudopotentials. \emph{Phys. Rev. B} \textbf{2001}, \emph{64}, 073106\relax
\mciteBstWouldAddEndPuncttrue
\mciteSetBstMidEndSepPunct{\mcitedefaultmidpunct}
{\mcitedefaultendpunct}{\mcitedefaultseppunct}\relax
\EndOfBibitem
\bibitem[Zhang(2014)]{Zhang2014}
Zhang,~Z. Spin–orbit DFT with analytic gradients and applications to heavy element compounds. \emph{Theoretical Chemistry Accounts} \textbf{2014}, \emph{133}, 1588\relax
\mciteBstWouldAddEndPuncttrue
\mciteSetBstMidEndSepPunct{\mcitedefaultmidpunct}
{\mcitedefaultendpunct}{\mcitedefaultseppunct}\relax
\EndOfBibitem
\bibitem[Blackford \latin{et~al.}(2002)Blackford, Petitet, Pozo, Remington, Whaley, Demmel, Dongarra, Duff, Hammarling, Henry, Heroux, Kaufman, Lumsdaine, Petite, Pozo, Remington, and Whaley]{Blackford2002}
Blackford,~L.~S. \latin{et~al.}  An updated set of basic linear algebra subprograms (BLAS). \emph{ACM Transactions on Mathematical Software} \textbf{2002}, \emph{28}, 135--151\relax
\mciteBstWouldAddEndPuncttrue
\mciteSetBstMidEndSepPunct{\mcitedefaultmidpunct}
{\mcitedefaultendpunct}{\mcitedefaultseppunct}\relax
\EndOfBibitem
\bibitem[Anderson \latin{et~al.}(1999)Anderson, Bai, Bischof, Blackford, Demmel, Dongarra, Du~Croz, Greenbaum, Hammarling, McKenney, and Sorensen]{Anderson1999}
Anderson,~E.; Bai,~Z.; Bischof,~C.; Blackford,~S.; Demmel,~J.; Dongarra,~J.; Du~Croz,~J.; Greenbaum,~A.; Hammarling,~S.; McKenney,~A.; Sorensen,~D. \emph{{LAPACK} Users' Guide}, 3rd ed.; Society for Industrial and Applied Mathematics: Philadelphia, PA, 1999\relax
\mciteBstWouldAddEndPuncttrue
\mciteSetBstMidEndSepPunct{\mcitedefaultmidpunct}
{\mcitedefaultendpunct}{\mcitedefaultseppunct}\relax
\EndOfBibitem
\bibitem[Correa(2025)]{Multi}
Correa,~A.~A. Multi: Multidimensional arrays for C++. https://gitlab.com/correaa/boost-multi, 2025\relax
\mciteBstWouldAddEndPuncttrue
\mciteSetBstMidEndSepPunct{\mcitedefaultmidpunct}
{\mcitedefaultendpunct}{\mcitedefaultseppunct}\relax
\EndOfBibitem
\bibitem[Andrade \latin{et~al.}(2012)Andrade, Alberdi-Rodriguez, Strubbe, Oliveira, Nogueira, Castro, Muguerza, Arruabarrena, Louie, Aspuru-Guzik, Rubio, and Marques]{Andrade2012}
Andrade,~X.; Alberdi-Rodriguez,~J.; Strubbe,~D.~A.; Oliveira,~M. J.~T.; Nogueira,~F.; Castro,~A.; Muguerza,~J.; Arruabarrena,~A.; Louie,~S.~G.; Aspuru-Guzik,~A.; Rubio,~A.; Marques,~M. A.~L. Time-dependent density-functional theory in massively parallel computer architectures: the \textsc{octopus} project. \emph{Journal of Physics: Condensed Matter} \textbf{2012}, \emph{24}, 233202\relax
\mciteBstWouldAddEndPuncttrue
\mciteSetBstMidEndSepPunct{\mcitedefaultmidpunct}
{\mcitedefaultendpunct}{\mcitedefaultseppunct}\relax
\EndOfBibitem
\bibitem[Aurenhammer \latin{et~al.}(2013)Aurenhammer, Klein, and Lee]{doi:10.1142/8685}
Aurenhammer,~F.; Klein,~R.; Lee,~D.-T. \emph{Voronoi Diagrams and Delaunay Triangulations}; WORLD SCIENTIFIC, 2013\relax
\mciteBstWouldAddEndPuncttrue
\mciteSetBstMidEndSepPunct{\mcitedefaultmidpunct}
{\mcitedefaultendpunct}{\mcitedefaultseppunct}\relax
\EndOfBibitem
\bibitem[Lebedeva \latin{et~al.}(2019)Lebedeva, Strubbe, Tokatly, and Rubio]{Lebedeva2019-er}
Lebedeva,~I.~V.; Strubbe,~D.~A.; Tokatly,~I.~V.; Rubio,~A. Orbital magneto-optical response of periodic insulators from first principles. \emph{Npj Comput. Mater.} \textbf{2019}, \emph{5}\relax
\mciteBstWouldAddEndPuncttrue
\mciteSetBstMidEndSepPunct{\mcitedefaultmidpunct}
{\mcitedefaultendpunct}{\mcitedefaultseppunct}\relax
\EndOfBibitem
\bibitem[Kaneko \latin{et~al.}(2022)Kaneko, Murakami, Takayoshi, and Millis]{2022Kaneko}
Kaneko,~T.; Murakami,~Y.; Takayoshi,~S.; Millis,~A.~J. Second-order magnetic responses in quantum magnets: Magnetization under ac magnetic fields. \emph{Phys. Rev. B} \textbf{2022}, \emph{105}, 195126\relax
\mciteBstWouldAddEndPuncttrue
\mciteSetBstMidEndSepPunct{\mcitedefaultmidpunct}
{\mcitedefaultendpunct}{\mcitedefaultseppunct}\relax
\EndOfBibitem
\bibitem[Surh \latin{et~al.}(1991)Surh, Li, and Louie]{PhysRevB.43.4286}
Surh,~M.~P.; Li,~M.-F.; Louie,~S.~G. Spin-orbit splitting of GaAs and InSb bands near \ensuremath{\Gamma}. \emph{Phys. Rev. B} \textbf{1991}, \emph{43}, 4286--4294\relax
\mciteBstWouldAddEndPuncttrue
\mciteSetBstMidEndSepPunct{\mcitedefaultmidpunct}
{\mcitedefaultendpunct}{\mcitedefaultseppunct}\relax
\EndOfBibitem
\bibitem[Hemstreet \latin{et~al.}(1993)Hemstreet, Fong, and Nelson]{PhysRevB.47.4238}
Hemstreet,~L.~A.; Fong,~C.~Y.; Nelson,~J.~S. First-principles calculations of spin-orbit splittings in solids using nonlocal separable pseudopotentials. \emph{Phys. Rev. B} \textbf{1993}, \emph{47}, 4238--4243\relax
\mciteBstWouldAddEndPuncttrue
\mciteSetBstMidEndSepPunct{\mcitedefaultmidpunct}
{\mcitedefaultendpunct}{\mcitedefaultseppunct}\relax
\EndOfBibitem
\bibitem[Kleinman(1980)]{PhysRevB.21.2630}
Kleinman,~L. Relativistic norm-conserving pseudopotential. \emph{Phys. Rev. B} \textbf{1980}, \emph{21}, 2630--2631\relax
\mciteBstWouldAddEndPuncttrue
\mciteSetBstMidEndSepPunct{\mcitedefaultmidpunct}
{\mcitedefaultendpunct}{\mcitedefaultseppunct}\relax
\EndOfBibitem
\bibitem[Kleinman and Bylander(1982)Kleinman, and Bylander]{Kleinman1982}
Kleinman,~L.; Bylander,~D.~M. Efficacious Form for Model Pseudopotentials. \emph{Phys. Rev. Lett.} \textbf{1982}, \emph{48}, 1425–1428\relax
\mciteBstWouldAddEndPuncttrue
\mciteSetBstMidEndSepPunct{\mcitedefaultmidpunct}
{\mcitedefaultendpunct}{\mcitedefaultseppunct}\relax
\EndOfBibitem
\bibitem[{van Setten} \latin{et~al.}(2018){van Setten}, Giantomassi, Bousquet, Verstraete, Hamann, Gonze, and Rignanese]{VANSETTEN201839}
{van Setten},~M.; Giantomassi,~M.; Bousquet,~E.; Verstraete,~M.; Hamann,~D.; Gonze,~X.; Rignanese,~G.-M. The PseudoDojo: Training and grading a 85 element optimized norm-conserving pseudopotential table. \emph{Computer Physics Communications} \textbf{2018}, \emph{226}, 39--54\relax
\mciteBstWouldAddEndPuncttrue
\mciteSetBstMidEndSepPunct{\mcitedefaultmidpunct}
{\mcitedefaultendpunct}{\mcitedefaultseppunct}\relax
\EndOfBibitem
\bibitem[Andrade \latin{et~al.}()Andrade, Simoni, Ping, Ogitsu, and Correa]{Andrade2025}
Andrade,~X.; Simoni,~J.; Ping,~Y.; Ogitsu,~T.; Correa,~A.~A. in preparation\relax
\mciteBstWouldAddEndPuncttrue
\mciteSetBstMidEndSepPunct{\mcitedefaultmidpunct}
{\mcitedefaultendpunct}{\mcitedefaultseppunct}\relax
\EndOfBibitem
\bibitem[Stamenova \latin{et~al.}(2016)Stamenova, Simoni, and Sanvito]{PhysRevB.94.014423}
Stamenova,~M.; Simoni,~J.; Sanvito,~S. Role of spin-orbit interaction in the ultrafast demagnetization of small iron clusters. \emph{Phys. Rev. B} \textbf{2016}, \emph{94}, 014423\relax
\mciteBstWouldAddEndPuncttrue
\mciteSetBstMidEndSepPunct{\mcitedefaultmidpunct}
{\mcitedefaultendpunct}{\mcitedefaultseppunct}\relax
\EndOfBibitem
\bibitem[Giannozzi \latin{et~al.}(2017)Giannozzi, Andreussi, Brumme, Bunau, Buongiorno~Nardelli, Calandra, Car, Cavazzoni, Ceresoli, Cococcioni, Colonna, Carnimeo, Dal~Corso, de~Gironcoli, Delugas, DiStasio, Ferretti, Floris, Fratesi, Fugallo, Gebauer, Gerstmann, Giustino, Gorni, Jia, Kawamura, Ko, Kokalj, Küçükbenli, Lazzeri, Marsili, Marzari, Mauri, Nguyen, Nguyen, Otero-de-la Roza, Paulatto, Poncé, Rocca, Sabatini, Santra, Schlipf, Seitsonen, Smogunov, Timrov, Thonhauser, Umari, Vast, Wu, and Baroni]{Giannozzi_2017}
Giannozzi,~P. \latin{et~al.}  Advanced capabilities for materials modelling with \textsc{Quantum Espresso}. \emph{Journal of Physics: Condensed Matter} \textbf{2017}, \emph{29}, 465901\relax
\mciteBstWouldAddEndPuncttrue
\mciteSetBstMidEndSepPunct{\mcitedefaultmidpunct}
{\mcitedefaultendpunct}{\mcitedefaultseppunct}\relax
\EndOfBibitem
\bibitem[Kresse and Furthmüller(1996)Kresse, and Furthmüller]{KRESSE199615}
Kresse,~G.; Furthmüller,~J. Efficiency of ab-initio total energy calculations for metals and semiconductors using a plane-wave basis set. \emph{Computational Materials Science} \textbf{1996}, \emph{6}, 15--50\relax
\mciteBstWouldAddEndPuncttrue
\mciteSetBstMidEndSepPunct{\mcitedefaultmidpunct}
{\mcitedefaultendpunct}{\mcitedefaultseppunct}\relax
\EndOfBibitem
\bibitem[Kresse and Hafner(1993)Kresse, and Hafner]{PhysRevB.47.558}
Kresse,~G.; Hafner,~J. Ab initio molecular dynamics for liquid metals. \emph{Phys. Rev. B} \textbf{1993}, \emph{47}, 558--561\relax
\mciteBstWouldAddEndPuncttrue
\mciteSetBstMidEndSepPunct{\mcitedefaultmidpunct}
{\mcitedefaultendpunct}{\mcitedefaultseppunct}\relax
\EndOfBibitem
\bibitem[Bousquet \latin{et~al.}(2011)Bousquet, Spaldin, and Delaney]{PhysRevLett.106.107202}
Bousquet,~E.; Spaldin,~N.~A.; Delaney,~K.~T. Unexpectedly Large Electronic Contribution to Linear Magnetoelectricity. \emph{Phys. Rev. Lett.} \textbf{2011}, \emph{106}, 107202\relax
\mciteBstWouldAddEndPuncttrue
\mciteSetBstMidEndSepPunct{\mcitedefaultmidpunct}
{\mcitedefaultendpunct}{\mcitedefaultseppunct}\relax
\EndOfBibitem
\bibitem[d’Avezac \latin{et~al.}(2007)d’Avezac, Marzari, and Mauri]{dAvezac2007}
d’Avezac,~M.; Marzari,~N.; Mauri,~F. Spin and orbital magnetic response in metals: Susceptibility and NMR shifts. \emph{Phys. Rev. B} \textbf{2007}, \emph{76}\relax
\mciteBstWouldAddEndPuncttrue
\mciteSetBstMidEndSepPunct{\mcitedefaultmidpunct}
{\mcitedefaultendpunct}{\mcitedefaultseppunct}\relax
\EndOfBibitem
\bibitem[Jensen and Mackintosh(1991)Jensen, and Mackintosh]{jensen1991rare}
Jensen,~J.; Mackintosh,~A. \emph{Rare Earth Magnetism: Structures and Excitations}; International Series of Monographs on Physics; Clarendon Press, 1991\relax
\mciteBstWouldAddEndPuncttrue
\mciteSetBstMidEndSepPunct{\mcitedefaultmidpunct}
{\mcitedefaultendpunct}{\mcitedefaultseppunct}\relax
\EndOfBibitem
\bibitem[Varsano \latin{et~al.}(2009)Varsano, Espinosa-Leal, Andrade, Marques, di~Felice, and Rubio]{Varsano2009}
Varsano,~D.; Espinosa-Leal,~L.~A.; Andrade,~X.; Marques,~M. A.~L.; di~Felice,~R.; Rubio,~A. Towards a gauge invariant method for molecular chiroptical properties in TDDFT. \emph{Phys. Chem. Chem. Phys.} \textbf{2009}, \emph{11}, 4481\relax
\mciteBstWouldAddEndPuncttrue
\mciteSetBstMidEndSepPunct{\mcitedefaultmidpunct}
{\mcitedefaultendpunct}{\mcitedefaultseppunct}\relax
\EndOfBibitem
\bibitem[Andrade(2010)]{Andrade2010}
Andrade,~X. Linear and non-linear response phenomena of molecular systems within time-dependent density functional theory. Ph.D.\ thesis, University of the Basque Country, Spain, 2010\relax
\mciteBstWouldAddEndPuncttrue
\mciteSetBstMidEndSepPunct{\mcitedefaultmidpunct}
{\mcitedefaultendpunct}{\mcitedefaultseppunct}\relax
\EndOfBibitem
\bibitem[Cho and Scheffler(1996)Cho, and Scheffler]{PhysRevB.53.10685}
Cho,~J.-H.; Scheffler,~M. Ab initio pseudopotential study of Fe, Co, and Ni employing the spin-polarized LAPW approach. \emph{Phys. Rev. B} \textbf{1996}, \emph{53}, 10685--10689\relax
\mciteBstWouldAddEndPuncttrue
\mciteSetBstMidEndSepPunct{\mcitedefaultmidpunct}
{\mcitedefaultendpunct}{\mcitedefaultseppunct}\relax
\EndOfBibitem
\bibitem[Stoeckl \latin{et~al.}(2021)Stoeckl, Swatek, and Wang]{10.1063/9.0000202}
Stoeckl,~P.; Swatek,~P.; Wang,~J.-P. Magnetocrystalline anisotropy of $\alpha^{\prime\prime}$–Fe16N2 under various DFT approaches. \emph{AIP Advances} \textbf{2021}, \emph{11}, 015039\relax
\mciteBstWouldAddEndPuncttrue
\mciteSetBstMidEndSepPunct{\mcitedefaultmidpunct}
{\mcitedefaultendpunct}{\mcitedefaultseppunct}\relax
\EndOfBibitem
\bibitem[He \latin{et~al.}(2017)He, Ma, Liu, Wang, and Chen]{10.1063/1.4992138}
He,~W.; Ma,~X.; Liu,~Z.; Wang,~Y.; Chen,~L.-Q. Magnetic anisotropy energy of ferromagnetic shape memory alloys Ni2X(X=Fe, Co)Ga by first-principles calculations. \emph{AIP Advances} \textbf{2017}, \emph{7}, 075001\relax
\mciteBstWouldAddEndPuncttrue
\mciteSetBstMidEndSepPunct{\mcitedefaultmidpunct}
{\mcitedefaultendpunct}{\mcitedefaultseppunct}\relax
\EndOfBibitem
\bibitem[Xu \latin{et~al.}(2024)Xu, Li, Huynh, Fadel, Huang, Sundararaman, Vardeny, and Ping]{Xu2024-lw}
Xu,~J.; Li,~K.; Huynh,~U.~N.; Fadel,~M.; Huang,~J.; Sundararaman,~R.; Vardeny,~V.; Ping,~Y. How spin relaxes and dephases in bulk halide perovskites. \emph{Nat. Commun.} \textbf{2024}, \emph{15}, 188\relax
\mciteBstWouldAddEndPuncttrue
\mciteSetBstMidEndSepPunct{\mcitedefaultmidpunct}
{\mcitedefaultendpunct}{\mcitedefaultseppunct}\relax
\EndOfBibitem
\bibitem[Koopmans \latin{et~al.}(2000)Koopmans, van Kampen, Kohlhepp, and de~Jonge]{PhysRevLett.85.844}
Koopmans,~B.; van Kampen,~M.; Kohlhepp,~J.~T.; de~Jonge,~W. J.~M. Ultrafast Magneto-Optics in Nickel: Magnetism or Optics? \emph{Phys. Rev. Lett.} \textbf{2000}, \emph{85}, 844--847\relax
\mciteBstWouldAddEndPuncttrue
\mciteSetBstMidEndSepPunct{\mcitedefaultmidpunct}
{\mcitedefaultendpunct}{\mcitedefaultseppunct}\relax
\EndOfBibitem
\bibitem[Kirilyuk \latin{et~al.}(2010)Kirilyuk, Kimel, and Rasing]{RevModPhys.82.2731}
Kirilyuk,~A.; Kimel,~A.~V.; Rasing,~T. Ultrafast optical manipulation of magnetic order. \emph{Rev. Mod. Phys.} \textbf{2010}, \emph{82}, 2731--2784\relax
\mciteBstWouldAddEndPuncttrue
\mciteSetBstMidEndSepPunct{\mcitedefaultmidpunct}
{\mcitedefaultendpunct}{\mcitedefaultseppunct}\relax
\EndOfBibitem
\bibitem[Simoni \latin{et~al.}(2017)Simoni, Stamenova, and Sanvito]{PhysRevB.96.054411}
Simoni,~J.; Stamenova,~M.; Sanvito,~S. Ab initio dynamical exchange interactions in frustrated antiferromagnets. \emph{Phys. Rev. B} \textbf{2017}, \emph{96}, 054411\relax
\mciteBstWouldAddEndPuncttrue
\mciteSetBstMidEndSepPunct{\mcitedefaultmidpunct}
{\mcitedefaultendpunct}{\mcitedefaultseppunct}\relax
\EndOfBibitem
\bibitem[Simoni \latin{et~al.}(2017)Simoni, Stamenova, and Sanvito]{PhysRevB.95.024412}
Simoni,~J.; Stamenova,~M.; Sanvito,~S. Ultrafast demagnetizing fields from first principles. \emph{Phys. Rev. B} \textbf{2017}, \emph{95}, 024412\relax
\mciteBstWouldAddEndPuncttrue
\mciteSetBstMidEndSepPunct{\mcitedefaultmidpunct}
{\mcitedefaultendpunct}{\mcitedefaultseppunct}\relax
\EndOfBibitem
\bibitem[Krieger \latin{et~al.}(2017)Krieger, Elliott, Müller, Singh, Dewhurst, Gross, and Sharma]{Krieger_2017}
Krieger,~K.; Elliott,~P.; Müller,~T.; Singh,~N.; Dewhurst,~J.~K.; Gross,~E. K.~U.; Sharma,~S. Ultrafast demagnetization in bulk versus thin films: an ab initio study. \emph{Journal of Physics: Condensed Matter} \textbf{2017}, \emph{29}, 224001\relax
\mciteBstWouldAddEndPuncttrue
\mciteSetBstMidEndSepPunct{\mcitedefaultmidpunct}
{\mcitedefaultendpunct}{\mcitedefaultseppunct}\relax
\EndOfBibitem
\bibitem[Pellegrini \latin{et~al.}(2022)Pellegrini, Sharma, Dewhurst, and Sanna]{PhysRevB.105.134425}
Pellegrini,~C.; Sharma,~S.; Dewhurst,~J.~K.; Sanna,~A. Ab initio study of ultrafast demagnetization of elementary ferromagnets by terahertz versus optical pulses. \emph{Phys. Rev. B} \textbf{2022}, \emph{105}, 134425\relax
\mciteBstWouldAddEndPuncttrue
\mciteSetBstMidEndSepPunct{\mcitedefaultmidpunct}
{\mcitedefaultendpunct}{\mcitedefaultseppunct}\relax
\EndOfBibitem
\bibitem[Krieger \latin{et~al.}(2015)Krieger, Dewhurst, Elliott, Sharma, and Gross]{Krieger_2015}
Krieger,~K.; Dewhurst,~J.~K.; Elliott,~P.; Sharma,~S.; Gross,~E. K.~U. Laser-Induced Demagnetization at Ultrashort Time Scales: Predictions of TDDFT. \emph{J. Chem. Theory Comput.} \textbf{2015}, \emph{11}, 4870--4874\relax
\mciteBstWouldAddEndPuncttrue
\mciteSetBstMidEndSepPunct{\mcitedefaultmidpunct}
{\mcitedefaultendpunct}{\mcitedefaultseppunct}\relax
\EndOfBibitem
\bibitem[Xu \latin{et~al.}(2021)Xu, Habib, Sundararaman, and Ping]{Xu2021-eo}
Xu,~J.; Habib,~A.; Sundararaman,~R.; Ping,~Y. Ab initio ultrafast spin dynamics in solids. \emph{Phys. Rev. B Condens. Matter} \textbf{2021}, \emph{104}, 184418\relax
\mciteBstWouldAddEndPuncttrue
\mciteSetBstMidEndSepPunct{\mcitedefaultmidpunct}
{\mcitedefaultendpunct}{\mcitedefaultseppunct}\relax
\EndOfBibitem
\bibitem[Xu and Ping(2024)Xu, and Ping]{Xu2024-cb}
Xu,~J.; Ping,~Y. Ab Initio Predictions of Spin Relaxation, Dephasing, and Diffusion in Solids. \emph{J. Chem. Theory Comput.} \textbf{2024}, \emph{20}, 492--512\relax
\mciteBstWouldAddEndPuncttrue
\mciteSetBstMidEndSepPunct{\mcitedefaultmidpunct}
{\mcitedefaultendpunct}{\mcitedefaultseppunct}\relax
\EndOfBibitem
\bibitem[Li \latin{et~al.}(2024)Li, Xu, Huynh, Bodin, Gupta, Multunas, Simoni, Sundararaman, Verdany, and Ping]{Li2024-lu}
Li,~K.; Xu,~J.; Huynh,~U.~N.; Bodin,~R.; Gupta,~M.; Multunas,~C.; Simoni,~J.; Sundararaman,~R.; Verdany,~Z.~V.; Ping,~Y. Spin dynamics in hybrid Halide perovskites - effect of dynamical and permanent symmetry breaking. \emph{J. Phys. Chem. Lett.} \textbf{2024}, \emph{15}, 12156--12163\relax
\mciteBstWouldAddEndPuncttrue
\mciteSetBstMidEndSepPunct{\mcitedefaultmidpunct}
{\mcitedefaultendpunct}{\mcitedefaultseppunct}\relax
\EndOfBibitem
\bibitem[Fern{\'a}ndez-Seivane \latin{et~al.}(2007)Fern{\'a}ndez-Seivane, Oliveira, Sanvito, and Ferrer]{Fernandez-Seivane_2007}
Fern{\'a}ndez-Seivane,~F.; Oliveira,~M.~A.; Sanvito,~S.; Ferrer,~J. On-site approximation for spin–orbit coupling in linear combination of atomic orbitals density functional methods. \emph{Journal of Physics: Condensed Matter} \textbf{2007}, \emph{19}, 489001\relax
\mciteBstWouldAddEndPuncttrue
\mciteSetBstMidEndSepPunct{\mcitedefaultmidpunct}
{\mcitedefaultendpunct}{\mcitedefaultseppunct}\relax
\EndOfBibitem
\bibitem[Isaacs and Wolverton(2018)Isaacs, and Wolverton]{PhysRevMaterials.2.063801}
Isaacs,~E.~B.; Wolverton,~C. Performance of the strongly constrained and appropriately normed density functional for solid-state materials. \emph{Phys. Rev. Mater.} \textbf{2018}, \emph{2}, 063801\relax
\mciteBstWouldAddEndPuncttrue
\mciteSetBstMidEndSepPunct{\mcitedefaultmidpunct}
{\mcitedefaultendpunct}{\mcitedefaultseppunct}\relax
\EndOfBibitem
\bibitem[Sharma \latin{et~al.}(2018)Sharma, Gross, Sanna, and Dewhurst]{doi:10.1021/acs.jctc.7b01049}
Sharma,~S.; Gross,~E. K.~U.; Sanna,~A.; Dewhurst,~J.~K. Source-Free Exchange-Correlation Magnetic Fields in Density Functional Theory. \emph{J. Chem. Theory Comput.} \textbf{2018}, \emph{14}, 1247--1253\relax
\mciteBstWouldAddEndPuncttrue
\mciteSetBstMidEndSepPunct{\mcitedefaultmidpunct}
{\mcitedefaultendpunct}{\mcitedefaultseppunct}\relax
\EndOfBibitem
\bibitem[Moore \latin{et~al.}(2025)Moore, Horton, Kaplan, Ashour, Griffin, and Persson]{PhysRevB.111.094417}
Moore,~G.~C.; Horton,~M.~K.; Kaplan,~A.~D.; Ashour,~O.~A.; Griffin,~S.~M.; Persson,~K.~A. Noncollinear ground states of solids with a source-free exchange correlation functional. \emph{Phys. Rev. B} \textbf{2025}, \emph{111}, 094417\relax
\mciteBstWouldAddEndPuncttrue
\mciteSetBstMidEndSepPunct{\mcitedefaultmidpunct}
{\mcitedefaultendpunct}{\mcitedefaultseppunct}\relax
\EndOfBibitem
\bibitem[Rivero \latin{et~al.}(2009)Rivero, Moreira, Scuseria, and Illas]{PhysRevB.79.245129}
Rivero,~P.; Moreira,~I. d. P.~R.; Scuseria,~G.~E.; Illas,~F. Description of magnetic interactions in strongly correlated solids via range-separated hybrid functionals. \emph{Phys. Rev. B} \textbf{2009}, \emph{79}, 245129\relax
\mciteBstWouldAddEndPuncttrue
\mciteSetBstMidEndSepPunct{\mcitedefaultmidpunct}
{\mcitedefaultendpunct}{\mcitedefaultseppunct}\relax
\EndOfBibitem
\bibitem[Rivero \latin{et~al.}(2008)Rivero, Moreira, Illas, and Scuseria]{10.1063/1.3006419}
Rivero,~P.; Moreira,~I. d. P.~R.; Illas,~F.; Scuseria,~G.~E. Reliability of range-separated hybrid functionals for describing magnetic coupling in molecular systems. \emph{The Journal of Chemical Physics} \textbf{2008}, \emph{129}, 184110\relax
\mciteBstWouldAddEndPuncttrue
\mciteSetBstMidEndSepPunct{\mcitedefaultmidpunct}
{\mcitedefaultendpunct}{\mcitedefaultseppunct}\relax
\EndOfBibitem
\bibitem[Mu\~noz \latin{et~al.}(2004)Mu\~noz, Harrison, and Illas]{PhysRevB.69.085115}
Mu\~noz,~D.; Harrison,~N.~M.; Illas,~F. Electronic and magnetic structure of ${\mathrm{LaMnO}}_{3}$ from hybrid periodic density-functional theory. \emph{Phys. Rev. B} \textbf{2004}, \emph{69}, 085115\relax
\mciteBstWouldAddEndPuncttrue
\mciteSetBstMidEndSepPunct{\mcitedefaultmidpunct}
{\mcitedefaultendpunct}{\mcitedefaultseppunct}\relax
\EndOfBibitem
\bibitem[Elhanoty \latin{et~al.}(2022)Elhanoty, Eriksson, Knut, Karis, and Gr\aa{}n\"as]{PhysRevB.105.L100401}
Elhanoty,~M.~F.; Eriksson,~O.; Knut,~R.; Karis,~O.; Gr\aa{}n\"as,~O. Element-selective ultrafast magnetization dynamics of hybrid Stoner-Heisenberg magnets. \emph{Phys. Rev. B} \textbf{2022}, \emph{105}, L100401\relax
\mciteBstWouldAddEndPuncttrue
\mciteSetBstMidEndSepPunct{\mcitedefaultmidpunct}
{\mcitedefaultendpunct}{\mcitedefaultseppunct}\relax
\EndOfBibitem
\bibitem[Wijewardane and Ullrich(2005)Wijewardane, and Ullrich]{PhysRevLett.95.086401}
Wijewardane,~H.~O.; Ullrich,~C.~A. Time-Dependent Kohn-Sham Theory with Memory. \emph{Phys. Rev. Lett.} \textbf{2005}, \emph{95}, 086401\relax
\mciteBstWouldAddEndPuncttrue
\mciteSetBstMidEndSepPunct{\mcitedefaultmidpunct}
{\mcitedefaultendpunct}{\mcitedefaultseppunct}\relax
\EndOfBibitem
\bibitem[Vignale and Kohn(1996)Vignale, and Kohn]{PhysRevLett.77.2037}
Vignale,~G.; Kohn,~W. Current-Dependent Exchange-Correlation Potential for Dynamical Linear Response Theory. \emph{Phys. Rev. Lett.} \textbf{1996}, \emph{77}, 2037--2040\relax
\mciteBstWouldAddEndPuncttrue
\mciteSetBstMidEndSepPunct{\mcitedefaultmidpunct}
{\mcitedefaultendpunct}{\mcitedefaultseppunct}\relax
\EndOfBibitem
\bibitem[Ullrich(2006)]{10.1063/1.2406069}
Ullrich,~C.~A. Time-dependent density-functional theory beyond the adiabatic approximation: Insights from a two-electron model system. \emph{The Journal of Chemical Physics} \textbf{2006}, \emph{125}, 234108\relax
\mciteBstWouldAddEndPuncttrue
\mciteSetBstMidEndSepPunct{\mcitedefaultmidpunct}
{\mcitedefaultendpunct}{\mcitedefaultseppunct}\relax
\EndOfBibitem
\bibitem[Andrade \latin{et~al.}(2009)Andrade, Castro, Zueco, Alonso, Echenique, Falceto, and Rubio]{Andrade2009}
Andrade,~X.; Castro,~A.; Zueco,~D.; Alonso,~J.~L.; Echenique,~P.; Falceto,~F.; Rubio,~A. Modified Ehrenfest Formalism for Efficient Large-Scale ab initio Molecular Dynamics. \emph{J. Chem. Theo. Comput.} \textbf{2009}, \emph{5}, 728–742\relax
\mciteBstWouldAddEndPuncttrue
\mciteSetBstMidEndSepPunct{\mcitedefaultmidpunct}
{\mcitedefaultendpunct}{\mcitedefaultseppunct}\relax
\EndOfBibitem
\bibitem[Sangalli \latin{et~al.}(2012)Sangalli, Marini, and Debernardi]{PhysRevB.86.125139}
Sangalli,~D.; Marini,~A.; Debernardi,~A. Pseudopotential-based first-principles approach to the magneto-optical Kerr effect: From metals to the inclusion of local fields and excitonic effects. \emph{Phys. Rev. B} \textbf{2012}, \emph{86}, 125139\relax
\mciteBstWouldAddEndPuncttrue
\mciteSetBstMidEndSepPunct{\mcitedefaultmidpunct}
{\mcitedefaultendpunct}{\mcitedefaultseppunct}\relax
\EndOfBibitem
\bibitem[Molina-S{\'a}nchez \latin{et~al.}(2020)Molina-S{\'a}nchez, Catarina, Sangalli, and Fernández-Rossier]{D0TC01322F}
Molina-S{\'a}nchez,~A.; Catarina,~G.; Sangalli,~D.; Fernández-Rossier,~J. Magneto-optical response of chromium trihalide monolayers: chemical trends. \emph{J. Mater. Chem. C} \textbf{2020}, \emph{8}, 8856--8863\relax
\mciteBstWouldAddEndPuncttrue
\mciteSetBstMidEndSepPunct{\mcitedefaultmidpunct}
{\mcitedefaultendpunct}{\mcitedefaultseppunct}\relax
\EndOfBibitem
\bibitem[Ebert(1996)]{Ebert_1996}
Ebert,~H. Magneto-optical effects in transition metal systems. \emph{Reports on Progress in Physics} \textbf{1996}, \emph{59}, 1665\relax
\mciteBstWouldAddEndPuncttrue
\mciteSetBstMidEndSepPunct{\mcitedefaultmidpunct}
{\mcitedefaultendpunct}{\mcitedefaultseppunct}\relax
\EndOfBibitem
\bibitem[Kuneš(2004)]{Kunes_2004}
Kuneš,~J. Magnetism and Magneto-Optics in DFT. \emph{Physica Scripta} \textbf{2004}, \emph{2004}, 116\relax
\mciteBstWouldAddEndPuncttrue
\mciteSetBstMidEndSepPunct{\mcitedefaultmidpunct}
{\mcitedefaultendpunct}{\mcitedefaultseppunct}\relax
\EndOfBibitem
\bibitem[Mason(2003)]{Mason2003}
Mason,~W. \emph{Comprehensive Coordination Chemistry II}; Elsevier, 2003; p 327–337\relax
\mciteBstWouldAddEndPuncttrue
\mciteSetBstMidEndSepPunct{\mcitedefaultmidpunct}
{\mcitedefaultendpunct}{\mcitedefaultseppunct}\relax
\EndOfBibitem
\bibitem[Lee \latin{et~al.}(2011)Lee, Yabana, and Bertsch]{Lee2011}
Lee,~K.-M.; Yabana,~K.; Bertsch,~G.~F. Magnetic circular dichroism in real-time time-dependent density functional theory. \emph{J. Chem. Phys.} \textbf{2011}, \emph{134}\relax
\mciteBstWouldAddEndPuncttrue
\mciteSetBstMidEndSepPunct{\mcitedefaultmidpunct}
{\mcitedefaultendpunct}{\mcitedefaultseppunct}\relax
\EndOfBibitem
\bibitem[Choi \latin{et~al.}(2014)Choi, Min, Lee, and Cahill]{Choi2014}
Choi,~G.-M.; Min,~B.-C.; Lee,~K.-J.; Cahill,~D.~G. Spin current generated by thermally driven ultrafast demagnetization. \emph{Nature Communications} \textbf{2014}, \emph{5}, 4334\relax
\mciteBstWouldAddEndPuncttrue
\mciteSetBstMidEndSepPunct{\mcitedefaultmidpunct}
{\mcitedefaultendpunct}{\mcitedefaultseppunct}\relax
\EndOfBibitem
\bibitem[Schneider \latin{et~al.}(2020)Schneider, Pfau, G\"unther, von Korff~Schmising, Weder, Geilhufe, Perron, Capotondi, Pedersoli, Manfredda, Hennecke, Vodungbo, L\"uning, and Eisebitt]{PhysRevLett.125.127201}
Schneider,~M.; Pfau,~B.; G\"unther,~C.~M.; von Korff~Schmising,~C.; Weder,~D.; Geilhufe,~J.; Perron,~J.; Capotondi,~F.; Pedersoli,~E.; Manfredda,~M.; Hennecke,~M.; Vodungbo,~B.; L\"uning,~J.; Eisebitt,~S. Ultrafast Demagnetization Dominates Fluence Dependence of Magnetic Scattering at Co $M$ Edges. \emph{Phys. Rev. Lett.} \textbf{2020}, \emph{125}, 127201\relax
\mciteBstWouldAddEndPuncttrue
\mciteSetBstMidEndSepPunct{\mcitedefaultmidpunct}
{\mcitedefaultendpunct}{\mcitedefaultseppunct}\relax
\EndOfBibitem
\bibitem[Tengdin \latin{et~al.}(2018)Tengdin, You, Chen, Shi, Zusin, Zhang, Gentry, Blonsky, Keller, Oppeneer, Kapteyn, Tao, and Murnane]{29511738}
Tengdin,~P.; You,~W.; Chen,~C.; Shi,~X.; Zusin,~D.; Zhang,~Y.; Gentry,~C.; Blonsky,~A.; Keller,~M.; Oppeneer,~P.~M.; Kapteyn,~H.~C.; Tao,~Z.; Murnane,~M.~M. Critical behavior within 20 fs drives the out-of-equilibrium laser-induced magnetic phase transition in nickel. \emph{Science advances} \textbf{2018}, \emph{4}, eaap9744\relax
\mciteBstWouldAddEndPuncttrue
\mciteSetBstMidEndSepPunct{\mcitedefaultmidpunct}
{\mcitedefaultendpunct}{\mcitedefaultseppunct}\relax
\EndOfBibitem
\bibitem[Roth \latin{et~al.}(2012)Roth, Schellekens, Alebrand, Schmitt, Steil, Koopmans, Cinchetti, and Aeschlimann]{PhysRevX.2.021006}
Roth,~T.; Schellekens,~A.~J.; Alebrand,~S.; Schmitt,~O.; Steil,~D.; Koopmans,~B.; Cinchetti,~M.; Aeschlimann,~M. Temperature Dependence of Laser-Induced Demagnetization in Ni: A Key for Identifying the Underlying Mechanism. \emph{Phys. Rev. X} \textbf{2012}, \emph{2}, 021006\relax
\mciteBstWouldAddEndPuncttrue
\mciteSetBstMidEndSepPunct{\mcitedefaultmidpunct}
{\mcitedefaultendpunct}{\mcitedefaultseppunct}\relax
\EndOfBibitem
\bibitem[Mathias \latin{et~al.}(2012)Mathias, La-O-Vorakiat, Grychtol, Granitzka, Turgut, Shaw, Adam, Nembach, Siemens, Eich, Schneider, Silva, Aeschlimann, Murnane, and Kapteyn]{doi:10.1073/pnas.1201371109}
Mathias,~S.; La-O-Vorakiat,~C.; Grychtol,~P.; Granitzka,~P.; Turgut,~E.; Shaw,~J.~M.; Adam,~R.; Nembach,~H.~T.; Siemens,~M.~E.; Eich,~S.; Schneider,~C.~M.; Silva,~T.~J.; Aeschlimann,~M.; Murnane,~M.~M.; Kapteyn,~H.~C. Probing the timescale of the exchange interaction in a ferromagnetic alloy. \emph{Proceedings of the National Academy of Sciences} \textbf{2012}, \emph{109}, 4792--4797\relax
\mciteBstWouldAddEndPuncttrue
\mciteSetBstMidEndSepPunct{\mcitedefaultmidpunct}
{\mcitedefaultendpunct}{\mcitedefaultseppunct}\relax
\EndOfBibitem
\bibitem[Ulrichs and Razdolski(2018)Ulrichs, and Razdolski]{PhysRevB.98.054429}
Ulrichs,~H.; Razdolski,~I. Micromagnetic view on ultrafast magnon generation by femtosecond spin current pulses. \emph{Phys. Rev. B} \textbf{2018}, \emph{98}, 054429\relax
\mciteBstWouldAddEndPuncttrue
\mciteSetBstMidEndSepPunct{\mcitedefaultmidpunct}
{\mcitedefaultendpunct}{\mcitedefaultseppunct}\relax
\EndOfBibitem
\bibitem[Au \latin{et~al.}(2013)Au, Dvornik, Davison, Ahmad, Keatley, Vansteenkiste, Van~Waeyenberge, and Kruglyak]{PhysRevLett.110.097201}
Au,~Y.; Dvornik,~M.; Davison,~T.; Ahmad,~E.; Keatley,~P.~S.; Vansteenkiste,~A.; Van~Waeyenberge,~B.; Kruglyak,~V.~V. Direct Excitation of Propagating Spin Waves by Focused Ultrashort Optical Pulses. \emph{Phys. Rev. Lett.} \textbf{2013}, \emph{110}, 097201\relax
\mciteBstWouldAddEndPuncttrue
\mciteSetBstMidEndSepPunct{\mcitedefaultmidpunct}
{\mcitedefaultendpunct}{\mcitedefaultseppunct}\relax
\EndOfBibitem
\bibitem[Iacocca \latin{et~al.}(2019)Iacocca, Liu, Reid, Fu, Ruta, Granitzka, Jal, Bonetti, Gray, Graves, Kukreja, Chen, Higley, Chase, Le~Guyader, Hirsch, Ohldag, Schlotter, Dakovski, Coslovich, Hoffmann, Carron, Tsukamoto, Kirilyuk, Kimel, Rasing, St{\"o}hr, Evans, Ostler, Chantrell, Hoefer, Silva, and D{\"u}rr]{Iacocca2019}
Iacocca,~E. \latin{et~al.}  Spin-current-mediated rapid magnon localisation and coalescence after ultrafast optical pumping of ferrimagnetic alloys. \emph{Nature Communications} \textbf{2019}, \emph{10}, 1756\relax
\mciteBstWouldAddEndPuncttrue
\mciteSetBstMidEndSepPunct{\mcitedefaultmidpunct}
{\mcitedefaultendpunct}{\mcitedefaultseppunct}\relax
\EndOfBibitem
\bibitem[Mizukami \latin{et~al.}(2010)Mizukami, Tsunegi, Kubota, Oogane, Watanabe, Naganuma, Ando, and Miyazaki]{Mizukami_2010}
Mizukami,~S.; Tsunegi,~S.; Kubota,~T.; Oogane,~M.; Watanabe,~D.; Naganuma,~H.; Ando,~Y.; Miyazaki,~T. Ultrafast demagnetization for Ni80Fe20 and half-metallic Co2MnSi heusler alloy films. \emph{Journal of Physics: Conference Series} \textbf{2010}, \emph{200}, 042017\relax
\mciteBstWouldAddEndPuncttrue
\mciteSetBstMidEndSepPunct{\mcitedefaultmidpunct}
{\mcitedefaultendpunct}{\mcitedefaultseppunct}\relax
\EndOfBibitem
\bibitem[X.~Andrade(2025)]{inq_gitlab}
X.~Andrade,~e.~a. INQ: TDDFT numerical simulations on HPC and GPUs. \url{https://gitlab.com/NPNEQ/INQ}, 2025; Accessed: 2025-07-17\relax
\mciteBstWouldAddEndPuncttrue
\mciteSetBstMidEndSepPunct{\mcitedefaultmidpunct}
{\mcitedefaultendpunct}{\mcitedefaultseppunct}\relax
\EndOfBibitem
\end{mcitethebibliography}

\end{document}